# Research Opportunities in Plasma Astrophysics

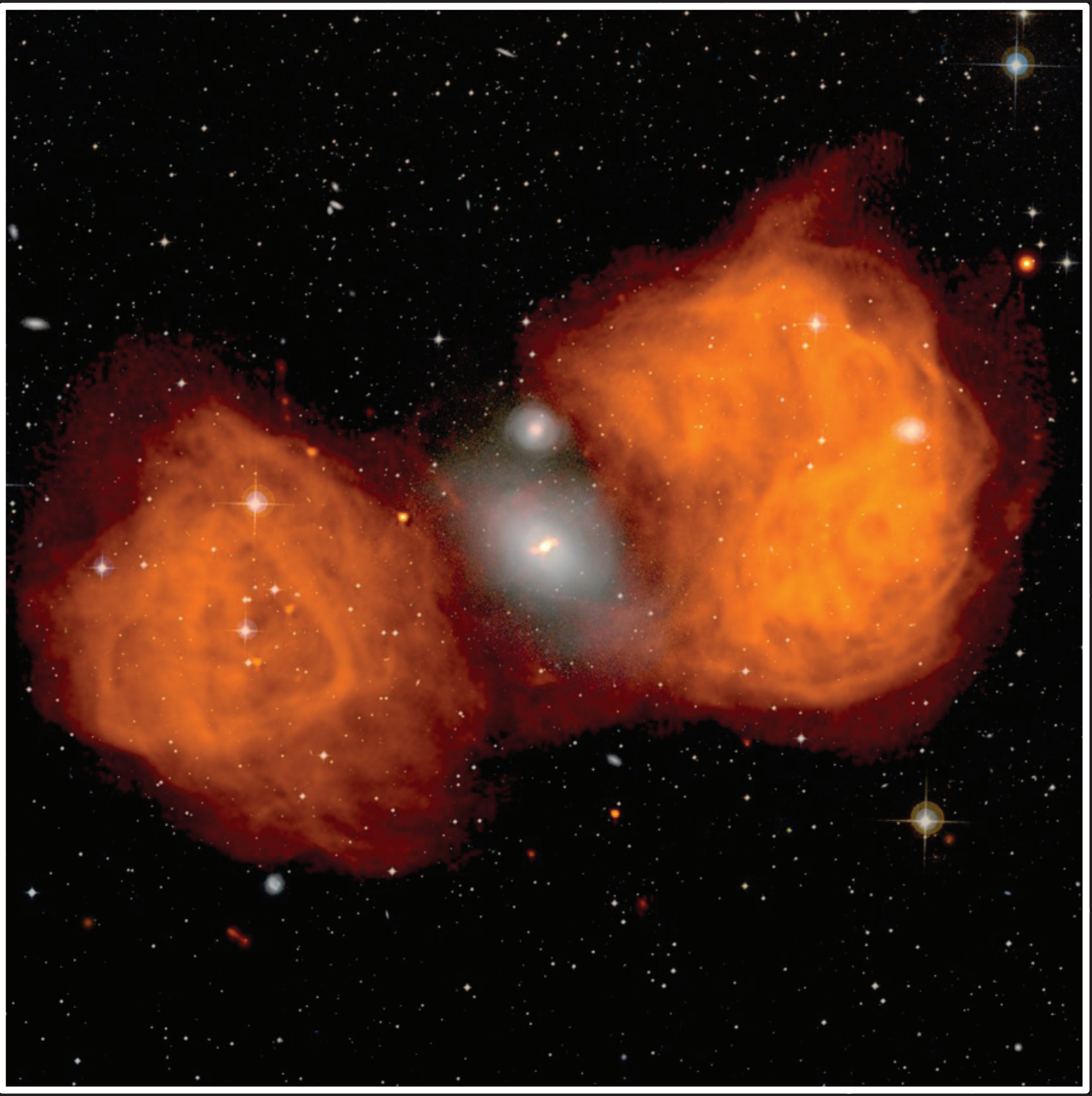

*Report of the Workshop on Opportunities in Plasma Astrophysics*
*Princeton, New Jersey — January 18-21, 2010*

*ON THE COVER*

*More than 100 million years ago, the giant elliptical galaxy, NGC1316 (center of the image), began devouring its small northern neighbor. The complex radio emission, associated with this encounter (called Fornax A, shown in orange) was imaged using the Very Large Array in New Mexico. This image shows the radio emission superimposed on an optical image (STScI/POSS-II). The radio emission consists of two large radio lobes, each about 600,000 light years across. As NGC 1316 cannibalizes the smaller galaxy, it strips away material that spirals toward a black hole at the center of the giant elliptical, which produces rings and asymmetries in NGC 1316. Friction from the in-falling matter generates a ten-million-degree plasma surrounding the black hole that emits an enormous amount of light and X-rays. By a magnetic focusing mechanism, not yet understood, high-energy particles are "beamed" away from the hot plasma in opposite directions – faint radio leakage (shown by the short, orange jets near the galaxy's heart) hint at this process. The flow of material is essentially invisible until the particles smash into tenuous material some 500,000 light years from the galaxy, producing the strong, orange radio-emission. Slow changes in the "beam's" direction as well as the dynamical influence of the magnetic field produce the intricate patterns we see in the radio lobes. These patterns have persisted over tens of millions of years.*

*(Image courtesy of NRAO/AUI and J. M. Uson.)*

# Research Opportunities in Plasma Astrophysics

## REPORT OF THE WORKSHOP ON OPPORTUNITIES IN PLASMA ASTROPHYSICS

**WOPA Committee:**

STUART BALE, University of California, Berkeley
AMITAVA BHATTACHARJEE, University of New Hampshire
FAUSTO CATTANEO, (Co-Chair) University of Chicago
JAMES DRAKE, University of Maryland
HANTAO JI, (Co-Chair) Princeton Plasma Physics Laboratory
MARTY LEE, University of New Hampshire
HUI LI, Los Alamos National Laboratory
EDISON LIANG, Rice University
MARC POUND, University of Maryland
STEWART PRAGER, (Co-Chair) Princeton Plasma Physics Laboratory
ELIOT QUATAERT, University of California, Berkeley
BRUCE REMINGTON, Lawrence Livermore National Laboratory
ROBERT ROSNER, (Co-Chair) University of Chicago
DMITRI RYUTOV, Lawrence Livermore National Laboratory
EDWARD THOMAS, Jr., Auburn University
ELLEN ZWEIBEL, University of Wisconsin-Madison

**PUBLICATION**

**Editor:** PATTI WIESER, Princeton Plasma Physics Laboratory

**Design/Layout:** GREG CZECHOWICZ, Princeton Plasma Physics Laboratory

**Editorial Assistants:** MARCO LANAVE and RAPHAEL ROSEN, Princeton Plasma Physics Laboratory

**Conference Assistance:** CAROL AUSTIN, TERRY GREENBERG, and JENNIFER JONES
Princeton Plasma Physics Laboratory

**The workshop and report production were supported by:**

U.S. Department of Energy, Office of Fusion Energy Sciences
NASA Heliophysics and Astrophysics Science Divisions
National Science Foundation
American Physical Society, Topical Group in Plasma Astrophysics
American Physical Society, Division of Plasma Physics
Center for Magnetic Self-Organization in Laboratory and Astrophysical Plasmas

*This Report is available on the WOPA web site at: http://www.pppl.gov/conferences/2010/WOPA/index.html*

# TABLE OF CONTENTS

# EXECUTIVE SUMMARY

## INTRODUCTION

Plasma pervades the universe at all measurable scales. At the very small scale, coupled processes in plasmas determine the behavior of the solar system. The Sun rotates, generates magnetic fields, and ejects mass in part because of plasma processes. The ejected plasma expands as the solar wind toward the Earth, becoming turbulent and hot as it travels. It then encounters and becomes trapped in the Earth's magnetic field, causes shocks, and produces magnetic substorms, aurora, and other plasma phenomena. This Sun-Earth system spans the short distance of $10^{-4}$ light years. Jumping ten orders of magnitude in size, extra-galactic jet systems are among the largest plasma structures in the universe. They begin with a rotating, accretion disk surrounding a supermassive black hole. Plasma transport processes determine the rate of accretion of matter onto the black hole, while producing the most luminous source of energy in the universe. The rotating plasma also launches a collimated jet that travels distances in the range of one million light years, ending in confined lobes of plasma. Dynamics of astrophysical systems at all scales between the solar system and jets are similarly regulated by plasma physics.

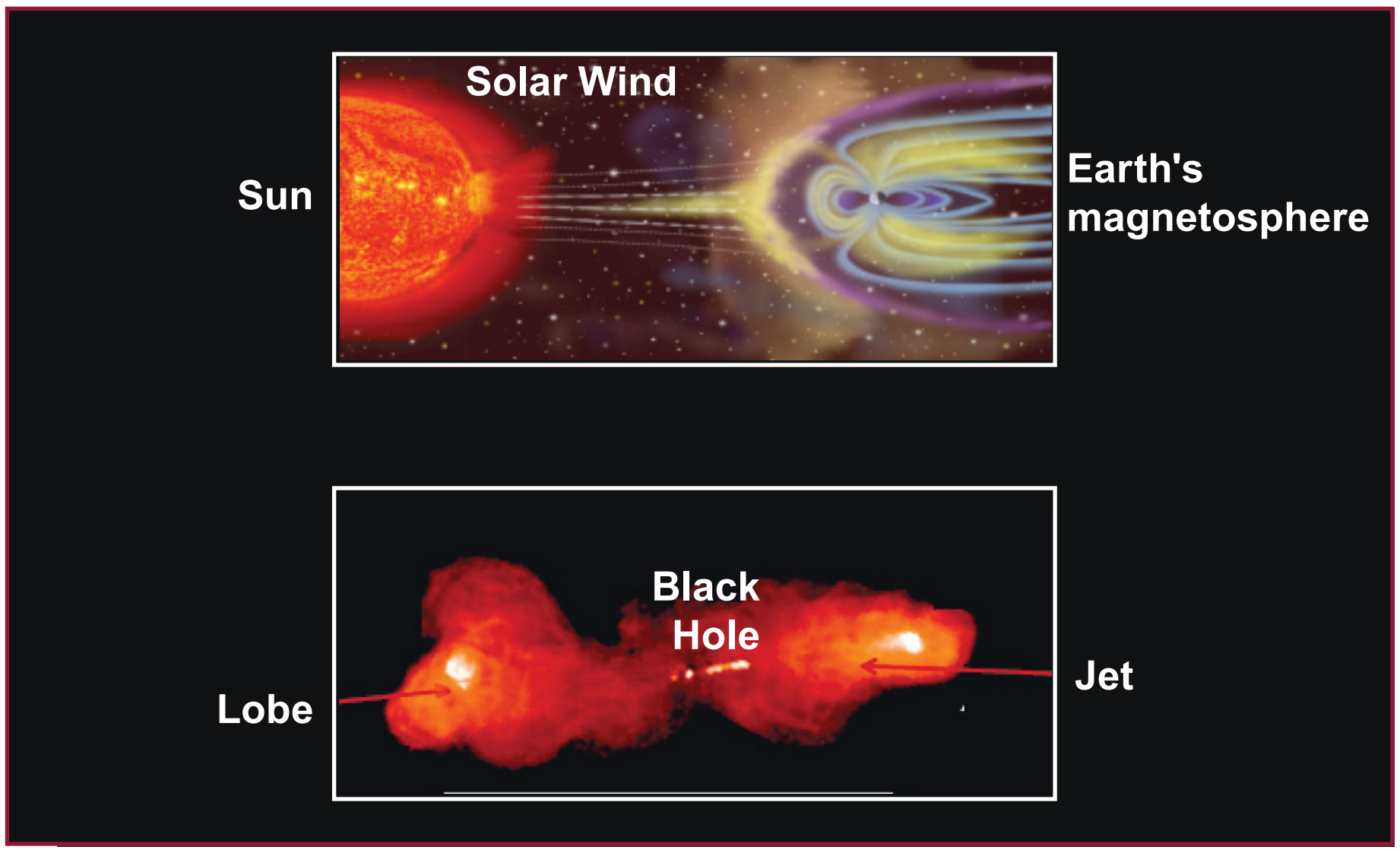

***Plasma structures span all spatial scales in the universe, from the small scale of the solar system (shown in the top figure as an artist's sketch of solar wind spanning ~$10^{-4}$ light years) to the large scale of extra-galactic jets, ~$10^6$ light years (shown in the bottom figure from observation).***

The study of plasmas beyond the Earth's atmosphere is here denoted as *plasma astrophysics*. This definition encompasses the usual realm of astrophysics (beyond the solar system), but also the domain of space physics (the Sun, the Heliosphere, and the magnetospheres of the Earth and the planets). The power of plasma astrophysics is that the same fundamental plasma processes appear in many different venues. For example, magnetic reconnection can drive not only magnetic substorms in the Earth's magnetosphere, but also flares on the surfaces of distant stars. The usual distinction between astrophysics and space physics disappears when viewed through the unifying lens of plasma physics.

Plasma astrophysics is positioned for rapid advance resulting from huge strides in astronomical observations, plasma technology, diagnostics, and plasma computation, combined with the maturity of plasma physics. Satellite and ground-based observations are set to measure plasma processes long invisible to us, from the solar interior to accretion disks. In-situ measurements of local plasma properties, including both field and particle information, have been expanded into every corner of our solar system: Earth's magnetosphere, the solar wind, other planets' magnetospheres, and the boundaries of our solar system. Multiple, coordinated satellites have greatly improved spatial and temporal resolution of magnetospheric measurements. Remote-sensing observations from both space and ground have moved beyond the traditional visible wavelengths to almost every wavelength band from far infrared emission from cold, partially ionized plasmas during star formation, to hard X-ray emission from extremely hot relativistic plasmas around supermassive black holes. High-power lasers now produce new plasma regimes with high energy density. These laser-produced, warm or hot, dense plasmas are similar to the interiors of giant planet cores and to the plasma that surrounds compact objects.

Plasma diagnostics can now measure, often remotely and non-perturbatively, a huge range of key particle and field quantities in the laboratory, both at the large scale and the small scale characteristic of turbulence. Modern techniques include laser scattering, laser-induced fluorescence, laser Faraday rotation, active spectroscopy using injected neutral atoms, miniaturized insertable probes, and electron cyclotron imaging techniques, to name a few. These provide new windows to detailed properties of magnetic fields, electric fields, electron and ion densities, plasma flow, and aspects of particle distribution functions. Advances in computation are revolutionizing how we study the complex behavior of plasmas. The surge in available computational power is being coupled with expansion of physics captured in computational models. Many plasma phenomena are governed by coupling between the large scale of the plasma system and the small scale characterized by microscopic plasma quantities (such as the particle gyroradius). New computational models can now treat this coupling, whether in multi-fluid treatments or new approaches that solve kinetic equations.

The opportunities represented by these technical advances can only be fully realized through a coordinated effort that brings together the communities of astrophysicists and laboratory plasma physicists. These communities involve observers, laboratory experimentalists, theorists, and computational physicists. Recent years have seen very significant beginnings of such coordinated efforts.

These beginnings indicate the large potential for accelerated progress and the need for an articulation of the major scientific challenges and opportunities in plasma astrophysics. Such an articulation would also express the unity and coherence of plasma astrophysics as a scientific discipline. Since it merges multiple areas of expertise, its unity can be overlooked, as reflected in the absence of a clear funding home for plasma astrophysics in the U.S.

To express the challenges, opportunities, and coherence of plasma astrophysics, more than 100 scientists were involved in preparing and participating in the Workshop on Opportunities in Plasma Astrophysics held in January 2010. The workshop, preceded by preparatory efforts of ten topical working groups, was a grass-roots effort organized by the plasma astrophysics community.

This effort brought together observers, experimentalists, computational plasma physicists, and theorists from universities, national laboratories, government research institutions, and private industry, including several scientists from outside the U.S. It also encompassed physicists studying magnetized plasmas and those studying high energy density plasmas. The breadth of participation uncovered cross-cutting opportunities previously unappreciated. This document reports the results from the workshop.

## MAJOR QUESTIONS AND TOPICS

There are two approaches to articulating the challenges and opportunities: through plasma processes or through astrophysical systems. Individual plasma processes affect multiple systems, and individual systems encompass multiple processes. We have taken both approaches. Each of the ten working groups focused on one of the following processes that express the physics challenges (listed here in random order):

- Magnetic reconnection
- Collisionless shocks and particle acceleration
- Waves and turbulence
- Magnetic dynamos
- Interface and shear instabilities
- Angular momentum transport
- Dusty plasmas
- Radiative hydrodynamics
- Relativistic, pair-dominated and strongly magnetized plasmas
- Jets and outflows

Discussion of these topics, and their links to astrophysics, constitutes the bulk of this report. From these studies, we then extracted ten major, system-based questions for plasma astrophysics (listed here in random order):

**How do magnetic explosions work?**
Astrophysical plasmas exhibit spontaneous "explosions," such as solar flares, stellar flares, and substorms in planetary magnetospheres. These events accelerate particles to high energy, and affect radio communications on Earth. The explosions are driven by magnetic reconnection, the physics of which must be unraveled to understand why energy contained in magnetic fields of stars and planets is released explosively, and how this energy is so efficiently converted to particle energy.

**How are cosmic rays accelerated to ultra-high energies?**
Energetic particles bombard the Earth from space with energies up to $10^{20}$ electron volts, enormously more energetic than those achieved in the most powerful accelerator in the laboratory.

The energy spectrum of the particles fits a power law, yet the source of the strong acceleration remains a mystery. Possible sources include plasma shocks initiated by supernova explosions, as well as magnetic reconnection and plasma turbulence.

**What is the origin of coronae and winds in virtually all stars, including the Sun?**
Most stars have hot coronae with temperatures exceeding a million degrees. A hot wind blows from the Sun into the interstellar medium. A major challenge is how these nearly collisionless plasmas can be heated by waves, turbulence, shocks, and magnetic reconnection. This challenge requires advances in understanding the nature of the anisotropic turbulence in these plasmas.

**How are magnetic fields generated in stars, galaxies, and clusters?**
The universe appears to be magnetized at nearly all observable spatial scales. Stars, galaxies, galaxy clusters, and generally accretion disks, all contain magnetic fields. The fields often vary in time, sometimes in cycles. Understanding the origin and dynamics of these fields is a major puzzle, and is key to explaining accretion, stellar evolution, and galaxy evolution. The behavior of the fields depends on the plasma physics of dynamos, turbulence, reconnection, and flows.

**What powers the most luminous sources in the universe?**
The accretion of matter onto supermassive black holes generates prodigious fluxes of radiation, so intense that radiation pressure can dominate gravity. Understanding this process, which underlies the most luminous sources that light up the universe, is a challenge in plasma physics. Accretion is sufficiently rapid that it must involve plasma processes that enhance the accretion rate, such as plasma instabilities, turbulence, and transport.

**How is star and planet formation impacted by plasma dynamics?**
Despite progress in understanding star formation, the role of magnetic fields, especially in angular momentum transport, are still not understood. How rotating gas, plasma, and charged dust lose their angular momentum to collapse to form stars and planets remains an unsolved problem. Answers to this problem require advances in the physics of dusty plasmas, magnetic reconnection, and magnetized turbulence.

**How do magnetic fields, radiation, and turbulence impact supernova explosions?**
Supernovae and gamma-ray bursts are nature's grandest explosions that originate from the collapse of massive stars. Recent observations suggest that these explosions may be highly anisotropic and may be turbulent. Investigating the roles of magnetic fields, radiation, and turbulence on the explosions is a fundamental challenge. Magnetic fields are particularly important in the so-called magnetars.

**How are jets launched and collimated?**
Powerful jets of plasma are observed in a variety of astrophysical systems. They emanate from compact objects ranging from protostars to supermassive black holes, and can deposit large amounts of energy in the surrounding medium over large distances. Magnetic fields are believed to govern the process of jet formation and collimation. But it is not yet known how the jets are launched and why they survive stably across large distances (up to extragalactic scales). The behavior of jets encompasses a broad range of plasma phenomena, including instabilities, transport, dynamo, and reconnection.

**How is the plasma state altered by strong magnetic fields?**
When magnetic energy density is higher than the rest mass energy density of plasmas, the dynamics are altered significantly, as found in environments such as pulsars and relativistic jets. The ultra-strong magnetic field can produce electron-positron pairs, and alter plasma wind propagation and dissipation of plasma energy. Fundamental plasma processes, such as magnetic reconnection, are changed in these exotic environments.

**Can magnetic fields affect cosmic structure formation?**
Magnetic fields are observed in galaxy clusters, raising the question of whether the fields play a fundamental role in determining the topology and properties of these very large-scale structures. Magnetic fields may also be important in understanding the baryon dynamics in cosmic structure formation. These questions extend the applications of many of the basic plasma processes investigated at smaller scale to these new environments.

These questions are manifestly of major significance in astrophysics. What might not have been so obvious is that their answers largely reside in plasma physics. This report is dedicated to describing the challenges, opportunities, and impact of the following plasma research areas (listed here in random order):

*Magnetic reconnection* — a change in the topology of magnetic fields in plasmas — can drastically alter plasma transport and plasma structures, as well as convert magnetic energy to particle energy. Reconnection is thought to be everywhere in the universe, particularly underlying processes such as stellar flares.

*Collisionless shocks* accompany explosive events, such as supernova, and can accelerate particles to high energies. The behavior of shocks, and dissipation of their energy, in astrophysical plasmas in which collisions between particles are rare, is quite different and more complex than for the better-understood case with strong collisions.

*Plasma turbulence* is ubiquitous since most astrophysical plasmas have substantial free energy — such as nonuniformities in density, temperature, fields, and velocity space distributions — to excite turbulence. Turbulence affects the macroscopic behavior of plasmas in many ways, and has profound influences on essentially all the astrophysical questions articulated above.

The *dynamo* problem has long been a major astrophysical puzzle: how does the mechanical energy associated with a flowing plasma give rise to large scale magnetic field growth from a small seed field? This question is motivated by the prevalence of magnetic field in the universe. The dynamo problem also encompasses the questions of how magnetic field is sustained in the presence of dissipation, and why it varies in time in various astrophysical venues.

*Interface and shear instabilities* arise in both high energy density plasmas and magnetized plasmas in the presence of sheared flow. Their onset, behavior, and nonlinear development underlie a wide range of astrophysical phenomena including stellar wind flow around planetary and pulsar magnetospheres, the transitional region from solar wind to the interstellar medium, photoevaporated molecular clouds, supernova explosions, and blast waves in supernova remnants.

*Momentum transport* in plasmas is often observed to occur at a rate faster than can be explained from viscosity due to collisions between particles. Believed to be driven by plasma instabilities or turbulence, an explanation of momentum transport is required to understand how rotating matter collapses into gravitational potential wells. This question is key to understanding the accretion of matter onto compact objects (from protostars to black holes), the formation of stars, and the formation of planets.

*Dusty plasmas* consist of relatively massive charged dust particles, and thereby have properties altered from the conventional plasma of lighter ions and electrons. As breeding grounds for planets and stars, dusty plasmas have influence at both the solar system and galactic scales.

*Radiation hydrodynamics* — the interaction of radiation with matter — introduces an additional player (radiation) to the plasma environment. It is important for the formation of the largest stars (through photo-ionization and heating of the gas surrounding young stars), the behavior of accretion disks (when radiation pressure is substantial), the structure of molecular clouds (due to photoevaporation), the atmospheres of extrasolar planets (influenced by radiation from stars), and supernova explosions of massive stars (where radiation pressure exceeds material pressure).

*Relativistic, pair-dominated and strongly magnetized plasmas* are somewhat exotic systems that are ubiquitous in the high energy universe. When particles are moving at relativistic speeds, when electron-positron pairs dominate the plasma (rather than electrons and ions of unequal masses) and when the magnetic field is ultra-strong (altering collisions, ionization, radiation, and atomic structure), the behavior of the plasma changes significantly. These conditions can occur in violent astrophysical phenomena such as pulsar winds, gamma-ray bursts, jets in active galactic nuclei, microquasars, neutron star atmospheres, and the first few seconds of the early universe.

*Jets and outflows* represent a type of plasma structure whose formation and behavior challenges many aspects of plasma physics. They are observed to emanate from a wide range of compact objects, and might have a significant impact on cosmic structures.

These ten topics in plasma physics are separately described in the chapters that follow. The opportunities articulated in this report both advance research in plasma topical areas and lead toward resolution of the major astrophysical questions cited earlier.

## IMPACTS AND CONCLUSIONS

These research opportunities will additionally have impact in three areas beyond these targeted major questions. First, much of the physics overlaps with central challenges for fusion energy, both magnetic and inertial (the purview of the DOE). For example, in magnetically confined plasmas, magnetic reconnection drives sawtooth oscillations and disruptions in tokamaks, momentum transport determines the rotation of toroidal plasmas, plasma turbulence regulates transport in nearly all magnetic fusion plasmas, and particle acceleration and heating by waves and instabilities occurs both deliberately and spontaneously. In inertially confined plasmas, Rayleigh-Taylor instabilities influence the symmetry of implosion and radiative hydrodynamics influences the burn phase. Many of the advances in plasma astrophysics have evolved directly from plasma physics developed for fusion energy, and plasma physics learned through astrophysical applica-tions is increasingly influencing aspects of fusion plasma physics. Second, development of plasma physics targeted to astrophysics advances basic plasma physics (the purview of the NSF and DOE). The wide range, sometimes extreme, of scales, particle energies, field strengths, and overall plasma parameters encountered in astrophysics extend greatly the scope and depth of plasma physics. Third, plasma astrophysics is crucial to the guidance and interpretation of observational missions (the purview of NASA). From satellite observations of the magnetosphere to the heliosphere and other galaxies, guidance from plasma astrophysics is needed to determine what quantities to measure and to make scientific sense of those measurements. And for some missions, the primary goal is to directly study a particular plasma process. That is, plasma astrophysics is needed in the preparation stage of missions, to optimize its value, and in the operational stage, to reap the benefits. For example, understanding aspects of magnetic reconnection, waves, and turbulence is essential to the success of the Magnetospheric Multiscale Mission, the Solar Dynamics Observatory, the Solar Orbiter, and the Solar Probe Plus. Understanding aspects of accretion and particle acceleration is key to reap the full benefit of the Nuclear Spectroscopic Telescope, the James Webb Space Telescope, Large Synoptic Survey Telescope, Atmospheric Čerenkov Telescope Array, and the International X-ray Observatory.

The report articulates many general and specific research opportunities. The content of each of the following chapters was written by a topical working group (see Appendix D for a list of the membership of each working group). Being a community report, no effort is made to rank the many opportunities that for each area were developed by the specific topical experts. That exercise is beyond the province of this activity. Thus, this report produces no recommendations except one. We recommend that the plasma astrophysics program in the U.S. be strengthened in structure and coordination across DOE, NSF, and NASA, to embrace the unity, coherence, and opportunities of the field. A strengthened program of plasma astrophysics greatly aids the missions of these agencies. One intention of this report, in addition to the immediate scientific value of the effort, is to provide motivation and justification for deeper consideration of the funding strategy for plasma astrophysics.

# CHAPTER 1:
## MAGNETIC RECONNECTION

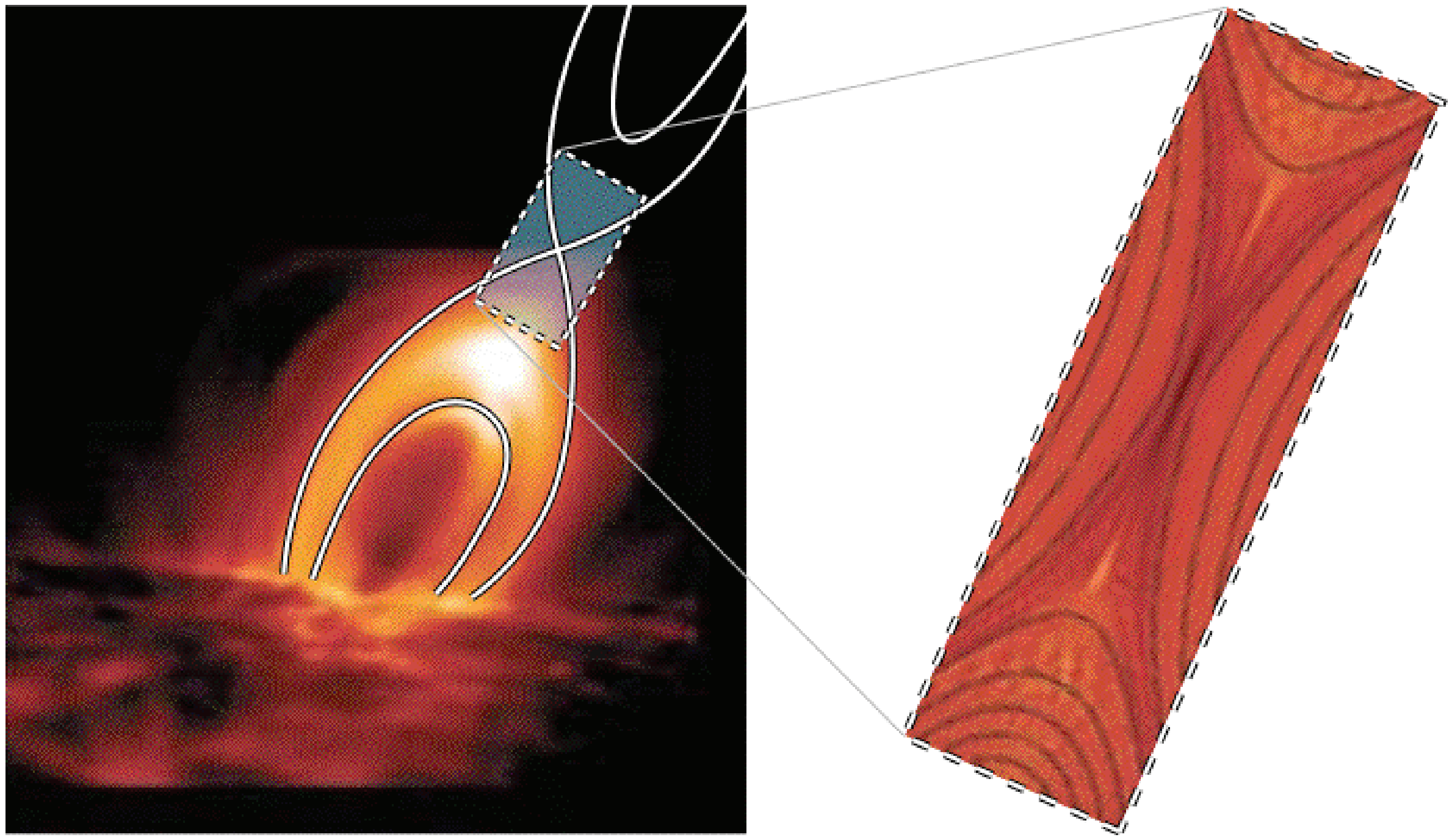

ON PREVIOUS PAGE

*An X-ray image of a solar coronal structure from Japan's Hinode spacecraft (left) is overlaid with an illustration of the magnetic field lines and the reconnection region (dotted box) that produces it. Computer simulations of electron populations in reconnection regions (right) have been key in deciphering more about how this event unfolds.* (Left image is courtesy of the Japanese Space Agency, right image is courtesy of James Drake, University of Maryland, and Michael Shay, University of Delaware.)

# CHAPTER 1: MAGNETIC RECONNECTION

## INTRODUCTION

Magnetic reconnection is the dominant process for dissipating magnetic energy in the universe. Therefore, it has dynamical importance in a broad range of space and astrophysical phenomena. Magnetic reconnection also has intrinsic scientific interest because the release of magnetic energy in a macroscopic system is linked to the dynamics of a narrow boundary layer, the "dissipation region," where dissipation facilitates the breaking of magnetic field lines. The complexity of reconnection and its scientific challenge comes from the extreme disparity of spatial scales between the global system and the dissipation region. Further, the dissipation region is often turbulent with electrons and ions displaying weakly coupled, complex dynamics. The ongoing scientific issues related to magnetic reconnection span an enormous range of topics. In this chapter we focus on five: the rate of magnetic reconnection; the onset of magnetic reconnection; cross-scale coupling in large systems; reconnection-driven particle heating and acceleration; and reconnection in extreme astrophysical environments. For each we discuss the science issues, recent progress and open issues, and strategies for reaching scientific closure.

## KEY SCIENTIFIC CHALLENGES

### The Rate of Magnetic Reconnection

A key question about magnetic reconnection in astrophysical and laboratory plasmas is why the reconnection rate, or flux transfer rate, is so fast in comparison with the rate predicted by classical magnetohydrodynamic (MHD) theory. In the Sweet-Parker model, the length of the dissipation region is determined by the macroscopic system size, while the much shorter width is controlled by weak dissipation. Balancing plasma inflow with Alfvénic outflow in such a high-aspect-ratio dissipation region yields rates of reconnection that are far smaller than those inferred from observations. The exploration of reconnection in collisionless and nearly collisionless plasma has therefore focused on the structure and dynamics of the dissipation region. Important progress has been made through numerical simulations, observations from satellites, and dedicated laboratory plasma experiments. Two-fluid effects, resulting from the fundamentally different behavior of ions and electrons at small spatial scales, are important within the dissipation region.

Recent research has focused on what provides the dissipation necessary to break the frozen-in condition in nearly collisionless plasma, and what controls the width and length of the dissipation region. In balancing the reconnection electric field, we have not established the relative roles of turbulence and laminar momentum transport, as described by the off-diagonal pressure tensor. NASA's Magnetospheric and MultiScale Mission (MMS), a four-spacecraft satellite mission, is designed to explore the fine-scale structure of the electron dissipation region and to answer these questions. Parallel measurements can be made in laboratory experiments and explored in 3-D reconnection simulations. Is the reconnection rate determined solely by the microphysics of the dissipation region, or also by the global boundary conditions? Such questions can be addressed in ongoing laboratory reconnection experiments.

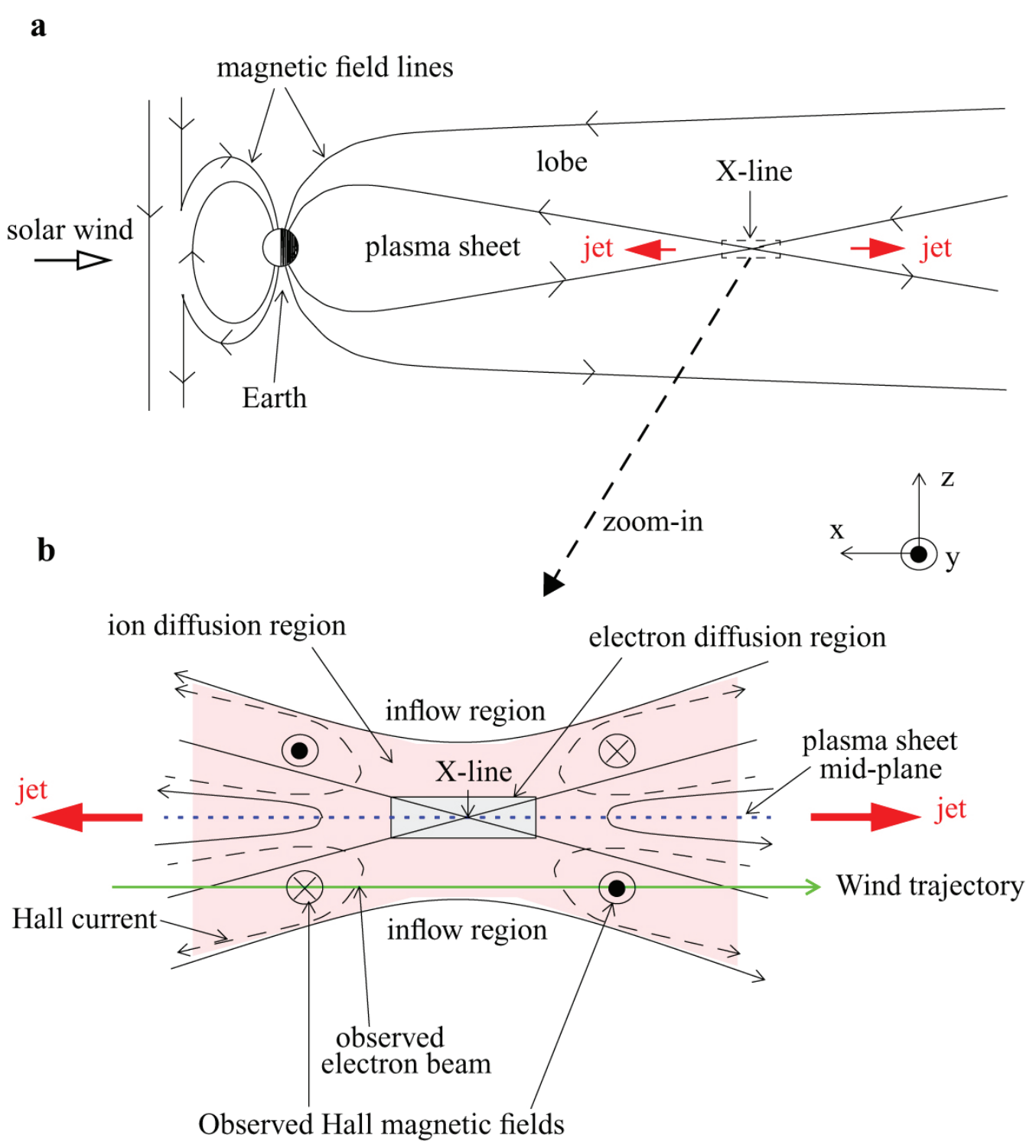

***Schematics of magnetic reconnection in the Earth's magnetosphere based on observations of the Wind spacecraft (Oieroset et al., Nature 412, 414, 2001). In (a) is the geometry of the magnetosphere as a whole. In (b) is a blowup around the reconnection region showing the inflow (above and below), the Alfvénic outflow jets (red arrows) and the electron (grey shading) and ion (tan shading) diffusion regions, where the particles are, respectively, demagnetized. The currents in the plane of reconnection (dashed lines) produce the characteristic quadrupole Hall magnetic fields that have been documented with satellite observations and laboratory experiments.***

## The Onset of Magnetic Reconnection

Magnetic reconnection often occurs in explosive events where magnetically stored energy is released on Alfvénic timescales. Examples are solar flares, magnetospheric substorms, and tokamak sawteeth. There has been much progress in understanding 2-D, steady-state reconnection, but the time-dependent problem is not well understood. For explosive energy release to occur, there must be a period of slow accumulation of magnetic energy followed by a sudden transition to fast reconnection. Reconnection can also take place in a quasi-steady fashion (e.g., at the Earth's magnetopause and in the solar wind). Two distinct models have been proposed, based on the local versus global properties of the magnetic equilibria.

Evidence indicates that reconnection onset in the Earth's magnetosphere (where direct measurements are available) is linked to the width of the current layer compared with the ion Larmor radius or skin depth. The magnetotail current sheet thins down to these scales prior to substorm onset. NASA's THEMIS spacecraft observations suggest that at least some substorms are triggered by the onset of collisionless reconnection in the midtail region. Further, current sheets wider than the ion skin depth in the solar wind compress at the Earth's bow shock and then undergo reconnection in the magnetosheath. In the solar corona, where collisions are likely to be important at onset, it has been proposed that fast reconnection sharply begins when the Sweet-Parker current layer approaches the ion Larmor radius or skin depth. There is some observational support from stellar data for this suggestion. In dedicated reconnection experiments in the laboratory, a strong increase in the rate of reconnection occurs when the Sweet-Parker current layer width falls below the ion inertial length. Bursts of fast reconnection are observed when the current layer width approaches the ion Larmor radius. Finally, although the current layer width has been a major focus of the effort to understand reconnection onset, it remains unknown if there are other local parameters that determine whether reconnection can take place.

In the corona, global rather than local dynamics may also drive the onset. The breakout model involves the reconnection of overlying magnetic field lines, and subsequent reconnection and expansion of flux tubes. In the sawtooth crash and in the recent reconnection experiments, observations suggest that the onset is spatially localized and then spreads — so that even in cases where the magnetic field geometry is mainly 2-D, 3-D effects can be important.

**Cross-scale Coupling in Large Systems**

During the past decade, much of the theoretical, computational, and laboratory research on the basic physics of magnetic reconnection focused on relatively small systems in order to understand fundamental issues regarding the rate and structure of a single reconnection site. A variety of two-fluid and kinetic models predict fast reconnection rates that are weakly dependent on the system size or dissipation mechanism. However, it is unclear how these results will extend to large-scale systems relevant to most astrophysical plasmas, where the separation between the dissipation and macroscopic scales is enormous, of order $10^8$ in stellar flares and $10^{10}$ in black hole accretion disks.

Given the huge range of scales, what controls the interaction between the dissipation scales and the global MHD evolution? Do the largest scales dictate the structure of the dissipation region or vice versa? Theory and simulations indicate that a single reconnection layer in large-scale systems may spontaneously break up into multiple interacting reconnection sites through the formation of secondary magnetic islands (or plasmoids). The behavior is similar in both collisional and kinetic regimes. An ongoing challenge is to understand how many islands are formed under various parameter regimes, and how these islands modify the global reconnection rate and particle acceleration. Finally, astrophysical systems typically have many sources of turbulence, which is predicted to facilitate reconnection at a rate insensitive to the dissipation.

As theory and simulation move forward on these questions, satellite observations — from the Earth's magnetosphere, the solar wind, and the solar corona — are providing important constraints. In the solar wind, the reconnecting current sheets are often highly planar and the structure of the outflow jet is remarkably smooth in contrast to some recent simulations showing highly distorted current sheets with embedded flux ropes around the x-line. To relate simulation results to astrophysical contexts, it is important to understand how the highly structured and transient plasma features near the dissipation region evolve downstream.

**Reconnection-driven Heating and Particle Acceleration**

The magnetic energy released during magnetic reconnection is converted into high-speed flows, heat, and energetic particles (typically with energy spectra in the form of power laws). While the convective flows in reconnection outflows have been widely documented by in situ satellite measurements in the Earth's magnetosphere and the solar wind — and are well described by the MHD jump conditions — the corresponding mechanisms for bulk plasma heating and particle acceleration remain poorly understood. Unraveling the mechanisms for heating and particle acceleration are essential to understanding the role that reconnection plays in heating the coronae of stars and accretion disks, and in driving their supersonic winds; in driving the relativistic jets from black holes and other compact objects; in powering giant radio galaxies; and in producing the cosmic ray spectrum. There is strong evidence within the heliosphere that the fraction of en-ergy going into bulk heating and channeled into the energetic component during reconnection are not universal. In the case of impulsive flares, roughly equal amounts of released energy appear in the form of thermal ions and electrons, and energetic ions and electrons, with recent over-the-limb observations suggesting that the pressure of the energetic electron component can approach that of the magnetic field. Thus, converting magnetic energy into the energetic electron and ion components can be highly efficient. On the other hand, even the largest solar wind reconnection events, which span hundreds of Earth radii, exhibit bulk ion heating, but no energetic ion component and no electron heating. The control parameters producing these stark differences have not yet been identified, making predictions of particle acceleration by reconnection problematic.

Because of their large parallel mobility, early models of reconnection-driven electron acceleration were based on the parallel electric fields that develop near a single magnetic x-line. However, parallel-electric-field models are inconsistent with the large numbers of energetic electrons seen in impulsive flares. In impulsive flares the electrons typically develop a thermal component (up to 10s of keV) and a power law tail up to several MeV with a spectral index that varies with the flare intensity . Observation and modeling  indicate that in the complex magnetic fields on the Sun, reconnection forms multiple magnetic islands that may be volume filling. Particle heating and acceleration should be explored in this context. In a multi-island environment, both parallel electric fields and Fermi acceleration in contracting islands drive electron acceleration. The exploration of this multi-island reconnecting environment, which requires modeling reconnection in a 3-D kinetic system, remains in its infancy and is limited by available computational resources.

Ion heating in reconnection-driven Alfvénic outflows has been well documented during reconnection in the Earth's magnetosphere and the solar wind. Strong ion heating has been measured during laboratory reconnection events with high-mass ions gaining greater energy than low-mass ions. The spectrum of energetic ions from impulsive flares extends up to the GeV/nucleon range in X-class flares, and the spectra of protons, as well as trace ions, have been well-documented by in situ satellite measurements in the solar wind. As in the case of electrons, super-Alfvénic ions may be efficiently accelerated during the contraction and merger of magnetic islands.

**Reconnection in Extreme Astrophysical Environments**

Much of the work on magnetic reconnection has been for systems of relatively tenuous, low-energy-density, optically thin environments, which can be represented as a collection of non-relativistic charged particles whose numbers are conserved. However, magnetic reconnection also has been frequently invoked in astrophysical systems, especially in high-energy astrophysics, whose parameters can be extreme. Examples include accretion disks and their coronae, and large-scale magnetospheres, jets, gamma-ray bursts (GRBs), pulsar magnetospheres and pulsar winds, and flares in soft gamma repeaters (SGRs). These systems require a number of new effects, including special relativity, radiation and pair creation.

Special relativistic effects are important when the reconnecting magnetic field is so strong that the magnetic energy density exceeds the rest-mass energy density of the upstream plasma. The corresponding Alfvén speed and the reconnection outflow velocity approach the speed of light. This situation is typical of astrophysical pair plasmas such as in radio-pulsar magnetospheres, which are believed to develop a large-scale equatorial current sheet beyond the light cylinder. Reconnection in this current sheet may produce the observed pulsed high-energy emission and coherent radio emission. Relativistic reconnection in pair plasmas may also power the giant magnetar flares in soft-gamma repeaters and the relativistic jets of gamma-ray bursts and active galactic nuclei (AGN).

Radiation may affect reconnection through several mechanisms. Radiative cooling can affect the energy balance and hence the dynamics of the reconnection layer, with different radiation cooling mechanisms acting in different astrophysical situations. In powerful events, the dissipated energy density and the plasma temperature are so high that radiation pressure enters the dynamics and can dominate the plasma pressure. Compton-drag resistivity due to electron-photon collisions — as opposed to electron-proton collisions —is important in several high-energy astrophysics situations.

Some astrophysical systems (magnetars, central engines of GRBs, and supernovae) are believed to possess super-strong magnetic fields that exceed the critical quantum field of about $4 \times 10^{13}$ Gauss, corresponding to the lowest electron Landau level energy equal to the electron rest mass. The magnetic energy density of such a strong magnetic field is so large that when converted to radiation energy, it results in temperatures greater than the electron rest-mass energy of 0.5 MeV, and electron-positron pair creation occurs. The associated increase in the optical depth also makes it highly collisional. How these processes affect the overall reconnection dynamics is unknown.

## MAJOR OPPORTUNITIES

**Multi-island Magnetic Reconnection and Particle Acceleration**
Observations in the magnetosphere and corona, and recent computer simulations, suggest that over a broad range of collisionality, reconnection at single large x-lines break up into multiple interacting islands or flux ropes. In systems with a guide field, reconnection can take place at a variety of layers so that these magnetic islands may be volume filling. The conversion of energy in such a reconnecting system represents a paradigm shift from the single x-line picture of reconnection. Can we understand particle acceleration in such a complex, turbulent environment and therefore understand the extraordinary efficiency of particle acceleration in flares? What is the role of reconnection in producing the cosmic ray spectrum?

A number of existing satellites (Themis, Cluster) provide in situ data in the magnetosphere. Remote observations of solar flares from the Rhessi, Stereo, SDO, and Hinode spacecraft give evidence for fine-scale structure during impulsive flares. Future missions — in particular, Solar Probe Plus — will produce key in situ measurements in the low-beta regime close to the Sun, enabling the diagnosis of flare energy release and solar wind reconnection. Two-dimensional simulations of reconnection and particle acceleration provide important information on the spontaneous development and evolution of secondary magnetic islands and the physics of particle acceleration. However, exploring 3-D multi-island reconnection is a significant challenge — even on the largest of present-day computers — because of the large separation of scales intrinsic to reconnection. The ongoing exploration of sawtooth events in reversed-field pinch experiments provides important data, although it is difficult to diagnose the dynamics of these events.

Progress could be made by assembling observations from a variety of satellite datasets to focus on the structuring and time variability of reconnection-driven energy release. A dedicated laboratory experiment, designed to explore 3-D multi-island reconnection and particle acceleration, would provide critical data to support ongoing NASA missions and to give closure on important problems in space and astrophysical systems. We recommend the formation of a National Working Group to explore possible geometries, parameters, and diagnostics required for such an experiment. Following the development of a cost-effective design, this Group would seek broad support for the construction of the experiment.

**Magnetic Reconnection in Extreme Environments**

In many astrophysical environments, including pulsar and magnetar magnetospheres and jets, reconnection is in the strongly relativistic regime. Radiation pressure, cooling, and drag, as well as pair production, may also enter the dynamics close to the source regions of jets and GRBs, and in magnetar magnetospheres. Thus, reconnection dynamics in these High-Energy-Density (HED) systems will be very different from that typically explored in heliospheric applications.

Satellite observations (Chandra, Fermi, and Swift) provide important constraints on jet structure and dynamics, and the driver mechanisms of GRBs and magnetar flares. In Poynting-flux dominated jets, the energy from fields dominates kinetic energy and can act as a source for energetic particles powering measured synchrotron emission. Because of the remoteness of most sources, however, the lack of resolution continues to limit progress. The understanding of relativistic reconnection is rapidly developing, but at present the ability to benchmark these results is limited.

A major effort to take advantage of advances in parallel computing to explore the strongly relativistic regime, and to implement a radiative transport model and pair production, should be pursued. In parallel, we suggest taking advantage of facilities in the HED community by pursuing a laser-driven reconnection experiment to explore reconnection in the relativistic regime. The laser facility at the University of Rochester has already demonstrated the compression of magnetic fields to $10^7$ G that produce Alfvén speeds $c_A/c \sim 10\text{-}100$. The possibility of compressing an initial anti-parallel magnetic field configuration should be explored. Such a system would drive reconnection with outflows in the relativistic regime and could be used to benchmark the parallel modeling effort.

**Explosive Onset of Magnetic Reconnection**

A fundamental question is why magnetic reconnection occurs as an explosion. The observations are nearly universal — the explosive onset of reconnection occurs in laboratory fusion experiments, magnetospheric substorms, and solar and stellar flares. On the other hand, solar wind reconnection appears to be quasi-steady. What are the differences between these systems? Is there a universal mechanism for reconnection onset? Since the delayed onset of reconnection controls the storage of energy prior to release, answering this question may provide information on the size of energy releases necessary to predict, for example, the size distributions of solar flares.

Some laboratory experiments exhibit the spontaneous onset of magnetic reconnection. While the onset occurs when the width of current layers approaches kinetic scales — consistent with predictions of theory and simulations — the measurements suggest that the onset is intrinsically 3-D.

Whether this result is generic remains an open issue. Recent advances have been made in understanding the onset of the sawtooth crash. THEMIS observations in the magnetosphere suggest that substorm onset is linked to magnetic reconnection in the midtail. Simulation studies of reconnection onset in the magnetotail and solar corona are ongoing. In the solar case, researchers are pursuing local criteria based on the transition from collisional to collisionless reconnection and geometrically based models.

Establishing a consortium of scientists studying reconnection onset in a wide variety of laboratory and astrophysical systems would help focus the present largely disconnected efforts. Establishing this consortium also would help raise the importance of the topic and thus encourage a broader effort in this area. Enhanced theoretical and modeling support for ongoing laboratory experiments would help maximize the scientific benefit of these experiments.

## IMPACTS AND MAJOR OUTCOMES

Magnetic reconnection plays a central role in the dynamics of a wide variety of astrophysical systems. Magnetic fields and associated reconnection processes dominate the dynamics of the coronae of the stars and accretion disks. Magnetic reconnection is required for the dynamo to amplify the magnetic fields that are seen throughout the universe and is likely to be dynamically important in accretion discs. The role of magnetic reconnection in jets, magnetar flares, and gamma-ray bursts continues to be explored. Over the past decade, there has been substantial progress in identifying the mechanisms driving the surprisingly fast release of magnetic energy seen in explosive phenomena in the laboratory and space. However, key areas remain poorly understood, and are critical for understanding and modeling the rich variety of astrophysical systems.

Within the heliosphere, there is substantial evidence for strong plasma heating and particle acceleration during magnetic reconnection. However, unlike with shocks, there is no standard model of particle acceleration from reconnection. The absence of an accepted predictive model makes modeling of remote astrophysical systems, where observations provide less detail, challenging. Developing more powerful computers, combined with dramatic advances by RHESSI, Hinode, and SDO in remote sensing of solar dynamics, fosters an exciting scientific climate where significant progress is likely — especially if a suitable laboratory experiment can be designed to further benchmark ideas from computations, theory, and remote observations. Can a standard model for reconnection-based particle acceleration be developed? Such a model would dramatically change our ability to reliably model astrophysical systems.

Much progress on magnetic reconnection has been based on low-energy-density systems found within the heliosphere. In the extreme environments of magnetar and pulsar magnetospheres — and the source regions of jets and gamma-ray bursts — radiation, relativistic effects, and pair production are likely to alter conventional reconnection scenarios. Understanding how reconnection dynamics changes in such environments is required to address even the most basic issues. Reconnection experiments based on intense lasers, combined with the development of new computational models, would facilitate significant progress in this area.

## CONNECTIONS TO OTHER TOPICS

Reconnection is a fundamental process describing the behavior of magnetic fields in plasma. Reconnection and its consequences are therefore a crucial element for describing other processes in magnetized plasma. The generic state of astrophysical fluids is often turbulent. In accretion disks, for example, the magnetorotational instability drives the disks into a turbulent state that facilitates the angular momentum transport required for accretion. Magnetic reconnection is a dynamically important player. Magnetic reconnection also is an essential element during magnetic-field generation via the dynamo. How do physics-based models of reconnection, which greatly differ from the MHD treatment, modify accretion and the turbulent dynamo?

Launching astrophysical jets and outflows requires strong magnetic stresses. Simulations of jets exhibit magnetic-field reversals that may reconnect and alter jet formation. Jets are also believed to be sources of high-energy particles. Observations suggest that both shocks and magnetic reconnection are efficient sources of energetic particles. What is the relative importance of the two acceleration processes in jets and other astrophysical phenomena? What are the astrophysical objects where one or another process dominates, and how do they couple? Particles accelerated during reconnection could act as seed particles for shock acceleration.

# CHAPTER 2:

## COLLISIONLESS SHOCKS AND PARTICLE ACCELERATION

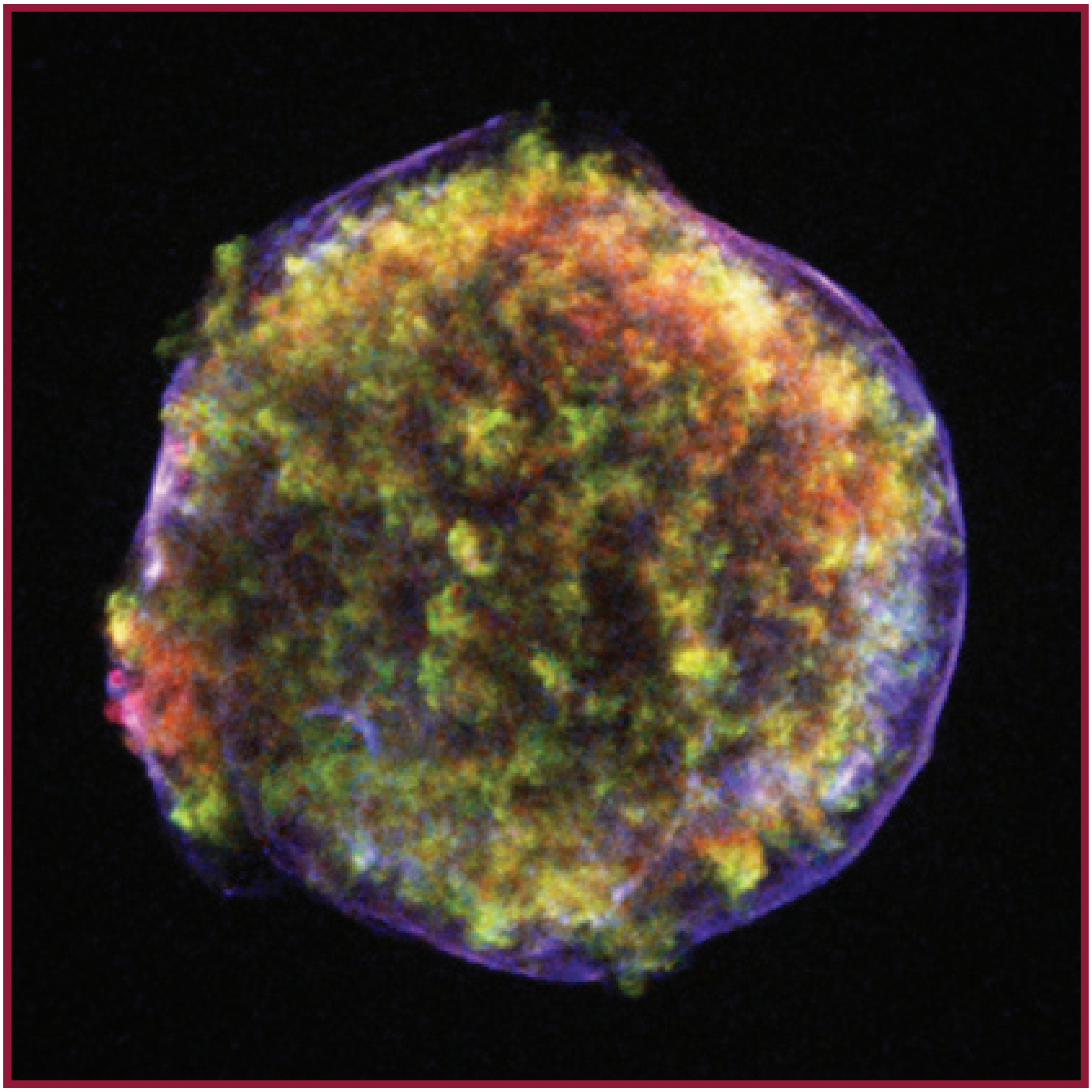

*ON PREVIOUS PAGE*

*CHANDRA X-ray image of Tycho's supernova of 1572. Synchrotron emission by relativistic electrons at the outer shock is seen in blue.* *(Courtesy: NASA/CXC/Rutgers/J. Warren & J.Hughes et al.)*

# CHAPTER 2: COLLISIONLESS SHOCKS AND PARTICLE ACCELERATION

## INTRODUCTION

Shocks generally result from the collision of two flows. This collision occurs frequently in the cosmos from the interaction between the small-scale flows of the heliosphere to the interaction between large-scale flows characteristic of galactic clusters and the jets in active galactic nuclei (AGN). Shocks observed in the solar wind include planetary bow shocks, the shocks driven by coronal mass ejections (CMEs), the shocks formed by the collision of fast and slow wind from adjacent regions of the Sun, and the solar wind termination shock. Shocks in the galaxy and beyond include those driven by supernovae explosions, galactic winds, AGN jets, accretion onto compact objects, galactic motions in galaxy clusters, and gamma-ray bursts (GRBs). These shocks span a huge domain of spatial scale, strength, and plasma parameter space. Because particle densities are generally very low throughout most of the cosmos, the mean free path due to Coulomb collisions between charged particles is typically large compared with the shock spatial scales of interest. Therefore, most shocks are collisionless and the interaction between far upstream and far downstream plasmas is mediated by electromagnetic fields. An interesting exception is charge-exchange coupling of the interacting upstream and downstream plasmas to the atoms present in partially ionized plasma. Despite their diversity, collisionless shocks share common characteristics: they are observed to accelerate nonthermal particles efficiently, and to generate and amplify magnetic fields, in addition to decelerating supersonic flows.

Shocks convert a fraction of the ordered kinetic energy density of the upstream flow to the higher entropy per unit mass downstream flow by dissipative processes occurring in the shock layers. As a result of the nonlinear plasma processes involved in the shock layers, and their diversity due to the broad range of possible plasma parameters, a general understanding of the physics of shock structure is challenging. The interaction between the upstream and downstream plasmas involves (i) the ambient magnetic field and its obliquity relative to the shock normal, (ii) an electric field parallel to the shock normal associated with charge separation in an ion-electron plasma (which can contribute to reflecting inflowing ions), (iii) a rich variety of possible streaming instabilities that excite electromagnetic fluctuations, which in turn couple the flows by scattering the individual ions and electrons, and produce an effective resistivity and viscosity in the shock layers, and (iv) particle acceleration from thermal energies up to relativistic energies. Each of these processes and instabilities, affecting the different particle species, is characterized by its own length scale parallel to the shock normal so that a particular shock will exhibit multiple length scales characteristic of the relevant processes and instabilities.

## KEY SCIENTIFIC CHALLENGES

### Are Shocks in the Cosmos Well Described as Planar and Stationary?

Theoretical models of shocks are often based on the simplifying assumptions that they are stationary, and planar on the length scale of the shock structure. The plasma kinetic process-es responsible for the dissipation clearly are not planar and stationary. However, the planar stationary "shock" resulting from an average over a large appropriately chosen ensemble of shocks provides a reasonable representation of the shock structure and physics. Community intuition about shock structure is often based on these ensemble-averaged shocks. However, observations and numerical simulations reveal interesting time dependence (sometimes periodic) and important spatial variations along a complex warped surface. Warps with a length scale similar to the turbulence correlation scale in the solar wind have been observed in interplanetary traveling shocks using multi-spacecraft measurements. Warps change the local magnetic obliquity of the shock, which, for example, affects particle injection and acceleration (see next section of this chapter). A major source of variations along the shock surface is inhomogeneity of the upstream plasma, especially density variations. The density variations create surface warps and inhomogeneous bulk flows downstream, which can drive turbulence and magnetic field amplification. An interesting temporally periodic feature of quasi-perpendicular shock structure revealed by simulations is shock reformation. In general, the mechanism of shock reformation is unclear. However, for large Mach numbers it appears to come from overstable proton reflection by an unsteady shock potential, which results in periodic dissipation and a periodic variation in shock speed and location. Although shock reformation can have important consequences for particle injection, for example, it is challenging to detect with spacecraft measurements.

**Understanding Particle Injection and Diffusive Acceleration at Shocks**

An important channel of shock dissipation is particle acceleration by a combination of first-order Fermi acceleration and shock drift acceleration known as diffusive shock acceleration (DSA). This mechanism is responsible for most of the energetic particle populations in the heliosphere, the majority of galactic cosmic rays, and presumably many of the other energetic particle populations in the cosmos. At higher energies the mechanism is conceptually straightforward, although the nature and excitation of electromagnetic fluctuations and their impact on particle scattering and transport is not well understood (see next section of this chapter). The major uncertainty in applying the mechanism to specific shocks and their associated energetic particles is the rate at which upstream thermal particles are injected into the process. This uncertainty undermines the predictive power of diffusive shock acceleration and is presumably responsible, in part, for the huge variation in observed ion intensities in solar energetic particle events. The injection rate is dependent on the detailed electromagnetic structure of the shock, which determines the rate at which incoming particles are reflected or scattered back upstream, and it appears to be very sensitive to the local magnetic obliquity. For quasi-perpendicular shocks, thermal particles are not able to scatter sufficiently to initiate diffusive shock acceleration before the magnetic field sweeps them through the shock. Determining the injection mechanism is nontrivial. Even after years of investigations of the Earth's bow shock based on International Sun-Earth Explorer (ISEE) and Cluster data, the origin of the field-aligned beams that initiate the ion-acceleration process is unknown.

Finally, the lower injection rate of electrons when compared with ions — in spite of the higher speed of electrons — is not well understood.

**Understanding Magnetic Field Amplification at Shocks**

Generally, the ambient magnetic field fluctuations in the solar wind and interstellar space are not sufficient to yield efficient diffusive shock acceleration. However, the accelerating particles are a high-energy manifestation of the interpenetrating upstream and downstream plasmas. The streaming of these particles relative to the upstream flow excites the cyclotron-resonant hydromagnetic streaming instability at lower proton intensities, the non-resonant current-driven instability at higher proton intensities, or variations of these instabilities. The hydromagnetic instability, which maximizes for wave propagation parallel to the ambient magnetic field, is generally evident as an enhancement in the upstream hydromagnetic fluctuation power at quasi-parallel shocks in interplanetary space. These shocks can inject solar wind ions into the acceleration process. The waves often grow to large amplitude, are compressed at the shock, modify the shock structure, and provide effective particle scattering downstream. They also modify the compression ratio sensed by the accelerating particles. Upstream of the Earth's bow shock, where wave magnetic amplitudes are comparable with the ambient field strength, the magnetosonic wave compressive front is often observed growing to a Short Large Amplitude Magnetic Structure (SLAMS). This structure seems to be excited by the free energy released by ions scattered in enhanced numbers from the structure back toward the shock. Other compressive wave fronts form "shocklets," which generate whistler precursors. The details of many of these processes, including their initiation, are not well understood, particularly the nonlinear evolution of the excited hydromagnetic waves. They should be pursued by analytical and numerical investigations.

At quasi-perpendicular shocks, the streaming instability is not as effective. Particle transport across the average field is primarily by a random walk of field lines, which leads to small scattering mean free paths parallel to the shock normal and steep spatial gradients. This configuration is unstable to a version of the Rayleigh-Taylor instability, as the ion pressure gradient decelerates the upstream plasma. The resulting warped field lines in the shock precursor presumably reduce the magnetic obliquity, increase the field strength, and increase injection rates and acceleration efficiency. This scenario is speculative and requires further calculations and simulations.

The magnetic field amplification by the non-resonant current-driven instability is well established theoretically and observationally through analyzing X-ray images of supernova remnants. This result is crucial to the theory for the origin of galactic cosmic rays up to the "knee" at supernova remnant (SNR) shocks (see next section of this chapter).

## MAJOR OPPORTUNITIES

**A Major Initiative to Understand the Acceleration of Cosmic Rays**

The discovery of cosmic rays a century ago marked the beginning of space science, led to the discovery of many new subatomic particles, and ushered in the Space Age. Incredibly, we still do not fully understand the origins of these particles. The most promising source of the majority of cosmic rays up to the "knee" appears to be diffusive shock acceleration at the shocks driven by supernovae remnants. Shocks can produce power-law energy spectra. Recent X-ray and $\gamma$-ray observations show that electrons and protons are accelerated to TeV energies in su-pernovae remnants, and theory has shown that magnetic fields are amplified at strong shocks to magnitudes that enable them to accelerate protons to the knee. However, many questions remain. They include: what are the injection rates of electrons and protons at the shock; why do cosmic ray electrons and ions have different power-law spectral indices; why are cosmic ray anisotropies so small; what is the source of the cosmic rays beyond the knee; and which nearby sources accelerate the highest energy electrons. With key observations available from HESS, FERMI, AUGER, PAMELA, and many other spacecraft, balloon, and ground-based experiments, and the development of powerful numerical simulations, the time is ripe to concentrate on the remaining pieces of the puzzle. Understanding the origin of cosmic rays requires an interdisciplinary approach focused on the structure of supernovae remnants and their shock, the process of diffusive shock acceleration, and the galactic propagation of cosmic rays.

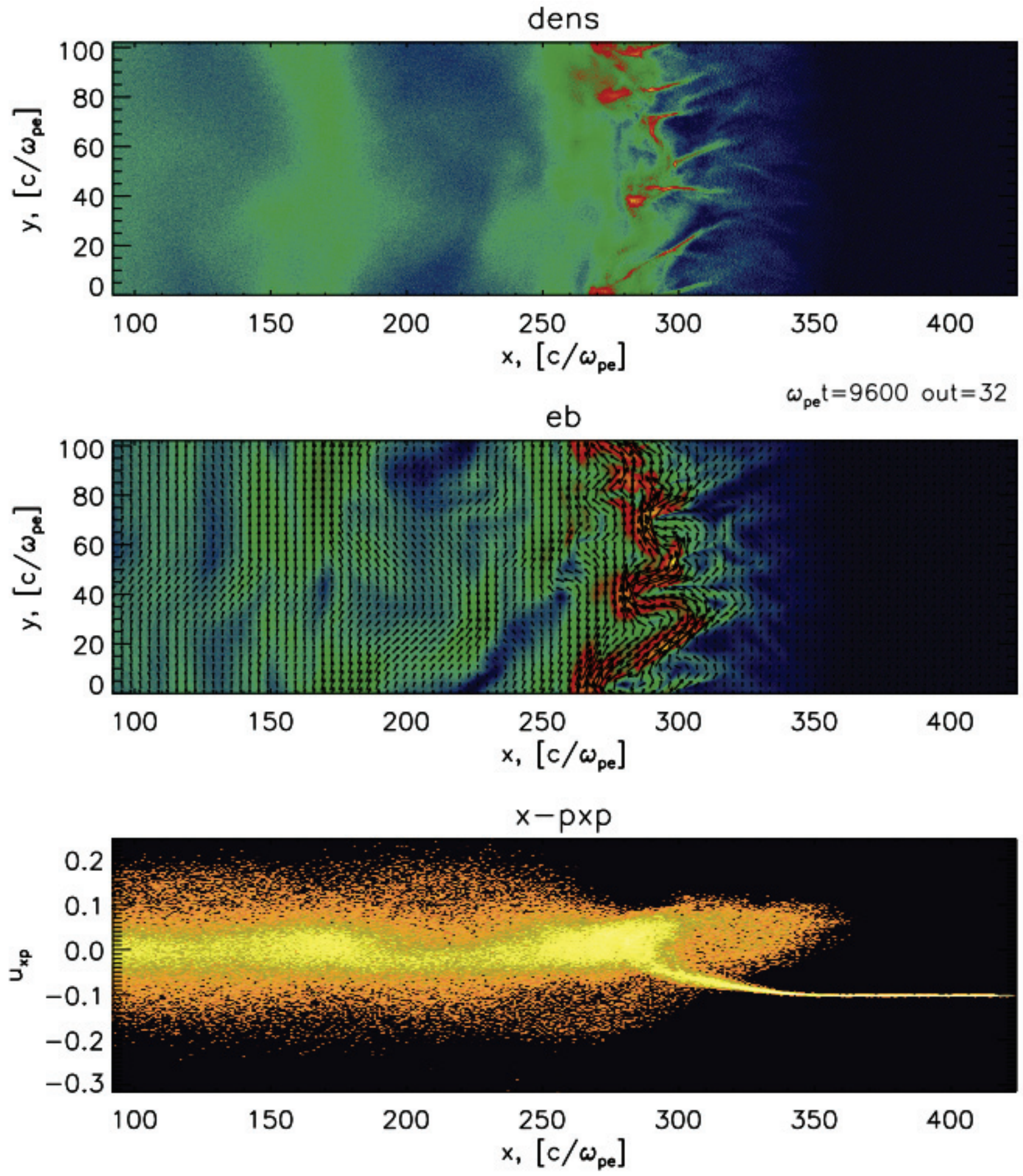

***Structure of nonrelativistic shock with Mach number 15, propagating from left to right. Initial magnetic field is oriented at 75 degrees to the shock normal. Top panel: density of plasma. Middle panel: magnetic energy, including vectors of magnetic field in plane; bottom panel: x-momentum phase space of ions, showing reflection from the shock in the "shock foot" region. Such high Mach number shocks display significant corrugation of the shock surface, and periodic reformations driven by the ion dynamics in the foot region.***

**Renewed Investigation of Shock Structure and Formation in the Laboratory**

With relatively primitive plasma machines and diagnostics, shocks were identified in laboratory devices in the 1960s and 1970s, a period described as the First Golden Age of shock studies. However, the slow response time of the ions, the influence of the chamber walls on the particle distributions, and the primitive diagnostics made it difficult to establish whether the "shock" was fully formed. Between this period and the present, shock studies suffered as funding decreased for magnetic-pinch fusion — the configuration in which most of the shocks had been formed. However, with the advent of new High Energy Density (HED) facilities and other facilities designed to study basic plasma physics, there are new opportunities to revive laboratory simulations of collisionless shocks. It should be possible in about a year to produce a perpendicular collisionless shock with Alfvén Mach number ~4, and with spatial and temporal scales large enough to include ~4 shocked proton Larmor radii and gyrations. These limitations will certainly improve, and also allow for the study of quasi-parallel shocks. Along with numerical simulations, such shock experiments will enable quantitative studies of shock formation timescales, proton and electron (and possibly a minor ion) dissipation processes, and electron acceleration. Such studies will improve our understanding of shocks observed in space.

**A Study of the Connection Between Astrophysical and Heliospheric Shocks**

The solar wind plasma $\beta$ ($\beta$ = ratio of plasma to magnetic pressure), including the important contribution of interstellar pickup protons to the pressure beyond about 10 astronomical units (AU) from the Sun, increases from values $\beta \ll 1$ near the Sun, through values $\beta \sim 1$ near Earth orbit, to values $\beta \gg 1$ in the outer heliosphere. Voyagers 1 and 2 recently crossed the nearly perpendicular (at the locations of the Voyager traversals) solar wind termination shock, whose downstream pressure was inferred to be dominated by shock-heated interstellar pickup protons. The NASA IBEX Mission is currently measuring energetic neutral atoms to probe the global morphology of the termination shock and the other heliospheric boundaries. In several years the European Space Agency (ESA) spacecraft Solar Orbiter and NASA's Solar Probe Plus will explore for the first time the inner heliosphere inside 0.3 AU. Solar Probe Plus will reach a distance of 0.05 AU from the Sun. With these missions we shall have a complete heliospheric shock laboratory featuring observed shocks with a very wide range of plasma beta and magnetic obliquity, and a reasonable range of Mach number. This diverse collection of shocks allows us to investigate several outstanding questions in shock physics, including the magnitudes of ion and electron injection rates, the influence of upstream fluctuations on shock warps and magnetic field amplification, the structure of nearly perpendicular shocks and their efficacy for particle acceleration, the degree of distinction between quasi-parallel and quasi-perpendicular shocks for high intensities of upstream turbulence, and the huge observed variation in the intensities of energetic particles accelerated by apparently similar shocks. These shocks will provide a broad variety of cases to compare with numerical simulations. Furthermore, they will provide scalings for injection rates, upstream turbulence, and other quantities as functions of the Mach number, which should allow us to learn much about astrophysical shocks at higher Mach numbers.

## IMPACTS AND MAJOR OUTCOMES

Cosmic rays have a pressure comparable with the interstellar gas and magnetic field in our galaxy, and presumably other galaxies. Establishing their origin would be a tremendously ex-citing accomplishment, which would highlight the interconnected roles of energetic particles, shocks, and supernovae in determining the structure of galaxies and other astrophysical objects.

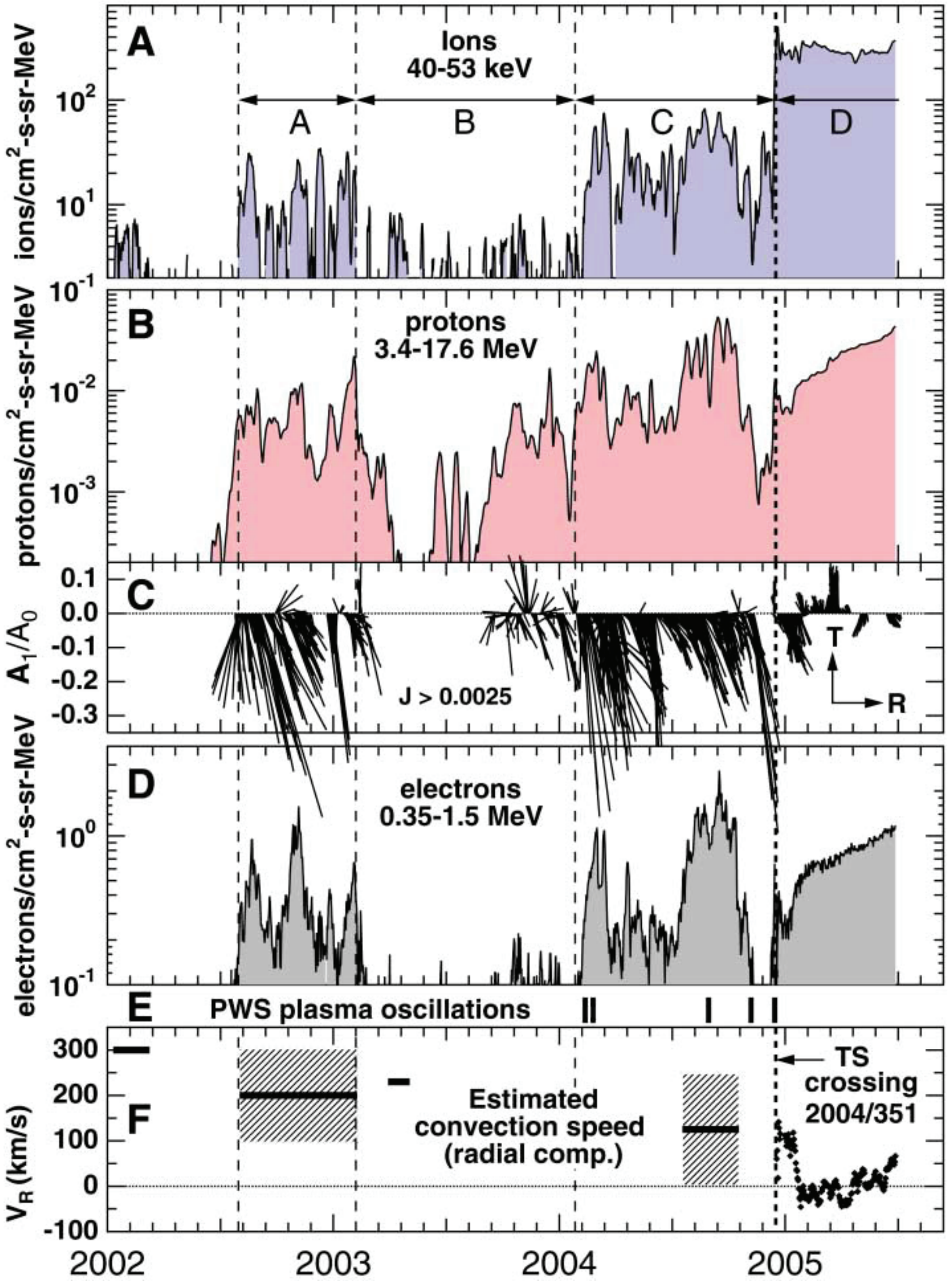

***Voyager 1 observations in 2002-2005 prior to, and during, the traversal of the solar wind termination shock, denoted by the vertical dashed line dividing Periods C and D. Panels A, B, and D show lower-energy proton and electron intensities, Panel C shows the first-order anisotropy vectors of the protons in Panel B, Panel E shows the time periods when plasma oscillations were detected, and Panel F shows the estimated radial component of the plasma velocity. The panels reveal a very irregular and probably very extended foreshock populated by protons and electrons escaping upstream from the shock and/or heliosheath. Courtesy of Decker et al. (Science, 309, 2020, 2005).***

A major unknown aspect of thermal plasma acceleration at a collisionless shock is the injection rate as a function of plasma parameters. The injection rate, which is not predicted by diffusive shock acceleration theory, is clearly dependent on the electromagnetic structure of the plasma shock transition. The opportunity to investigate shock structure and possibly injection rates as a function of magnetic obliquity in several current and planned laboratory plasma experiments, with supporting numerical simulations, is sure to significantly advance our understanding of shock structure and particle acceleration. Finally, the ongoing in situ observational and theoretical studies of shocks in the heliosphere — and their associated energetic particle populations — will provide insights into their structure and specific shock processes such as injection. The solar wind termination shock, recently encountered by Voyagers 1 and 2 and currently viewed remotely by IBEX, is challenging our ideas about shocks; in a few years Solar Orbiter and Solar Probe Plus will encounter shocks close to the Sun in a domain with small plasma beta unlike any we have observed previously. These observations of heliospheric shocks, with the support of numerical simulations and analytical theory, will provide scalings and insights into the nature of astrophysical shocks.

## CONNECTIONS TO OTHER TOPICS

Much of the earlier material in this chapter describes the unstable growth of magnetic fluctuations at a shock (and their associated velocity fluctuations, density fluctuations, and plasma heating) into large amplitude structures as an intrinsic feature of collisionless shocks. Since a shock can generally be viewed as a large-amplitude magnetosonic wave, this is perhaps not surprising. Although the growth of the magnetic fluctuations may be couched in terms of wave growth and quasi-linear theory, the importance of nonlinear wave-particle and wave-wave interactions is apparent, particularly for the strong shocks expected in interstellar space. The question arises whether these fluctuations evolve to a turbulent state in which the initial quasi-linear associations between velocities and wave vectors are lost to the turbulence characteristics. This must certainly be the case for sufficiently strong shocks. It also raises the question whether the distinction between quasi-perpendicular and quasi-parallel shocks has any meaning in a turbulent shock. Would a turbulent shock perhaps be amenable to a simpler theoretical description? This topic, at least for strong shocks, is clearly connected to Waves and Turbulence (Chapter 3). As is evident in this chapter's subsection titled, "Understanding Magnetic Field Amplification at Shocks," the amplification of magnetic field upstream of strong collisionless shocks also connects this topic to Magnetic Dynamos (Chapter 4). Finally, the common association of shocks with jets and outflows (e.g., the solar wind termination shock), in particular relativistic outflows (e.g., the shocks occurring in relativistic jets from AGNs), connects this topic to Jets and Outflows (Chapter 10), and to Relativistic, Pair-dominated and Strongly Magnetized Plasmas (Chapter 9).

# CHAPTER 3:
## WAVES AND TURBULENCE

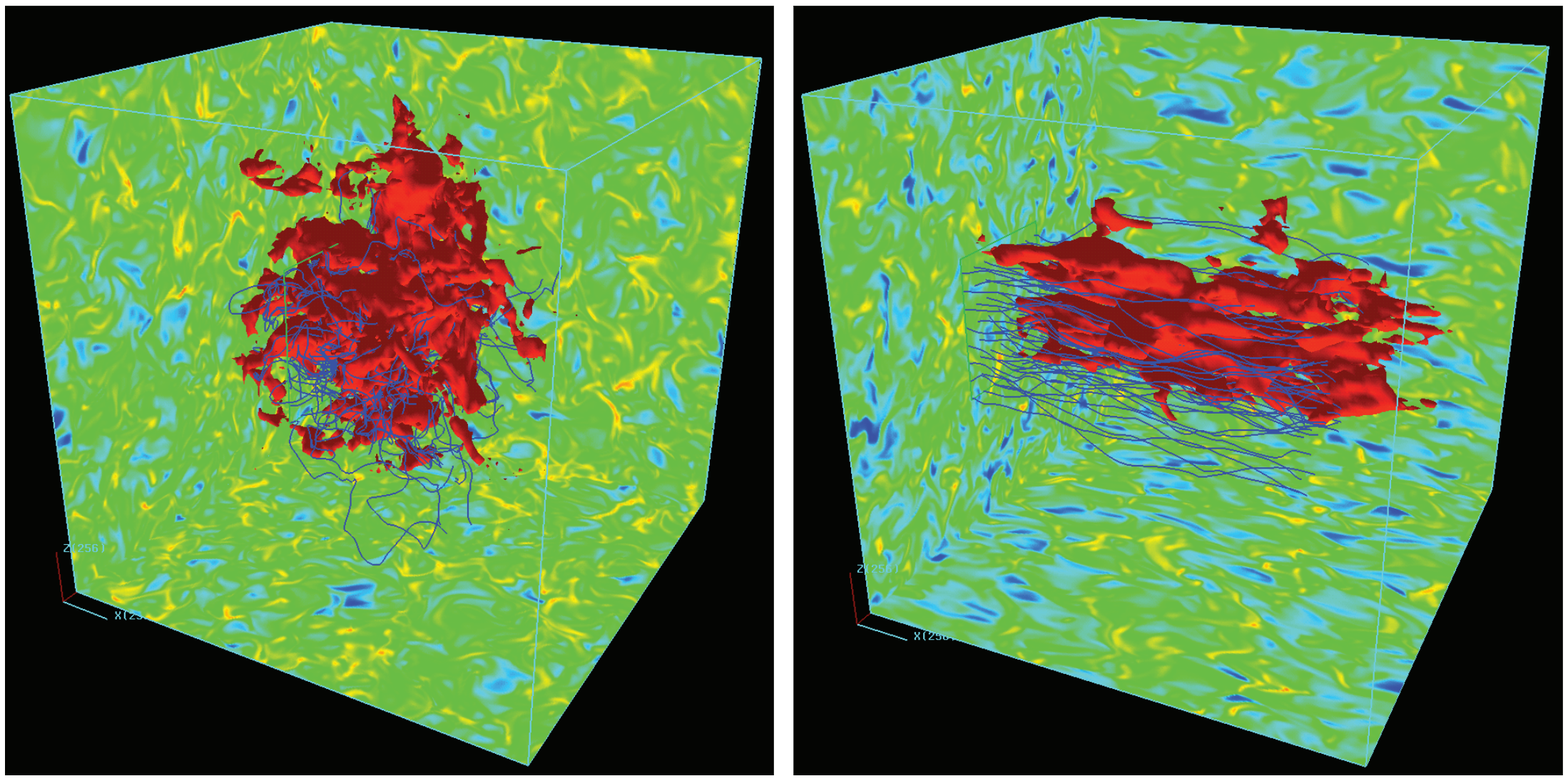

*ON PREVIOUS PAGE*

*The logarithm of the density on the faces of the computational volume for driven turbulence with a weak (left) and strong (right) magnetic field, the blue lines are representative magnetic field lines, and (to illustrate mixing of the fluid) the red isosurface shows a passive contaminant originally confined to a cylinder at the center of the box. Both models are computed at a resolution of $256^3$. (Image courtesy of James Stone of Princeton University.)*

# CHAPTER 3: WAVES AND TURBULENCE

## INTRODUCTION

Waves and turbulence are ubiquitous in space and astrophysical plasmas. Like fluid turbulence, plasma turbulence is one of the most important, unsolved problems of classical physics, and has significant implications for nearly every other topic discussed in this Workshop. The subject is treated extensively in many monographs and over a million papers, and it would be neither possible nor desirable to list all the important questions and strategies, and tackle them in this brief narrative. Our focus is limited to identifying some key questions, broadly framed, whose answers can be potentially transformative. While we will make connections with astrophysical objects such as accretion disks and the interstellar medium, we will treat in some depth only the solar corona and wind.

Although we focus on the solar corona and wind, it should be kept in mind that most stars have hot coronae, with temperatures exceeding a million degrees. The Sun not only has a hot corona, but also a hot wind stretching into the interstellar medium. Among the most important questions pertaining to the origins of the solar corona and wind are:

1. How are these nearly collisionless plasmas heated by waves and turbulence?

2. What is the nature of magnetohydrodynamics (MHD) and collisionless turbulence in these plasmas, which are permeated by magnetic fields?

3. What are the dissipation mechanisms, and their roles in particle acceleration and heating?

4. What are the effects of inhomogeneity and the role of coherent structures on waves and turbulence?

Following discussions on these key scientific questions, we then describe a few major opportunities in laboratory experiments, fluid and kinetic high-performance computing, and in situ and remote sensing observations. We conclude by summarizing the potential impacts of answering these key questions, and discussing connections to other topics.

## KEY SCIENTIFIC CHALLENGES

***What is the nature of MHD and collisionless turbulence in magnetized space and astrophysical plasmas?***

Plasma in the universe is magnetized and turbulent. Observations indicate that plasma fluctuations span a huge range of scales, from hundreds of kilometers to hundreds of parsecs. Observations of the solar wind and the interstellar medium (ISM) reveal qualitatively similar scaling laws of magnetic, velocity, and density fluctuations, which extend down to the ion gyro scales. There is significant debate as to whether there is a "universal" turbulent cascade in such systems.

At large scales compared to plasma microscales, MHD provides a good description of plasma dynamics. The plasma beta in space and astrophysical plasmas is often close to or greater than unity, distinguishing space and astrophysical plasmas from laboratory (including fusion) plasmas. The sonic Mach number is of the order of one, which distinguishes astrophysical turbulence from most terrestrial applications. Although such turbulence is compressible, incompressible one-fluid MHD is a useful point of departure. At small scales, which are close to or below the ion gyroradius ($\rho_i$) (or ion-sound scales in a plasma with $T_e > T_i$), plasma dynamics become much richer, as compressibility, two-fluid, and kinetic effects become important. Those scales are harder to address analytically. However, various two-fluid, gyrokinetic, and kinetic plasma modeling and simulations produce promising results. Observations of solar wind turbulence provide guidance in these studies.

Incompressible MHD turbulence exhibits certain limiting regimes, such as strong or weak and balanced or imbalanced. Different regimes may be present at different scales in the same system. For example, large-scale weak turbulence eventually becomes strong at small scales or globally balanced turbulence is locally imbalanced. It is important to understand under which conditions turbulence exhibits one of these regimes.

Weak MHD turbulence is dominated by Alfvén waves weakly interacting with each other. Its practical applications are limited, as turbulence in nature is typically strong. However, weak MHD turbulence permits fuller analytical treatment and serves as a test bed for fundamental ideas in the theory of MHD turbulence, such as anisotropy and a tendency to realize critical balance, locality, and self-similar energy cascades, and so on. Strong MHD turbulence assumes balance between linear wave propagation and nonlinear interaction. It lacks rigorous analytical treatment. Good physical models and numerical simulations are indispensable. Inherent local anisotropy is a fundamental property of strong MHD turbulence. Small-scale fluctuations are progressively more anisotropic at smaller scales as the balance between the linear and nonlinear interaction times is preserved independently of scale. This is the critical balance condition.

Both weak and strong turbulence can be balanced or imbalanced. Imbalance means that energy fluxes associated with Alfvén modes propagating in opposite directions along the guide magnetic field are unequal. Imbalanced turbulence has nonzero cross-helicity and it is not mirror-invariant. It is reasonable to believe that MHD turbulence occurring in nature and in the laboratory is typically imbalanced since it is generated by localized sources (e.g., solar wind or antennae in controlled experiments). Due to the constraints imposed by conservation laws, imbalance cannot be destroyed by MHD dynamics if dissipation is negligible. Imbalance seems to be an inherent property of strong MHD turbulence. Recent numerical results indicate that strong MHD turbulence is locally imbalanced even if it is balanced overall. It spontaneously produces correlated regions of imbalanced fluctuations of both positive and negative signs.

The various regimes of MHD turbulence can be described in terms of the shear-Alfvén modes, which are incompressible. Compressible effects are associated with the fast and slow modes, and with the entropy mode. These modes seem to be either strongly damped or dynamically unessential in the turbulence cascades. A variety of plasma processes could be responsible for their damping at various scales.

One point of view is that the limiting regimes are "universal." An observational example is that the ISM density fluctuation is Kolmogorov (i.e., proportional $k^{-5/3}$ over nine decades in wave number [$k$] space) and anisotropic, which stimulated the Goldreich-Sridhar theory of anisotropic turbulence. Another point of view argues against "universality," noting that the lack of universality occurs due to a number of reasons: the dependence of the turbulence on dimensionless physical parameters such as the plasma beta; the ratio of the magnitudes of the magnetic fluctuation to the background (or mean) magnetic field; and the nature of the driving and initial conditions. If MHD turbulence does not possess a single universal character, where do we go? It seems necessary to understand the cascaded ideal invariants in various parameters regimes, and that means understanding how turbulent relaxation processes operate in various parameters regimes. The spectra will be associated with fluxes of ideal invariants such as energy, while higher order statistics are associated with characteristic coherent structures. Better understanding of fast (less than an eddy time) and slow relaxation processes will thus clarify not only spectral variability, but also intermittency and its effects on topology of fields and flows, as well as turbulent dissipation.

Turbulence and inhomogeneity interaction represents a difficult problem, and methods for tackling it have been an active area of research, especially in fusion plasmas. Approaches include multiscale analysis leading to transport equations. These need to be tested against observations or very large and multiscale simulations for validation. This leads to extremely demanding computational problems. Fortunately, both computer simulations and theory are well positioned to make progress on these topics, which will immediately have an impact on applications as diverse as coronal heating, solar wind radial evolution, space weather, cosmic ray propagation, and galactic turbulence.

***What are the dissipation mechanisms and their roles in particle acceleration and heating?*** The inertial range of MHD turbulence in wave number space has a "break point" where the effects of dissipation typically leads to a steepening of the inertial range spectrum. In weakly collisional space and astrophysical plasmas, the beginning of the range of wave numbers where this occurs often has a two-fluid or kinetic origin (e.g., the ion skin depth or ion gyroradius [$\rho_i$]). While there is clear evidence of this dissipation regime in observations of plasma turbulence in the solar wind or the ISM, there are, as yet, no definitive theories. Theoretical and computational models based on two-fluid (or Hall MHD) equations, gyrokinetic equations, and fully kinetic models, have been put forward and compared with specific features of observations, with some successes. Several cases suggest that the dissipation range is itself multiscale, contains new power laws representing kinetic Alfvén and/or Whistler turbulence, and that wave damping or particle heating occurs at high wave numbers within this range.

For example, the shear Alfvén-wave turbulence becomes kinetic-Alfvén-wave turbulence when $k_\perp \rho_i \geq 1$. Then the ions decouple from the waves, and the electrons dominate the damping. As a result, the kinetic Alfvén waves do not undergo significant proton cyclotron damping in linear wave theory, but they do damp via Landau and transit-time damping. If kinetic Alfvén turbulence dissipates via Landau and transit-time damping, then the resulting turbulent heating should increase only the parallel component of the particle kinetic energy, thereby increasing the parallel temperature. On the other hand, in a number of systems such as the solar corona and solar wind, ions are observed to undergo perpendicular heating despite the fact that most of the fluctuation energy is believed to be in the form of low-frequency kinetic Alfvén wave fluctuations. The causes of such perpendicular ion heating are critical unsolved problems in the study of space and astrophysical turbulence.

Likely, there is not just one answer to questions about dissipation in turbulent astrophysical plasmas. In the so-called collisionless limit, there may be multiple mechanisms available, including those that operate in the parallel and perpendicular directions (of wave vector relative to the large-scale magnetic field). Most mechanisms have been identified traditionally within the context of homogenous linear Vlasov theory or other reduced kinetic or two-fluid models. A major theoretical problem is to understand this traditional approach's realm of accuracy. A complementary direction for theory is to look for inhomogeneous dissipation mechanisms, which may be associated with regions of strong magnetic or velocity shear, as well as regions of rapid variation of temperature or density. One promising candidate is dissipation in channels of strong magnetic shear, or electric current density. These are promising regions to look for strong inhomogeneous dissipation, even though magnetic reconnection may or may not be active in such regions. It is also reasonably well established (mainly, but not exclusively, from test particle simulations) that particles can be accelerated near current sheets and channels. The precise nature of this heating, especially its anisotropy, is currently being studied and discussed. Wave-particle interaction can also influence the suprathermal particle populations, and certain cases — such as cometary and interplanetary pick-up ion assimilation — are veritable laboratories for quantitative exploration and testing of the associated theories.

There are many different plasma environments in which the dissipation of MHD turbulence gives rise to particle energization that is in some way "preferential" (i.e., dependent on charge and mass, or anisotropic in physical space or velocity space). This has been seen not only in space and astrophysical plasmas, but also in laboratory experiments, including several types of fusion plasmas. Recent debate centers on the relative importance of various kinds of structure at the smallest dissipation scales. In many environments, however, the number of suggested mechanisms and structures is bewilderingly large. For example, the figure below illustrates a subset of the many different kinetic dissipation processes that have been proposed to explain the strong preferential heating of heavy ions observed in the solar corona. In many situations, ion heating is likely to be just the final stage of a multistep process of energy conversion among waves, turbulent motions, reconnection structures, and various kinds of distortions in the particle velocity distribution functions.

At present, there is no general understanding of how the smallest-scale turbulent fluctuations partition their energy between the different particle species. Simulation efforts are often focused, by necessity, on only one primary mechanism at a time. What is needed, however, is an objective assessment of the relative contributions from the large number of suggested dissipation processes. To do this, the scope of existing theory must be broadened to build true "sandbox models" that allow the most important processes to assert their dominance in the presence of many other competing processes. These broader models involve careful tradeoffs in that they may not have the computational rigor of the more focused models, but they would be able to answer a wider range of questions than the focused models. Examples of such tradeoffs could include: (1) not modeling the full dynamics of the turbulent eddies, but instead treating the cascade as a diffusion pro-cess in wave number space, and (2) parameterizing the results of nonlinear particle simulations in terms of net rates of heating. These broader models are likely to require increased collaboration among groups and increased community support for true working workshops (during which the specifics of these sandbox models are determined, and even the initial models are coded).

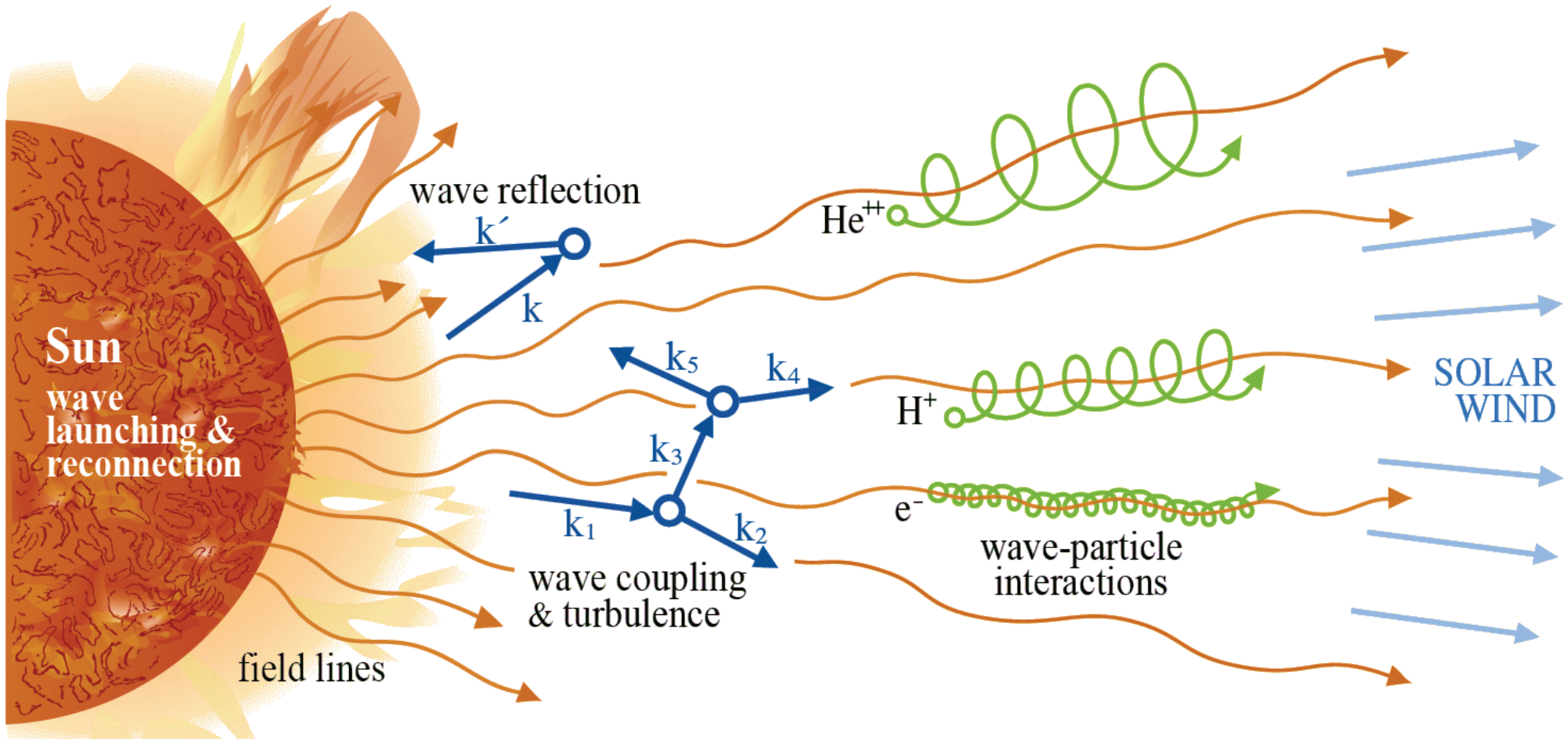

***Drawing illustrating various kinetic dissipation processes in the solar corona. Image courtesy of B. Chandran, M. Lee, and K. Donahue of the University of New Hampshire.***

### *What are the effects of inhomogeneity and interactions of turbulence with mean fields? What are the roles of coherent structures?*

Virtually all relevant or real instances of waves and turbulence involve inhomogeneity, which can drive turbulence (e.g., density and temperature gradients can drive drift-Alfvén turbulence, velocity gradients can drive magnetorotational turbulence). Inhomogeneity can reflect, modulate, and scatter waves (e.g., Alfvén waves in the solar wind) and can couple to velocity space structure (e.g., resonant Alfvén excitation by cosmic rays). Inhomogeneity in plasmas has the effect of coupling phenomena that are often separable in homogeneous plasmas. It can affect all of excitation (i.e., waves and instabilities), linear propagation and nonlinear transfer (i.e., scattering, reflection, and modulation), and dissipation (i.e., resonant absorption). Moreover, inhomogeneity can contribute to the relevant time scales in the problem, as in shear flow turbulence when in the rapid distortion limit.

Traditionally, MHD turbulence has been divided into distinct realms of turbulence that deal with issues such as cascades and structure functions, and "mean field" treatments that deal with dynamos, transport, and other mean field processes. This separation is increasingly seen as artificial. For example, it is now understood that a small-scale dynamo can alter or quench a large-scale dynamo and change the turbulence dynamics as well. Similarly, a strong mean shear flow can excite the magnetorotational instability (MRI) but also leave a "foot-print" on the dynamics of smaller scales, via rapid distortion. Mean field coupling, such as field amplification (i.e., dynamo) and turbulent resistivity (i.e., spatial transport or microscopic momentum exchange) will necessarily have an impact on turbulence and wave dynamics via enhanced dissipation, induced alignment, and nonlinear modification of cross-phases. Turbulence can either amplify or quench mean fields and flows, and so must be treated on an equal footing with them to satisfy relevant conservation laws.

Energetic particles are a ubiquitous means for exciting MHD and plasma Alfvénic turbulence. Of particular note and importance are cosmic rays, which can resonantly and nonresonantly excite the Alfvénic MHD turbulence that "confines" them to the shock, which in turn is thought to accelerate them by the direct shock acceleration mechanism. Many other examples exist as well. Energetic particles can excite waves by linear resonance, mediate nonlinear evolution via nonlinear wave-particle scattering (i.e., nonlinear Landau damping), and terminate excitation by nonlinear trapping. Strong wave-particle resonance leads to phenomenon such as structure formation, re-emission of waves, and frequency chirping, which are rich topical areas that merit further study and increased emphasis.

Understanding the coherent structures formed in plasma and MHD turbulence is a central theoretical, observational, and experimental issue. Coherent structures are likely to be central in understanding dissipation, and their formation and dynamics provide information critical to understanding cascade and relaxation processes. Based on hydrodynamic antecedents, one would expect that dissipation occurs mainly in coherent structures, although not exclusively in the most intense of these small-scale entities. Indeed, statistical intermittency of turbulence is connected with coherent structure formation. To a great degree the content of multi-fractal analysis and the study of higher order statistics is an effort to characterize quantitatively the nature of these structures. Recently, there has been considerable interest in understanding the relationship between observed near-discontinuous structure in the solar wind and MHD turbulence properties. These may not always be fossil classical MHD discontinuities, but might be produced by local turbulence cascade processes. Further study may be able to relate the current sheets and other discontinuities to rapid local relaxation due to turbulence. In this way, it is possible that relaxation might be related to a real-space picture of intermittency that is connected with observations and properties readily computed in simulations. Similar ideas may be applicable to the corona, as well as various observations showing strong evidence of structure that is probably perpendicular to the local magnetic field direction. Besides their relation to cascade, intermittency, and dissipation, coherent structures also might be related to larger scale topological issues such as the appearance, survivability, and structure of large-scale flux tubes. These, in turn, may be important in guiding or channeling energetic particle populations, in a way that random phase transport theory cannot capture. An example of this may be the well-known phenomenon of dropouts in solar energetic particle observations.

## MAJOR OPPORTUNITIES

### Laboratory Experiments

While laboratory experiments cannot typically match astrophysical parameters — either dimensional or dimensionless — they can contribute to our understanding of plasma physics phenomena relevant to astrophysical plasmas. Plasma waves, instabilities, and turbulence have been studied in detail in the laboratory for decades, and many concepts that emerged from these studies have been employed to explain space and astrophysical observations. For example, electromag-netic wave emission by plasmas due to mode conversion of Langmuir turbulence was first studied in the laboratory before being invoked to explain radio emission from pulsars. Similarly, properties of double layers, which are often formed due to electrostatic instabilities and turbulence, have a long history in laboratory experiments, and are a prominent candidate for auroral acceleration in the Earth's magnetosphere.

There are a number of opportunities pertaining to waves and turbulence in space and astrophysical plasmas that can be addressed by laboratory experiments, including fusion devices. These include the basic physics of nonlinear wave interactions and damping, important instabilities driven by anisotropy (e.g., mirror or fire-hose), and the properties of turbulent cascades driven at large scale either through driven flows or injected Alfvén waves. There are two complementary possibilities: "basic plasma devices," which have typically simple geometry, low temperatures (~10 eV) and high collisionality (except at low density), and very detailed probe dynamics; or fusion confinement devices that typically have more complicated geometry, high temperatures, and low collisionality, but are more difficult to diagnose even with sophisticated techniques. Since no laboratory experiments will match space or astrophysical parameters, a reasonable strategy is to identify physical processes that are common to both types of plasmas, and use theory and simulation to bridge the parameter gap. Presently, it appears possible to design a new basic plasma device that is weakly collisional (with system size comparable to the mean free path). This device would hold plasmas of moderate density and plasma beta of the order unity, with a large enough magnetic field to allow adequate separation between the system size and the ion gyroradius.

Laboratory experiments can also play a key role in testing the large-scale simulation codes that are of increasing importance in astrophysical research. For example, the kinetic Alfvén turbulence models for the solar wind are based on simulation codes developed to predict the behavior of laboratory plasmas. In particular, the measured decay of Alfvén wave energy in controlled laboratory experiments has provided a good test of theoretical damping models based on ion cyclotron and electron Landau damping. Predictions from kinetic simulation codes are being compared extensively with turbulence measurements in fusion plasmas such as tokamaks. This comparison can validate the code's ability to capture the physics of kinetic processes in collisionless plasmas, providing confidence in extending the simulation to plasmas of astrophysical interest, such as the solar wind or accretion disks. Anisotropic ion heating — and its isotopic dependence — is currently under study in reversed-field pinches by means of sophisticated diagnostic techniques. This may have important qualitative implications for analogous mechanisms of solar corona heating.

### High-performance Computing

Following Moore's law, which predicts the doubling of computing power every 18 months, doubling the Reynolds number and thus the grid resolution in three dimensions occurs every six years. Hence, direct numerical simulations (DNS) of turbulence advance slowly in the range of Reynolds and Lundquist number. This, and the fact that turbulent behavior may be dominated by spatially localized intermittent structures, are the driving forces behind one of the community's main objectives to develop a suite of methods that complement each other. These methods would incorporate realistic conditions that pertain to the many facets of plasma turbulence, from fluid to kinetic (both Lagrangian and Eulerian) models, exploring fundamental as well as more applied features in complex systems.

Enhancing the realism of fluid turbulence simulations to include, for example, a complex magnetic geometry, background inhomogeneity, or kinetic effects, requires enhancements in the mathematical models, and improvements in numerical algorithms and solvers. Parallel and algorithmic scalability are major challenges confronting fluid codes.

Kinetic turbulence simulations in plasma astrophysics have benefitted greatly in recent years from cross-fertilization between fusion and astrophysics. Traditional particle-in-cell (Lagrangian) as well as continuum (Eulerian) algorithms have enjoyed successes in describing important astrophysical phenomena. The challenge for kinetic turbulence simulations is to describe and resolve multiscale physics, which is mesocale and microscale physics.

Petascale and exascale computing initiatives, now under way at DOE and NSF, will be more effective and accessible to plasma astrophysicists if such initiatives recognize the strongly interdisciplinary character of plasma astrophysics. Plasma astrophysics is separate from the more traditional disciplines of astrophysics and plasma physics, with separate allocations tailored to the unique needs of the discipline.

**Opportunities For In Situ and Remote Sensing Observations**

Much of our knowledge about turbulence in distant astrophysical environments comes from remote observations that provide rather loose constraints on fluctuation properties. In astronomy, "turbulence" itself is often defined apart from its fluid dynamics roots, i.e., all that is often required is a collection of motions that are unresolved either spatially or temporally and have no clearly dominant frequency. In many cases, firm evidence for the existence of an actual turbulent cascade awaits direct in situ exploration. Substantial progress can be made, however, if the remote sensing observations are combined with theoretical modeling and extrapolations from existing in situ measurements. Theorists must be better informed about the kinds of measurements that exist (and what does not exist), and observers must be more aware that their data may be useful in advancing understanding in fields other than their own. In many cases there is insufficient communication among subfields in astrophysics, space physics, and laboratory plasma physics, such that there tends to be a "reinvention of the wheel" regarding analysis techniques and model code development.

In addition to well-publicized observations, there are also many existing plasma properties that have some kind of empirical constraints on their values, but have not been adequately "processed" or published in forms accessible to the theoretical community. More effort needs to be devoted to crosscutting analysis of archival data that may shed new light on important physical processes. In what follows, we discuss in greater depth the solar wind, widely recognized as a rich laboratory for turbulence studies.

The solar wind is the paradigm for more general stellar winds driven by magnetic activity. The solar wind flow at solar minimum is subdivided into high and low-speed streams, with speeds of around 750 km/s and 400 km/s, respectively (to be compared with the escape speed from the Sun, ~ 600 km/s). The Ulysses mission has shown that the fast wind is the basic outflow from the corona at solar minimum, while the much more irregular slow solar wind is confined to the equatorial regions, presumably arising from regions adjacent or inside the streamer belt. As the solar cycle progresses, the streamer belt expands in latitude so that, at activity maximum, the corona appears to be nearly uniformly distributed around the solar disk, while high-speed wind streams occur over a much smaller volume. The fast solar wind, with average speed around 750 km/s, originates from regions where the coronal electron temperature is lower. This inverse correlation between flow speed and coronal electron temperature, where the freezing in of minor ion charge states occurs, shows that the foundation of Parker's original theory of the solar wind (i.e., that high coronal electron temperatures and electron heat conduction drive the solar wind expansion) needs to be reconsidered. The Solar and Heliospheric Observatory (SOHO) satellite measurements of the very high temperatures of the coronal ions, together with the persistent positive correlation of in situ wind speed and proton temperature, suggest that other forces — namely magnetic mirror and wave-particle interactions — should also contribute strongly to the expansion of the outer corona.

SOHO observations have shown that the slow solar wind, which is confined to regions emanating from the magnetic activity belt and seems to expand in a bursty, intermittent fashion from the top of helmet streamers, expands continuously in X-rays. A third type of flow arises from larger eruptions of coronal magnetic structures, or coronal mass ejections (CMEs), which also lead to the acceleration of high-energy particles. As the solar activity cycle progresses, the simple fast-slow structure gives way to a more variable, but typically slower, solar wind at activity maximum. This apparently originates from the sparser coronal hole regions and quiet Sun, as well as from coronal active regions.

We believe that several fundamental plasma physical processes (i.e., waves and instabilities and turbulent cascades, as well as magnetic reconnection — another theme of this Workshop) operating on a vast range of temporal and spatial scales play a role in coronal heating and solar wind acceleration. Basic unanswered questions concern the storage, transport, and release of the mechanical energy required for coronal heating, the specific mechanism (s) for the conversion of energy between the magnetic field and thermal particles, and the dynamics of photospheric and coronal magnetic fields in the source regions of the solar wind. Issues pertaining to waves and turbulence strongly affected these questions.

***(1) What causes coronal heating and wind acceleration?*** The solar corona loses energy in the form of radiation, heat conduction, waves, and the kinetic energy of the solar wind flow. This energy must come from mechanical energy residing in photospheric convection — the solar magnetic field acting both to channel and store this energy in the outer atmospheric layers. However, the mechanisms by which the energy is transferred and dissipated to generate the hot corona, solar wind, and heliosphere throughout the Sun's activity cycle remain one of the fundamental unanswered questions in solar and heliospheric physics.

***(2) What causes the rapid acceleration of fast solar wind streams so close to the Sun?*** SOHO Ultraviolet Coronagraph Spectrometer (UVCS) observations using the Doppler dimming technique and interplanetary scintillation measurements indicate that the high-speed solar wind is rapidly accelerated near the Sun, reaching speeds of the order of 600 km/s within 10 $R_S$. Observations of comet C/1996Y1 confirm a most probable speed of about 720 km/s for the solar wind at 6.8 $R_S$. Such rapid acceleration appears to result from the extremely large and anisotropic effective temperatures in the lower corona, which have been measured by SOHO UVCS in coronal holes, though not directly for protons, the main solar wind constituent. These temperatures are much higher perpendicular to the magnetic field. The fast solar wind measured in situ shows what may be a relic of this anisotropy, smaller than that inferred from coronal observations, but persisting in the distance range from 0.3 to 5 AU. Proton, alpha particle, and minor ion distribution functions in the fast wind also present a nonthermal beam-like component whose speed is comparable to the local Alfvén speed. All these properties suggest that Alfvén or ion cyclotron waves play a major role in coronal heating and solar wind acceleration in high-speed wind.

***(3) Where are the different composition, plasma, and turbulence properties of fast and slow wind established?*** The fast solar wind flow is steady, with fluctuations in radial speed of order the 50 km/s, and the charge-state distributions indicate a low freezing-in temperature. The slow solar wind is variable, with higher but variable freezing-in temperatures. The composition of the fast and slow wind also differs, Mg and Fe being overabundant with respect to O in the slow wind. Solar wind protons and ions are, however, typically hotter in high-speed streams than in the slow wind. The difference between the fast and the slow solar wind extends to the shape of the particle distribution functions. The fast wind exhibits proton perpendicular temperatures that are slightly higher than the parallel temperatures. Proton distribution functions in the fast wind also present a beam accelerated compared to the main distribution by a speed comparable to the Alfvén speed, a feature shared by the alpha particles. Turbulence is also different in fast and slow streams. Fast streams contain fluctuations in transverse velocity and magnetic fields that are more strongly correlated in Alfvénic turbulence, a well-developed spectrum of quasi-incompressible waves propagating away from the Sun. In the slow wind no such preferred sense of propagation is observed, while larger density and magnetic field magnitude fluctuations are present, revealing a much more standard and evolved MHD turbulent state.

The three issues (1-3) do not exhaust the questions or important physical effects associated with the solar corona or wind. Our discussion has focused on issues pertaining mainly to waves and turbulence. We have omitted, for instance, any discussion of “velocity filtration” models that invoke nonthermal wings in particle distribution functions to account for coronal heating in a steady solar wind, or how coronal magnetic field structure orders the slow solar wind. We have also not discussed how impulsive events like nanoflares, microflares, or CMEs, in which magnetic reconnection is widely believed to play an important role, contribute to the intermittency observed in the solar wind.

## IMPACTS AND MAJOR OUTCOMES

The questions identified in the previous section are among the most important in experimental and theoretical studies of nonlinear waves and turbulence. Answering these questions and adopting some of the proposed solution strategies will have a broad and deep impact on plasma astrophysics and space science. We will be able to understand the nature of anisotropic turbulence in magnetized plasmas in the universe — how these plasmas develop on large scales and how they dissipate — and predict how they heat and accelerate ions and electrons. We will be able to predict how turbulence evolves in inhomogeneous plasmas and interacts with the background fields to which it is strongly coupled — invalidating the artificial separation between mean fields and turbulent fluctuations — and understand the conditions under which turbulence can amplify or quench mean fields and flows. We will understand the important role of coherent structures that spontaneously evolve out of turbulence, and how they affect cascades, relaxation, and dissipation processes in space and astrophysical plasmas.

## CONNECTIONS TO OTHER TOPICS

As mentioned in the Introduction, the topic of waves and turbulence touches upon and has significant implications for nearly every other topic in this Workshop. For example, reconnection in turbulent systems, despite some recent interesting results, is one of the least understood and important challenges in astrophysical plasma physics. Turbulent mechanisms for particle acceleration and heating are synergistic with shock and reconnection mechanisms. Turbulence has very important consequences for angular momentum transport, for the accretion process in stars, and for the conversion of magnetic energy to particle energy in jets and outflows. And without a better understanding of how turbulent fluctuations quench or amplify mean fields, which is at the heart of the dynamo problem, it is unlikely that we will understand how the universe is magnetized.

# CHAPTER 4:
## MAGNETIC DYNAMOS

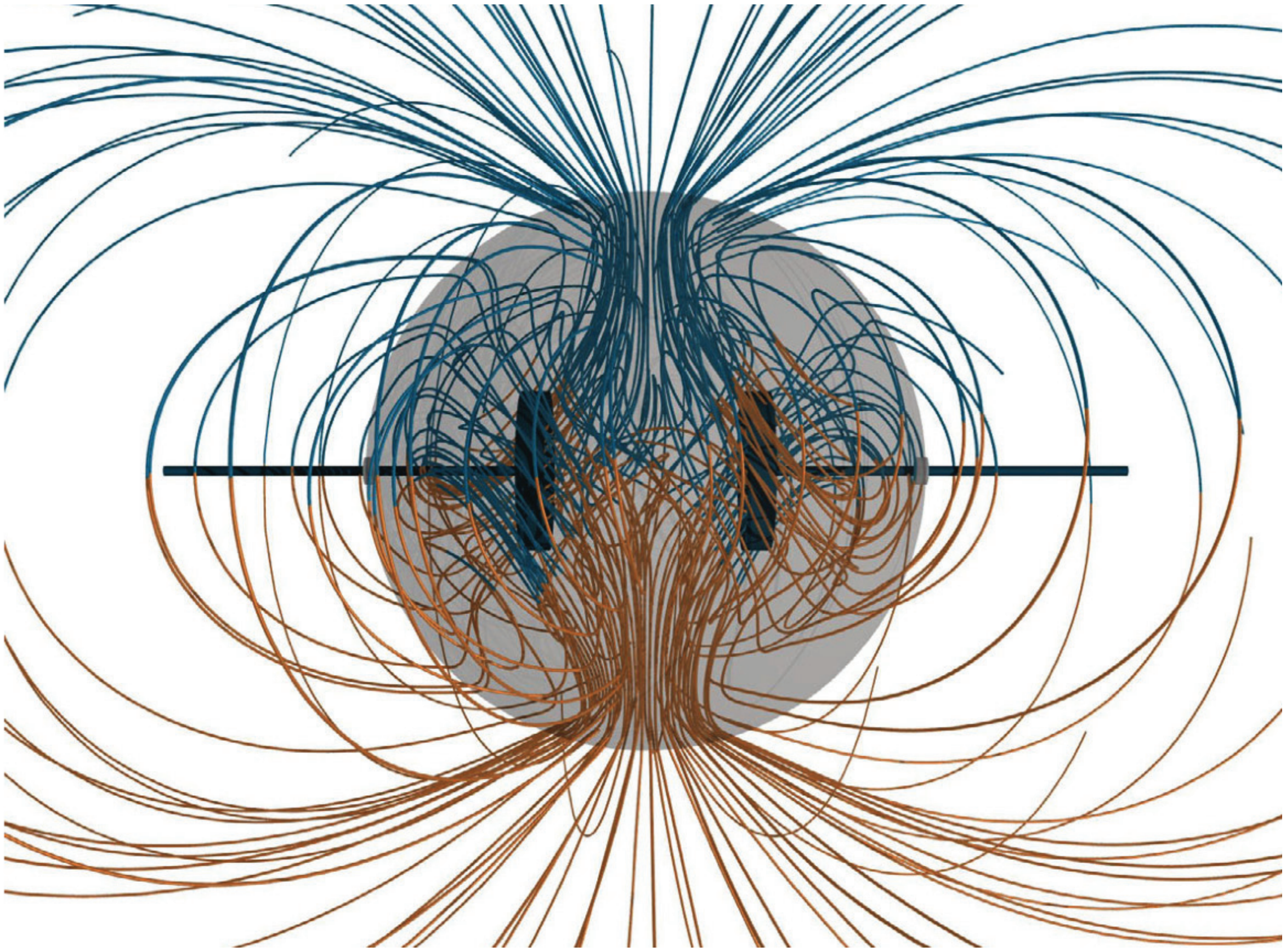

*ON PREVIOUS PAGE*

*Theoretically calculated magnetic field lines expected to be generated in the rotating, stirred, liquid sodium Madison Dynamo Experiment. In reality, dynamo activity is suppressed, possibly because of enhanced diffusion due to fluid turbulence. (Courtesy of Cary Forest, University of Wisconsin-Madison.)*

# CHAPTER 4: MAGNETIC DYNAMOS

## INTRODUCTION

A magnetic dynamo is a set of mechanisms that converts mechanical energy into magnetic energy, and sustains the magnetic field against dissipation. Dynamos produce the ordered, in some cases cyclic, magnetic fields observed in stars, galaxies, galaxy clusters, accretion disks, and jets. Understanding the origin of these fields, and being able to predict the dependence of their properties on the host system, are necessary to understand important aspects of stellar and galactic structure and evolution, and the nature of accretion. There also is a practical reason to study dynamos: the solar dynamo underlies solar magnetic activity, which drives space weather and affects the Earth's climate.

Astrophysical dynamos are generally flow dominated: gravitationally or thermally driven flow is the main energy reservoir. In accretion disks, disk galaxies, and some stars, differential rotation is the predominant form of kinetic energy. However, axisymmetric differential rotation alone cannot sustain the field. In most models of dynamos, small-scale turbulence provides additional induction. Turbulence can also accelerate the decay of the field, which is otherwise well frozen to the plasma.

A description of large-scale dynamo action by shear, turbulent induction, and turbulent decay was developed in the 1950s and 1960s. The theory is known as mean field electrodynamics (MFE) and the dynamo models are sometimes called $\alpha\omega$ dynamos after their parameterization of turbulent induction ($\alpha$-effect) and rotational shear ($\omega$-effect). MFE theory has been applied to interpret observations of solar, stellar, and galactic magnetic fields. Aspects of MFE also helped explain laboratory experiments, even though such plasmas are magnetically dominated rather than flow dominated. An $\alpha$-effect in a plasma-confinement experiment called a Reversed Field Pinch (RFP) was conjectured in the 1970s and confirmed in the 1990s. In the RFP, the $\alpha$-effect converts a poloidal to a toroidal field, and the toroidal field is sustained for longer than a resistive time. Flux conversion, also seen in spheromaks and during sawtooth crashes in tokamaks, represents the tendency of these magnetically dominated systems to relax to preferred states.

Serious challenges to MFE arose in the 1990s as a result of observation, theory, and numerical simulations. The solar interior differential rotation measured using helioseismology gives a $\omega$-effect, which causes the mean toroidal field to migrate away from the equator, opposite to what is observed in the Sun. On the theoretical side, analysis of the MFE equations and direct numerical simulation predicted that most of the energy in the magnetic field should lie at scales many orders of magnitude below what is observed in stars and galaxies.

Some ideas presently under development may address these problems. One idea is to follow the flow of magnetic helicity, which describes the linkage of magnetic field lines, and is conserved in magnetically closed, highly conducting systems. The MFE equations, however, do not properly conserve magnetic helicity. A modified theory, which couples turbulent kinetic helicity and the evolution of small-scale magnetic helicity to the evolution of the large-scale field, predicts the growth and saturation of a large-scale field seen in some numerical simulations. The theory also predicts that unless magnetic helicity can be ejected from the system, the growth of the large-scale field becomes very slow. How strong the field becomes with or without helicity ejection is a subject of current research.

Essentially nonlinear dynamos, in which the magnetic field is strong enough to affect the flow and has enough free energy to drive instability, are another approach to describing the saturated state. A simple, self-consistent model based on magnetic buoyancy and a shear flow provides an example of how a saturated, cyclic dynamo could operate. Further work is needed to show whether such dynamos can exist when turbulence is present, and to extend the models to astrophysical situations.

## KEY SCIENTIFIC CHALLENGES

Astrophysical dynamos operate under a vast range of physical conditions. Systems with dynamos vary over many orders of magnitude in the ratio of magnetic diffusion time to dynamical time (magnetic Reynolds number Rm), the ratio of viscous to magnetic diffusivity (magnetic Prandtl number Pm), the ratio of rotation period to eddy turnover time (Rossby number Ro), and collisionality, as well as in their dynamics, geometry, and the quality of available observations.

Two key challenges are essential in applying dynamo theory to astrophysics:

- How are large-scale, possibly cyclic, magnetic fields generated?
- How do dynamos operate in systems dominated by non-magnetohydrodynamics (MHD) effects?

Addressing these challenges will require interplay among theory, computation, experiment, and observation. The characteristic dimensionless parameters of astrophysical systems are too extreme to realize in either simulations or laboratory experiments (although it is possible in some cases to simulate experiments). Theoretical ideas are necessary to understand what to simulate and how to extrapolate, while observations provide ground truth, as well as opportunities to test predictions.

### Generation of Large-scale, Cyclic Fields

Solar, stellar, and galactic magnetic fields are coherent on scales significantly larger than the scales of turbulence in these systems. For example, in galaxies, supernova remnants tens of parsecs in size inject turbulence into the interstellar medium, but the magnetic field is coherent on scales of at least several kiloparsecs. The solar magnetic field, likewise, displays a dipole component and a coherent toroidal field, which is antisymmetric about the equator, but the largest scale of turbulent convection is a few tenths of solar radius. On the other hand, a significant body of analytical and numerical work suggests that in systems with large Rm, the magnetic power spectrum should peak at small scales. The underlying reason is that in order to amplify the field in a system of fixed volume, the field lines must lengthen in proportion to their amplification. This tends to produce a highly tangled field.

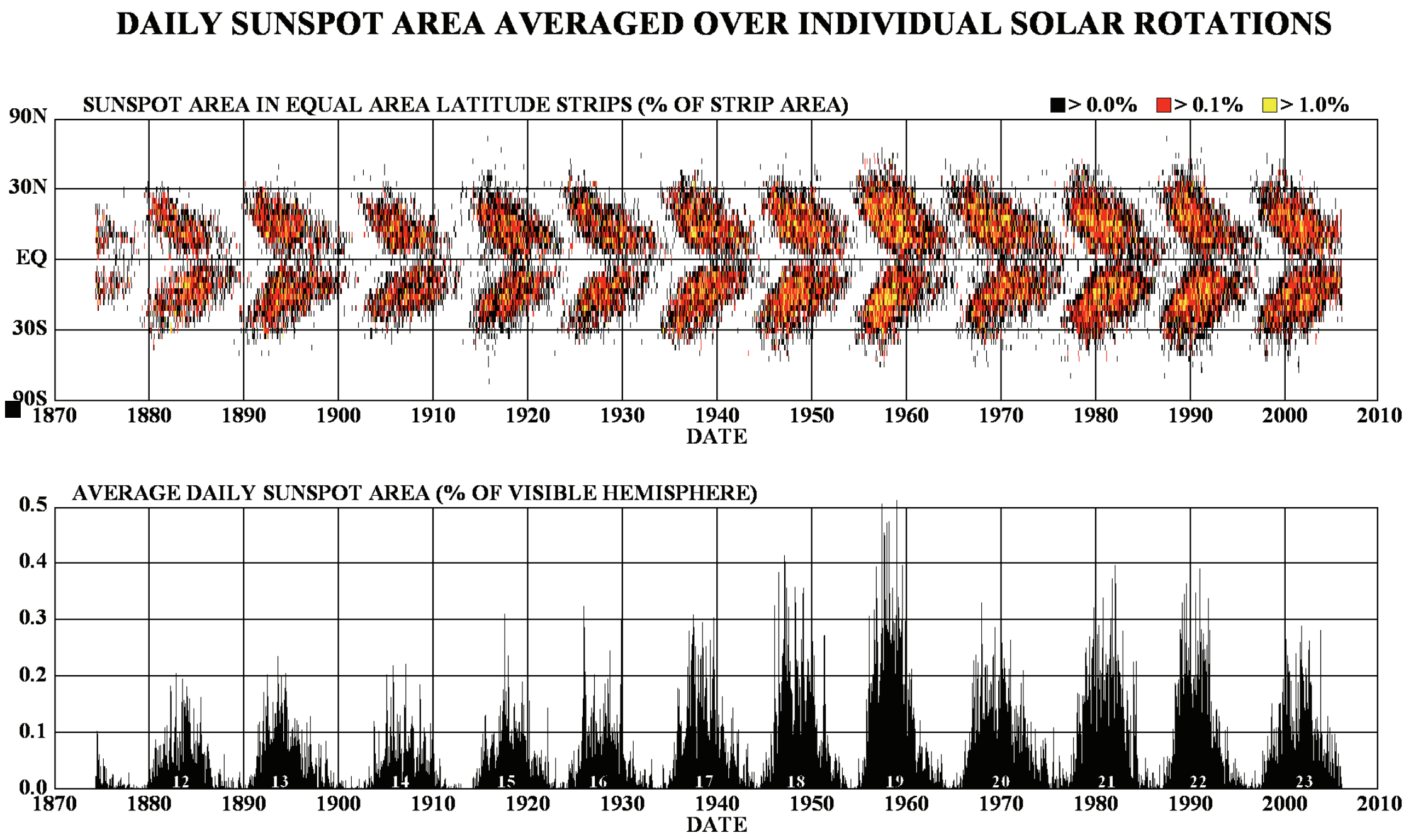

***Top panel: Famous "butterfly diagram" showing the distribution of sunspots on the solar surface over more than 80 years. Each pair of "butterflies" represents a complete solar cycle in sunspot number and magnetic polarity. Bottom panel: Area of the solar surface covered by sunspots over the same period, showing temporal regularity but considerable variation in cycle amplitude. The basic phenomenology of the solar cycle still defies detailed explanation. Image courtesy of Hathaway/NASA/MSFC, 2010/09.***

The cyclic nature of at least some magnetic fields is an important part of this challenge. Direct measurement shows that the 22-year solar cycle has existed, with minor interruptions, for millenia. Regular cycles on other stars also have been detected. Understanding how these cycles arise — and how they correlate with the structure of stars — is fundamental to understanding dynamos, and is an aspect of dynamo theory that can be tested observationally.

## Dynamos Beyond MHD

Although most studies of dynamos are based in resistive MHD, this approximation is inadequate for many astrophysical systems.

Dynamo action in the interiors of solar-type stars is well described by single-fluid, resistive magnetohydrodynamics. Flux escape, however, depends on physical processes in the outer atmosphere — processes that are collisionless. Many processes in interstellar and intergalactic gas, and in hot accretion disks surrounding compact objects, are also collisionless. In a collisionless plasma, stresses and transport coefficients can be anisotropic. This anisotropy affects instabilities in the medium, and causes new instabilities. The study of dynamos in collisionless plasmas is still in its infancy.

Effects beyond MHD can appear even in collisional systems. In highly luminous stars, and in some accretion disks, radiation pressure is significant and modifies turbulence. In a weakly ionized media such as cold interstellar gas or protostellar disks, ion-neutral friction leads to nonlin-

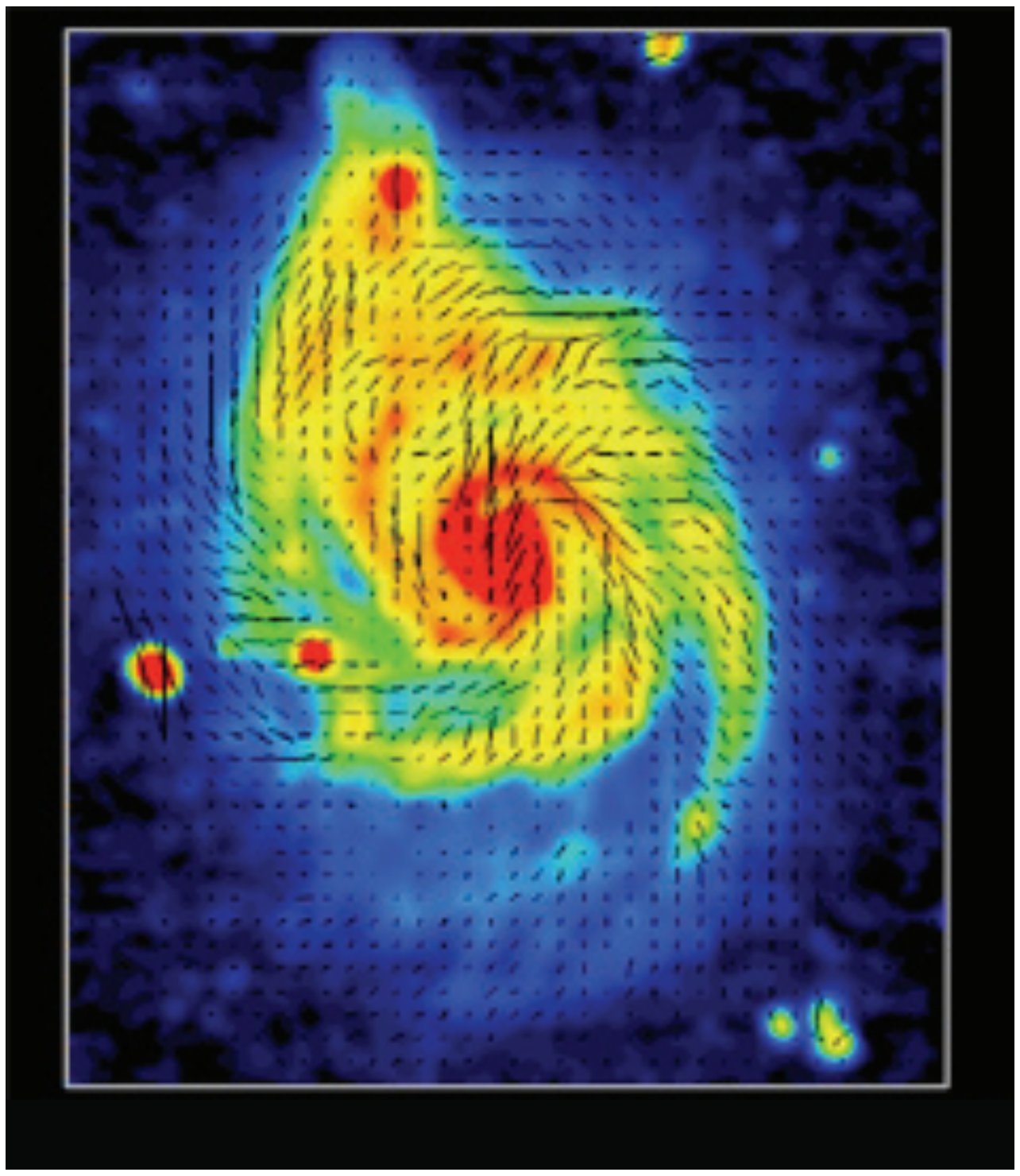

*Radio continuum brightness (color) and magnetic field orientation (line segments) in the spiral galaxy M51. Galactic dynamos must sustain an ordered magnetic field despite continual turnover of the interstellar medium. Copyright: MPIfR Bonn (R. Beck, C. Horellou & N. Neininger).*

ear transport of the magnetic field, can modify magnetic reconnection, and increases the length scale on which ions and electrons decouple, making the Hall effect more relevant. Relativistic cosmic rays, which are coupled to interstellar and intergalactic gas by scattering from small-scale fluctuations, provide buoyancy, stresses, heating, and viscosity. Although some of these effects are already included in dynamo models, or in the study of instabilities driven by buoyancy or differential rotation thought to be relevant to dynamos, no comprehensive treatment of dynamos in these regimes yet exists.

## MAJOR OPPORTUNITIES

High-priority observational programs include:

- Using existing and planned facilities to better characterize the solar magnetic field.
- Expanding the sample of magnetic fields detected on other stars.
- Producing more detailed, better sampled maps of the galactic magnetic field.
- Detecting magnetic fields in other galaxies at earlier cosmic times.
- Searching for an intergalactic field.

Two new solar observatories — the recently launched Solar Dynamics Observatory and the planned Advanced Technology Solar Telescope — will focus on solar activity and on dynamics of the solar convection zone. The Sloan Digital Sky Survey has catalogued many low-mass, active stars, allowing exploration of trends such as age-activity and rotation-activity relationships in stars. These and similar relationships are being followed up with the Kepler mission.

The capability to map galactic magnetic fields would greatly expand with two proposed facilities. The Square Kilometer Array, planned by an international consortium, has made cosmic magnetic fields a high priority. It would be able to detect synchrotron radiation from millions of galaxies and map the magnetic morphology of nearby ones in unprecedented detail. A proposed mission, CMB-Pol, is designed to detect gravitational waves generated during the Inflation era, but will (as a by-product) produce a high-resolution, ultrasensitive polarization map of the galactic magnetic field.

Ultrahigh energy cosmic rays and $\gamma$ rays provide additional constraints on the strength and structure of galactic and intergalactic magnetic fields. Detection of a pervasive intergalactic field would suggest that a top-down process that operates everywhere magnetized the universe.

**Laboratory Experiments**

Lab experiments, well supported by theory and simulation, can directly probe dynamo phenomena in a controlled and reproducible way. Such experiments currently lack a funding home at any agency.

Many important issues in dynamo theory could be probed by plasma experiments in the weak-magnetic-field, flow-dominated regime. This regime is nontraditional in the lab, but similar to natural plasmas. It would allow the effects of rotation, turbulence, compressibility, collisionality, various transport regimes, and boundary conditions to be probed for the first time in the laboratory. It would be possible to probe saturation mechanisms and distinguish different operating regimes, as well as study basic flow-related processes such as buoyancy-driven convection and large-scale circulation. Construction of one such experiment is already funded.

Studies of magnetic relaxation and the $\alpha$-effect in magnetically dominated laboratory experiments should continue. These experiments allow characterization of effects beyond MHD, the coupling of relaxation, the $\alpha$-effect, and momentum transport, and the role of magnetic stochasticity, and boundary conditions. These experiments, which are often motivated by fusion research, complement dedicated experiments needed to access the flow-dominated regime.

Liquid metal experiments complement plasma experiments by accessing the low-Pm regime encountered in cold, dense plasmas. These experiments can also probe mechanisms behind the $\alpha$- and $\beta$-effects, and the conditions necessary to achieve magnetic cycles.

**Theory and Simulation**

Theory and simulation can be used together to understand the extreme parameter regimes encountered in astrophysics, and to distill the results of simulations, observations, and experiments into simple theories. Theory and simulation support maximizes the impact of observations and experiments. Experiments and simulations can run in similar parameter regimes, offering opportunities to validate the codes and optimize the design of experiments. Extracting the maximum benefit from the observational programs described here, from the Sun to the intergalactic medium, will require extensive theoretical modeling and simulation. Much needs to be done numerically and theoretically to explore the vast parameter space of astrophysical dynamos.

Basic studies of magnetic field evolution in geometrically simple systems and studies of dynamos in global models of disks, stars, and galaxies are both useful. Analytic or semi-analytic models can address a large range of dimensionless parameters, and can control the input flow field and turbulent spectrum, but require approximations that need to be tested. Numerical models work within a narrower range of parameters and capture usually less than four decades of inertial range turbulence, but can address a wider range of nonlinear aspects of magnetic field evolution. Global models include the correct geometry and much of the important physics, but at a coarsely resolved level. Such models probe what geometric features, such as aspect ratio, are necessary to generate large-scale fields, better couple the dynamo process to the physics of the system, and can use natural boundary conditions.

## CHALLENGES FOR THE NEXT FIVE YEARS

Certain aspects of the key challenges in understanding dynamos should yield to progress in the next five years. There will be large bodies of data from the Solar Dynamics Observatory that better constrain magnetic fields and flows on the Sun, and data from Kepler and other satellites that constrain the parameter space of stellar dynamos. First results from the plasma dynamo experiment should be in hand. It will be time to put together detailed observing programs for cosmic magnetic fields, based on existing observations and modeling.

The time is right for a major thrust in dynamo modeling, requiring significantly expanded computational resources. Such a program would be similar in scope to efforts in climate modeling. It would include both global and local models, with the high degrees of spatial and temporal resolution necessary to resolve many decades of a turbulent spectrum. The most extensive models should be MHD because of the relative maturity of MHD dynamo theory, and the choice of models should be informed by theory and by smaller simulations. We also recommend committing significant resources to collisionless models that combine correct microscopic physics with the large-scale features of astrophysical systems, such as rotational shear and gravitational stratification. The program should be a community effort, with benchmarking studies to assess agreement between codes, and the computational output should be available to the community. Such a program is necessary to address the two major challenges of the subject: generation of large-scale, possibly cyclic magnetic fields (regular, like the Sun, or chaotic, like the Earth); and how dynamos operate in environments not described by simple MHD. A desirable outcome of these studies would be simple theories, or low-order dynamical models, that predict the main features of astrophysical magnetic fields. This goal is not necessarily achievable within the next five years, but it should remain within our sights.

Experimental progress demands development of a plasma dynamo experimental program. There are many steps toward fully realizing the potential of laboratory dynamo experiments: the development of advanced, fusion-research-grade, noninvasive diagnostics; the development of techniques to achieve buoyancy (a critical factor in many natural dynamos); and the development of technologies for driving flow in ways that best replicate astrophysical conditions. Simulation and modeling should be integrated with technological innovation to make the most efficient use of resources. The next generation of plasma dynamo experiments, informed by current and planned experiments — including possible upgrades to these experiments — also could be considered under the auspices of such a program. As the subject matures, a large experiment, operated as a user facility, could be a resource for the whole community.

## IMPACTS

The study of astrophysical dynamos as outlined through this combination of observation, experiment, simulation, and theory should be able to predict the overall structure, strength, any cyclic behavior, and the power spectrum of magnetic fields in astrophysical bodies. Because magnetic fields are a major constituent of the interstellar medium in galaxies and play a key role in angular momentum transport in stars and accretion disks, and because their origin in the universe is an unsolved cosmological problem, dynamo theory is fundamental to astrophysics. A few examples of problems where dynamo theory could have an impact are:

- Do interior dynamos directly drive coronal emission?
- Do accretion disks generate large-scale magnetic fields that significantly affect their angular momentum evolution and structure their jets and winds?
- How do magnetic fields affect star formation, including efficency, multiplicity, and initial mass function?
- How do stars, beginning with the protostellar phase, spin down?
- How is energy apportioned between magnetic fields, thermal gas, and cosmic rays in galaxies and galaxy clusters?

## CONNECTIONS TO OTHER TOPICS

The study of dynamos is closely bound up with other topics in plasma astrophysics, and there is natural synergism between them. There is a parallel synergism with other experiments through common needs for advanced diagnostics and other aspects of technology.

Magnetic reconnection, without which there can be no change in magnetic topology, is key to dynamos. Reconnection may suppress growth of the field at small scales and permit the escape of field through open boundaries. Because reconnection is an inherently small-scale process, effects beyond MHD may be important in reconnection layers even if not elsewhere in the dynamo medium.

There is interplay between dynamos and momentum transport. The torsional oscillations seen in solar surface rotation vary over the solar cycle and may be produced by magnetic torques. The flows established by the RFP dynamo may have an analog in flux conversion processes in astrophysics, such as launching jets by rotational shear. The structure of magnetic fields in accretion disks is fundamental to how they transport angular momentum.

Large-scale instabilities may place a role in the growth and evolution of large-scale magnetic fields. The structure of small-scale turbulence is also fundamental to dynamos, and dynamos affect turbulence.

# CHAPTER 5:

## INTERFACE AND SHEAR-FLOW INSTABILITIES

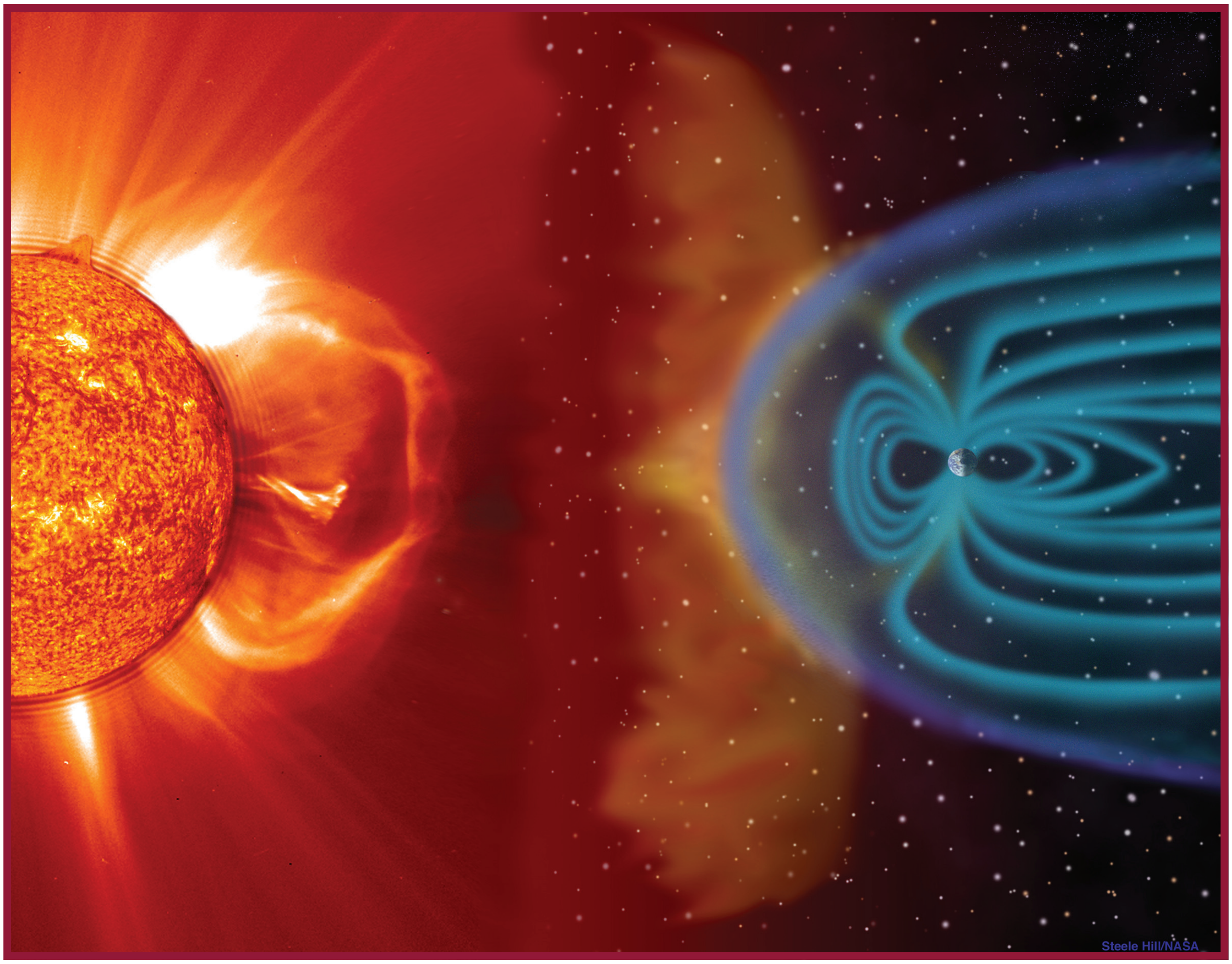

*ON PREVIOUS PAGE*

*A Coronal Mass Ejection makes its way from the Sun to the Earth. The plasma processes (such as shear instabilities) at the interface between solar wind and the Earth's magnetosphere (shown in blue) are important in determining entrainment efficiency of solar plasma into the magnetosphere as well as its dynamical responses. (Image courtesy of NASA/Steele Hill.)*

# CHAPTER 5: INTERFACE AND SHEAR-FLOW INSTABILITIES

## INTRODUCTION

Interface and shear-flow instabilities are ubiquitous in space physics and astrophysics. They play a critical role in systems such as the solar and stellar wind flow around the planetary and pulsar magnetospheres, the transitional region from solar wind to interstellar medium, photo-evaporated molecular clouds, supernova explosions, blast waves in supernova remnants, and many other systems.

Although a linear stage of any instability is important in defining the characteristic modes, a full effect of instabilities cannot be understood without getting to their nonlinear, and often turbulent, stage. This is where the key questions, challenges, and opportunities lie.

In this short summary, it is impossible to provide a comprehensive assessment of the status of research in the area of interface and shear-flow instabilities, and to identify all the possible points of future rapid growth in observations, simulations, and theory. We present only a few examples of the processes occurring at a broad range of scales and astrophysical objects, as well as their laboratory counterparts.

For example, the Kelvin-Helmholtz (KH) instability may play a vital role in energy transport into the magnetospheres of magnetized planets and other astrophysical bodies. Recent theoretical and multi-satellite observational studies at the Earth — especially with the Cluster constellation — have suggested that vortex merging and secondary reconnection triggered within shear-flow vortices may critically modulate the efficiency of these transport processes. Shear-flow instabilities, which generate surface waves, may also be important for energy transport in the solar corona.

On much larger scales, from a fraction of a parsec to megaparsecs, interface and shear-flow instabilities determine the properties of astrophysical jets, as well as their collimation, intermittent behavior, and length. The electric current flowing along the jet may cause intense kink instabilities, and the shear flow may play an important role in stabilizing them. Several laboratories, such as Imperial College London, the University of Nevada-Reno and the University of Washington in Seattle, have designed experiments to investigate how axial flows affect the stability of magnetically confined plasma columns. These experiments produce large aspect-ratio plasma jets, which have a significant axial flow that is sheared in some cases. More detailed investigations could provide insight into the possible stabilizing effect and, with proper scaling, would allow comparisons to astrophysical jets.

Instabilities driven by the normal acceleration of interfaces determine a broad class of astrophysical phenomena. These instabilities (primarily the Rayleigh-Taylor and Richtmyer-Meshkov instabilities) play a very important role in Type-II supernova explosions. Questions remain about explosion process details. Improved understanding would help form a reasonably complete picture.

## KEY SCIENTIFIC CHALLENGES

***Challenge 1: How does the magnetic field affect instabilities and further nonlinear behavior of astrophysical systems (accretion disks, supernovae, supernova remnants, jets and outflows, and dense molecular clouds)?***

**Existing Research Capabilities**

The critically important role of the magnetic field in the behavior of astrophysical systems has long been recognized and, in some cases, supported by direct observations with radio interferometers (CARMA, VLA, SMA), single-dish radio telescopes (GBT) and optical telescopes with polarimeters (CFHT).

Techniques have been developed for introducing dynamically significant magnetic fields to high-energy density (HED) physics experiments and directly measuring them with proton deflectometry. There exist laboratory facilities with both magnetized plasma jets (where velocity profiles and plasma structure are measured) and unmagnetized plasma jets (where plasma structure and global velocity are measured, including interactions with plastic).

Magnetorotational instability (MRI) is principally responsible for angular momentum transfer in the inner regions of disks around compact objects. Angular momentum transfer determines the accretion rate. In the last decade or so, most of our understanding of accretion-disk magnetohydrodynamics (MHD) has come from advances in theory and numerical simulations. Astronomical observations are only now catching up.

**Gaps**

Direct observations of astrophysical magnetic fields are rare. We lack a sufficient collecting area to do large survey work. Existing facilities could study a few examples of each kind of astrophysical system as typical principal investigator experiments. There is a need for detailed model predictions at the level of synthetic observations. In HED experiments, researchers have made only a few attempts to imitate the effects of magnetic fields in astrophysics, for instance, jets.

Thus far, laboratory experiments conducted with jets do not have some important ingredients of astrophysical objects, such as the presence of an axial magnetic field and significant external density. Theory and simulations have not explained paradoxical resiliency of jets; this may indicate incomplete physics. Numerical modeling does not include anisotropy of transport coefficients. In many cases, diagnostics are inadequate to make detailed comparisons with models.

Magnetized shocks in the supernova remnants do accelerate cosmic rays, but the cosmic rays must have a back effect on the structure and stability of these shocks. Little is known regarding ripple instabilities of the shock front in such a setting. A clear answer to the stability problem may help explain the complex structures observed in supernova remnants. It is unknown to what extent the ripple instability of the collisionless shock front depends on the shock's orientation with respect to the upstream magnetic field. So far, we have not observed shock stability or instability signatures. The presence of the ripple instability may introduce additional, mesoscopic-scale features in the structure of the shock front. These processes could be assessed in properly designed laboratory experiments with collisionless plasmas. To our knowledge there is no organized and supported activity in this direction.

While MRI has been reproduced in laboratory fluid experiments, it is more difficult to reproduce in scaled laboratory plasma experiments. An observational strategy also must be developed to identify the accretion's most salient features related to momentum transfer. The goal would be to differentiate models. Laboratory experiments may provide insight into such a strategy.

***Challenge 2: How can we understand the interaction of the solar wind with the geomagnetic field through the smallest scales? How do we characterize the role of kinetic effects and magnetic reconnection in the nonlinear development of KH instabilities and interchange instabilities, and their impact on the efficiency of plasma mixing, transport, and energy propagation?***

**Existing Research Capabilities**

Fluid-scale constellation (e.g., ESA-NASA Cluster) and local single-satellite missions (e.g., Geotail, THEMIS) have characterized the development of nonlinear KH instabilities at the magnetopause — including the potential importance of secondary magnetic reconnection inside nonlinear KH instability vortices — and of interchange instability in dipolarization fronts in the plasma sheet. Three-dimensional nonlinear simulations have examined these effects in the fluid domain in global magnetospheric simulations, or at the kinetic scale in simplified geometries and in limited spatial domains.

**Gaps**

Satellite observational capabilities for understanding nonlinear KH instabilities and interchange instabilities development are mostly limited to fluid scales and single-point measurements and generally exclude the kinetic regime. Extended phase missions often focus on science targets where measurements from a single platform or mission can yield closure. Targets requiring multiple missions, and hence multipoint measurements (e.g., as a "great observatory"), are often de-emphasized or lacking. Kinetic 3-D simulations are generally not possible at global magnetospheric scales. Local simulations are often limited by simplifications such as periodic boundary conditions, less realistic geometries, or by limited spatial or temporal resolution and limited domain. The coupling between the development of shear-flow structure and vortices and kinetic-scale processes in the intra-vortex reconnection regions is also not well understood.

The development of coherent structures within the time-dependent behavior of shocks remains poorly understood. The formation and reformation of coherent structures appears to play a key role, especially in quasi-parallel shocks, but the details of how these structures form, their timescales, and their relationship to shock processes at the fluid, electron, and ion scales are not well-known. Observational studies are confined mostly to single-point measurements. Recent observations also have suggested that coherent, wave-like structures may form on the shock front itself. This is unexpected and not understood.

***Challenge 3: Can we apply HED physics techniques — both experiments and simulations — to National Ignition Facility (NIF)-based laboratory astrophysics experiments relevant to interface and shear instabilities in radiation-controlled astrophysical systems?***

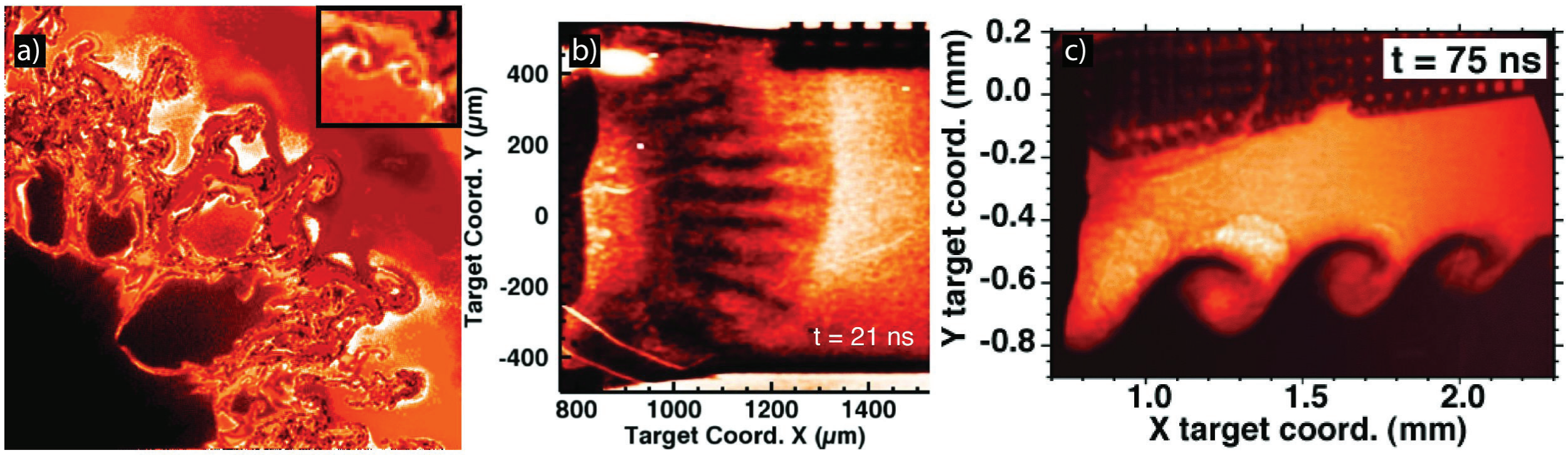

***(a) Results from a 2-D calculation of SN1987A showing the instability growth at the He-H interface of SN1987A. The Rayleigh-Taylor instability causes dense spikes of He to move outward and lower density bubbles of H to penetrate inward. The Kelvin-Helmholtz instability is also seen in these results (inset) caused by shear at the spike-bubble interface. Through laboratory experiments, we can study these instabilities individually. Experiments performed at the Omega Laser Facility investigate (b) the Rayleigh-Taylor instability and (c) the Kelvin-Helmholtz instability.***

## Existing Research Capabilities

There are numerous HED physics experiments on the laser-driven Rayleigh-Taylor (RT) and KH instabilities. Computer capabilities include: petaflop-scale computers; direct numerical simulation of 3-D RT with a Reynolds number up to 32,000 — just above the Re~20,000 turbulent mixing transition; very high-resolution 2-D simulations; sub-grid-scale models to capture the effects of unresolved scales; and multidimensional multiphysics astrophysics codes.

## Gaps

Smaller scale HED physics facilities cannot reach the regimes of the radiation-dominated processes. While there are many codes, they often do not meet the needs of specific problems. Smaller scale facilities cannot achieve a radiation-pressure-dominated regime. There are limited diagnostics on NIF, and funding and facility time are both currently sparse.

The RT instability plays a very important role in Type-II supernova explosions. Scaled laboratory experiments have already been used to explore relevant unstable flows and validate codes used to simulate supernova hydrodynamics. However, there is no consensus view that can explain the available data, including evidence of the very fast mixing of material from the deep interior into the outer layers of the progenitor.

From very early in the explosion, deformed shocks drive Richtmyer-Meshkov instability. It is not clear how this instability seeds and interacts with the RT growth that later dominates. Once perturbations have grown to large amplitudes, there can be a transition to an inherently three-dimensional turbulent mixing zone. This presents significant challenges for both computational and experimental approaches. Numerical simulations are limited in the effective Reynolds numbers they can attain, and techniques have not been developed for diagnosing turbulent HED laboratory systems.

In computations, we are unable to do fully developed 3-D turbulence with extended inertial range. The treatment of transitional flow is incomplete, and astrophysical codes are invalidated through routine application to relevant nonlinear-phase experiments.

## MAJOR OPPORTUNITIES

### *Challenge 1*

In observations, telescopes currently under construction will broaden the scope of research. The Atacama Large Millimeter Array (ALMA) telescope, with a large collecting area and very high resolution, will allow extensive surveys of many classes of objects. The Large Synoptic Survey Telescope (LSST) will open a new regime of time-domain astronomy to potentially investigate the evolution of jets, disks, and supernova remnants. In laboratory experiments, we can use the capabilities of HED physics facilities (NIF, Z, and Omega) combined with advanced diagnostics. More important than facilities are people: we must broaden collaboration among observers, modelers, and experimenters.

Magnetic fields typically have been neglected in astrophysical simulations of supernovae and laboratory experiments. It may be quite challenging to introduce a dynamically significant magnetic field in laser-driven experiments, but the reward could be significant, providing a test-bed for validation and verification for the MHD codes used in simulating magnetized supernovae.

Additional factors can potentially be introduced by the differential rotation of the magnetized progenitor. Rotation may also be responsible for a typically high peculiar velocity of the Type-II supernova remnant. Reproducing low-mode-number supernova instabilities in laboratory experiments would be a great step forward. It may be challenging, but not impossible, to design experiments that would imitate the explosion of a rotating star. These experiments may also shed light on the connection between the instability structure created during the rapid explosion and the structure observed much later in the remnant.

### *Challenge 2*

Current and future proposed constellation mission operations should be coordinated to cover ion, electron, and fluid scales at the magnetopause and plasma sheet to target nonlinear KH instability development and transport, "dipolarization front" interchange instabilities, and the role of such coherent structures in plasma dissipation and turbulence. Future missions include NASA MMS, JAXA-CSA SCOPE, and ESA Cross-Scale.

Future solar missions such as Solar Probe and Solar Orbiter may be important for establishing the role of these shear instabilities in the acceleration and properties of the fast and slow solar wind and energy transport during large-scale magnetic topology changes in solar-active regions. At the heliopause — and in rotating planetary and astrophysical magnetospheres — interchange instabilities may also be contemporaneously important with shear-flow instabilities.

In the case of the magnetosphere and heliosphere, collisionless effects strongly affect the detailed structure of shocks and other interfaces, making this whole area of research rich for plasma physics. The Cross-Scale and SCOPE missions may provide unique opportunities to study fundamental collisionless shock processes with multipoint measurements at the electron, ion, and fluid scales.

Collisionless shear-flow instabilities are often present in fusion devices and are believed to significantly affect their performance. This has been the focus of much effort in experimental, ana-lytical, and numerical studies. Most of the results pertain to a low-beta plasma, but in some cases, plasmas with beta approaching unity have been studied, most notably in mirror devices. Shear flows in collisionless plasmas — much like their hydrodynamic counterparts — may drive instabilities, but they may also lead to the stabilization of more virulent instabilities. In fusion devices, shear flows often take the form of zonal flows, which also exist in space and astrophysical plasmas. We believe the exchange of information among fusion scientists, astrophysicists, and space physicists could be more intense and lead to helpful cross-fertilization. Perhaps there could be an attempt to identify a set of problems of common interest (e.g., in the area of Earth's magnetosphere).

### *Challenge 3*

NIF can achieve new regimes that can be relevant to instabilities occurring in astrophysics. It can use larger and more complex geometries. More extensive participation is needed in interdisciplinary meetings, and collaborations with astronomers and astrophysicists in laboratory experiments should proceed from inception through execution. The field must take a long-term view in addressing the community integration and young-researcher-influx problems by establishing a program to involve astrophysics students and postdoctoral researchers in laboratory astrophysics. We also need more cross-fertilization in diagnostic development, etc., as well as conceptual guidance and simulation.

In supernova physics, the effects of spherical divergence may become important, and multiple mixing zones can begin to interact. However, most supernova-motivated laboratory RT experiments have focused on a single interface in planar geometry. Higher energies available on new facilities such as NIF should allow the community to significantly broaden the scope of its experiments. For example, we should create an experiment with an RT-unstable interface and a radiative shock to observe how radiation affects the instability. This could be related to red supergiant supernovae, where the reverse shock is strongly radiative and interacts with the RT-unstable shocked ejecta.

In the case where ablation pressure accelerates the interface, an "ablatively driven RT" may develop. This instability is the basis for a leading model of the formation of complex structures of photo-evaporated molecular clouds. For instance, the instability may be responsible for the formation of pillars in the Eagle Nebula. The dynamics of photo-evaporated clouds are likely strongly affected by magnetic fields, both regular and turbulent. Direct observations of the magnetic field in both the dense gas regions and in the ionized hydrogen regions that border them would help solve the mystery of molecular-cloud dynamics. Additional insights could come from developing an adequate laboratory platform for scaled simulation of the ablation instability, both with and without the magnetic field.

Exaflop-scale computing (considered feasible this decade) would reach another factor of 10 in Reynolds number, allowing unambiguous, fully developed, RT-turbulent mixing. An integrated "cradle to grave" stellar modeling approach would provide better coupling of stellar modeling to explosion modeling, and explosion modeling to remnant modeling. The field could collect a benchmark set of instability experiments defined and characterized in astrophysics-code-friendly terms and deliver them to the astrophysics community as a tool for code validation.

## IMPACT AND MAJOR OUTCOMES

### *Challenge 1*

Detailed observations of the magnetic field in astrophysical objects would certainly lead to breakthroughs in understanding their behavior. This could be accompanied by scaled laboratory experiments on a variety of facilities, both with low-density (collisionless) and high-density (usually collisional) plasmas. Specifically, we could confirm or reject models of supernova explosions, accretion disks, and photo-evaporated clouds that imply an important role of the magnetic field.

### *Challenge 2*

Constellation-type missions with many tens of nano-satellites would measure spatial and temporal characteristics of instabilities and turbulence in a collisionless plasma with an unprecedented level of detail, possibly exceeding the level attainable in laboratory experiments. This would lead to truly comprehensive models of collisionless turbulence, a subject of great intellectual value.

### *Challenge 3*

Broad use of new HED physics facilities (NIF, Z, and facilities to be built in other countries) for dedicated, astrophysics-related experiments would make it possible — in a scaled fashion — to reach the domain where radiation pressure and the magnetic field play dynamically significant roles. This would allow experimentalists to peek into the processes occurring in systems such as supernovae and accretion disks around black holes.

## CONNECTION TO OTHER TOPICS

Interface and shear instabilities are ubiquitous in astrophysics. They may set the stage for the further development of nonlinear motion and turbulence. On the other hand, they are affected by the processes occurring at micro-scale in the collisionless domain. They are directly connected to several chapters in this report.

Chapter #1 *(Magnetic Reconnection)*. Shear-flow instabilities affect reconnection physics, in particular, by shearing the magnetic field and preparing the stage for reconnection.

Chapter #2 *(Collisionless Shocks and Particle Acceleration)*. Microturbulence that determines the structure of collisionless shocks will certainly affect macroscopic (ripple) stability of such shocks.

Chapter #3 *(Waves and Turbulence)*. Interfacial instabilities often set the stage for further development of turbulence.

Chapter #6 *(Angular Momentum Transport)*. Shear-flow instabilities make transport possible.

Chapter #8 *(Radiative Hydrodynamics)*. In many cases, radiation strongly influences gravity-driven instabilities, particularly in accretion disks. Radiative drive is of prime importance for the stability of photo-evaporated fronts.

Chapter #10 *(Jets and Outflows)*. Both Kelvin-Helmholtz and Rayleigh-Taylor instabilities are important in this general area.

# CHAPTER 6:
# ANGULAR MOMENTUM TRANSPORT

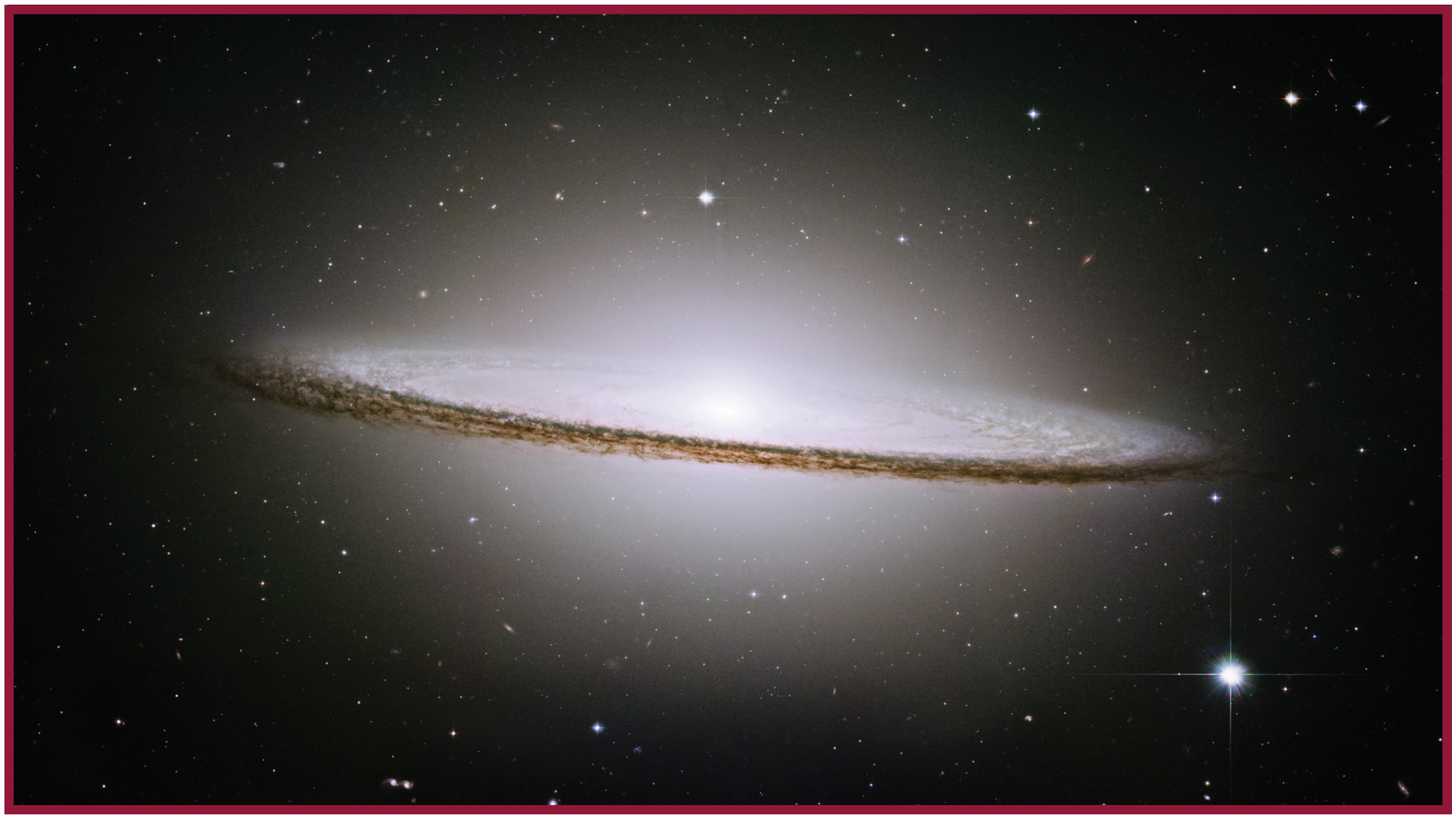

ON PREVIOUS PAGE

*An image of M104, the Sombrero Galaxy, taken with the Hubble Space Telescope. The galaxy consists of a bright white core of spherically distributed stars and a disky component of stars and gas, seen nearly edge on. The galaxy is 50,000 light-years across. Momentum transport shapes the properties of nearly all astrophysical plasmas, from the large scales of galaxies like M104, to the small-scale accretion disks around black holes.* (COURTESY OF NASA AND THE HUBBLE HERITAGE TEAM, STSCI/AURA.)

# CHAPTER 6: ANGULAR MOMENTUM TRANSPORT

## INTRODUCTION

The redistribution of angular momentum plays a critical role in most astrophysical systems. It is responsible for shaping much of the structure we see around us in the universe and for powering the brightest known sources of electromagnetic radiation. Disk galaxies — like our own Milky Way — form as plasma flows toward the center of gravitational potential wells established by dark matter, eventually settling into a disk when rotational support becomes significant. Stars form inside galaxies even today as gas clouds collapse, shed angular momentum, and eventually reach temperatures at which nuclear fusion begins. And planets — including the Earth — form as gas and rocks coalesce in the rotationally supported disk of debris surrounding a new star.

Rotation also plays a critical role in the death of stars and in the properties of the resulting compact objects (white dwarfs, neutron stars, and black holes). How massive stars explode as luminous "supernovae" after fusion ceases critically depends on the rotation of the star. The redistribution of angular momentum is also the power source for the brightest sources of electromagnetic radiation in the universe: accretion of plasma onto black holes can produce energy up to 50 times more efficiently than nuclear fusion.

Magnetic fields are one of the key ways of transporting momentum in astrophysical plasmas because these fields can transmit forces over relatively long distances. (For the same reason, gravity is important on galactic scales.) As a result, understanding angular momentum transport in astrophysical plasmas requires understanding the origin, destruction, and coherence of magnetic fields. It is thus intimately connected to the problems of astrophysical dynamos, magnetized turbulence, and magnetic reconnection.

## KEY SCIENTIFIC CHALLENGES

To highlight several key scientific questions related to angular momentum transport in astrophysical plasmas, we focus on some of the outstanding problems in the areas of stellar astrophysics and accretion disks. These areas were identified because of their broad astrophysical importance and because the importance of momentum transport by magnetic fields in ionized plasma has been well established.

### Stellar Astrophysics

The surface of the Sun rotates differentially, with a fast equator and slow poles. Measurements of millions of solar acoustic oscillations reveal that this rotation profile largely imprints through the convective envelope, but that there is a transition to nearly solid-body rotation in the stably stratified region below. The narrow boundary layer of shear between these two regions, called the tachocline, figures prominently in most models of the global solar dynamo. Although the differential rotation within the convective envelope is widely thought to arise, at least partially, from momentum transport by the turbulent flows, a basic understanding of how this occurs (and how such transport depends upon parameters like the rotation rate) remains elusive. The transition to solid-body rotation in the radiative region is likewise poorly understood, though momentum transport by magnetic fields and internal gravity waves is thought to be crucial.

Recent observations have revealed differential rotation in other stars as well. These observations, which typically involve either examining spot patterns at the stellar surface or subtle analyses of rotational line broadening effects, have generally been possible only during the last few years, but will continue to grow in number with the advent of space-borne photometers like Kepler and COROT. Such differential rotation must figure prominently in the evolution of these stars — through its role in magnetic dynamo action, for instance — and so demands detailed theoretical modeling.

The mean rotation rates of stars are not constant, but typically decrease in time as stellar winds and magnetic fields remove angular momentum, spinning down the star. How stellar spindown affects the interior rotation is, however, uncertain, and depends on how well magnetic fields couple the interior and exterior of the star. This uncertainty translates into significant uncertainty in how stars end their lives, and in the "birth" rotation and magnetic fields of white dwarfs, neutron stars, and black holes. For example, it is now observationally well-established that the most relativistic explosions in the universe — long-duration gamma-ray bursts — are associated with the collapse and explosion of massive stars at the end of their lives; but the rate of such gamma-ray bursts is only ~ 0.1 percent of the core-collapse supernova rate. This diversity in stellar death is almost certainly tied to diversity in the rotation and magnetic fields of massive stars, but this is not qualitatively understood, either theoretically or observationally.

**Accretion Disks**

Accretion is the inflow of matter toward a central gravitating object. In general, the inward accretion of matter requires an outward transport of angular momentum. Collisional viscosity is incapable of producing the level of angular momentum transport needed in astrophysical disks, and thus turbulent transport is required for accretion to proceed. In 1991, Balbus and Hawley showed that magnetorotational instability (MRI) is important in many astrophysical accretion flows. The MRI is an instability of a differentially rotating magnetized plasma in which the magnetic field is amplified on a timescale comparable to the rotation period of the disk. Magnetic tension exchanges angular momentum between fluid elements, allowing some of the plasma to flow inwards. Nonlinear simulations have shown that turbulence driven by MRI can be an effective angular momentum transport mechanism. However, many outstanding questions remain. Broadly speaking, one of the key challenges is to increase the realism of the physics in disk simulations, and to use such calculations to make predictions that can be quantitatively compared to observations. Comparisons between simulations and laboratory experiments may also provide a useful testing ground for models of turbulence in astrophysical disks.

Numerical studies of transport in disks were initially based on the equations of ideal magnetohydrodynamics (MHD), which are appropriate for a fully ionized collisional plasma. However, in cold and dense flows (e.g., planet-forming disks around young stars) the plasma may be mostly neutral, and the ideal MHD approximation does not apply. Instead, non-ideal processes such as the Hall effect and ambipolar diffusion are crucial. At the other extreme, the inflowing plasma near a black hole is sometimes so hot and rarified that it is effectively collisionless. Fully kinetic calcula-tions of angular momentum transport are required to understand such systems, for example the ~ 3.6 million solar mass black hole at the center of our galaxy; these have yet to be carried out. Simplified fluid models suggest that angular momentum transport in collisionless disks can be quite different from that predicted by MHD. A related question is whether turbulence (or reconnection) in these accretion disks can accelerate some of the particles to suprathermal energies. If so, this could significantly modify the spectrum of radiation from such systems.

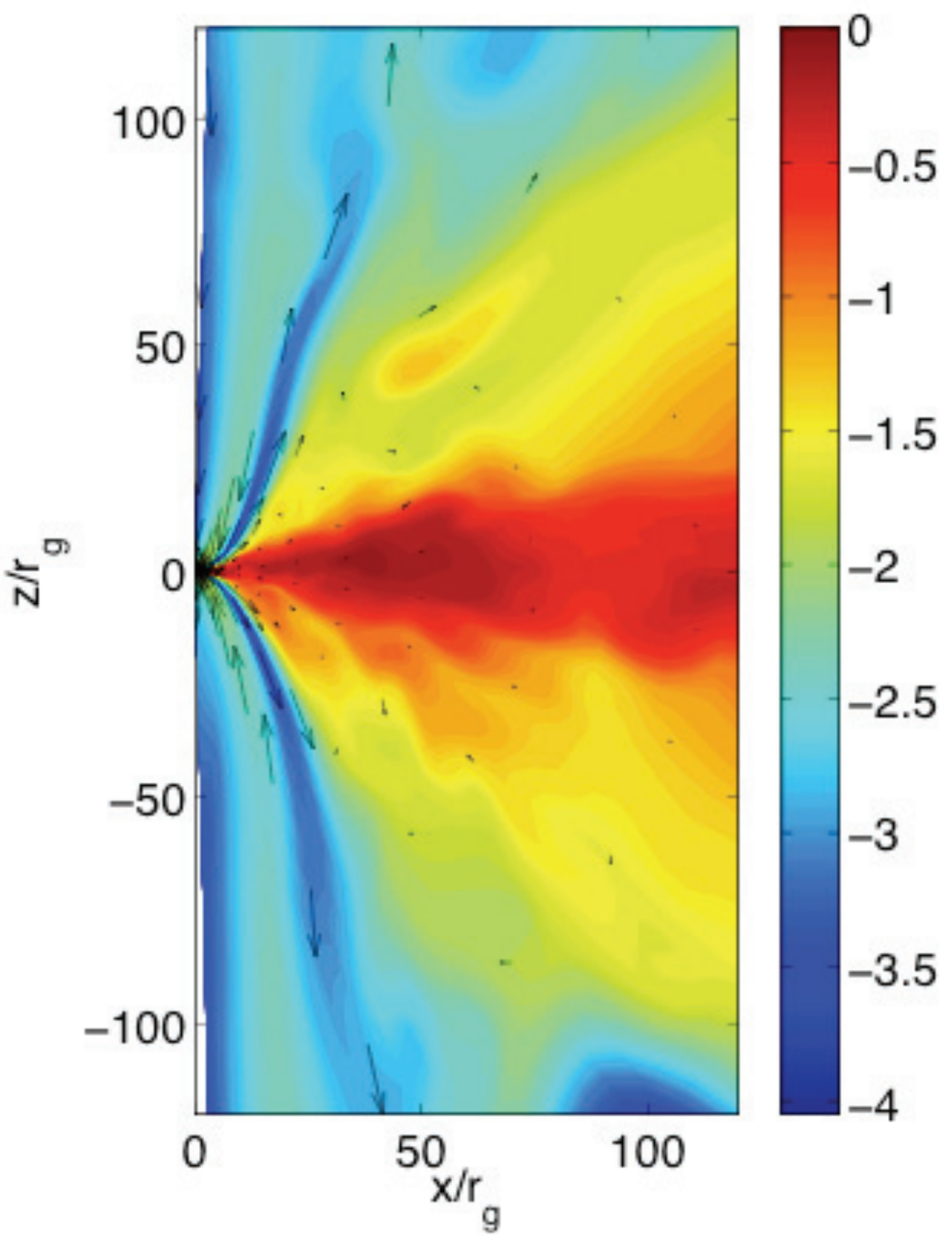

***Color plots of log density in arbitrary units in the quasi-steady state of an MHD simulation of black hole accretion. Arrows indicate the velocity direction and show the presence of an outflow near the disk's surface in the polar direction. Image courtesy of Prateek Sharma.***

For the most luminous accreting systems, the energy density in the radiation field close to the central black hole or neutron star greatly exceeds that internal energy density of the plasma. In such cases, the radiation field is thus dynamically important and must be self-consistently included in the dynamics. There has been significant progress in studies of the MHD of radiation-dominated flows using the flux-limited diffusion approximation and gray (frequency independent) transport. However, flux-limited diffusion breaks down right where it matters most: near the photosphere, where the photons we see originate. Moreover, frequency dependent transport (or at least, the inclusion of the effects of spectral lines) might be important for understanding radiation-driven outflows from disks, and relativistic effects will be important in the very innermost regions of disks around black holes. An enormous amount of work remains in studying radiation-dominated flows, and experiments may provide crucial insights into the dynamics in this regime, as well as allowing validation of the numerical methods.

## MAJOR OPPORTUNITIES

Astrophysical systems typically encompass an enormous range of spatial and temporal scales, with momentum transport likely involving many disparate scales. The physical processes that are relevant at one scale may not apply at others — for instance, some systems may be collisional in some places and collisionless elsewhere. No one computation or experiment is likely to model the physics of all these scales correctly in the foreseeable future. Analytic theory is thus critical for determining the im-portant physical processes and the important length and time scales that must be resolved. By then combining results from numerical modeling, ever-more-detailed astronomical observations, and laboratory experiments — each of which probe largely distinct parameter regimes — we hope to arrive at a comprehensive understanding of transport in astrophysical environments.

On the computational front, the ongoing increase in processor performance will likely afford continued progress. That increase will ultimately allow simulations that model higher-resolution domains with greater and greater fidelity. The resolution requirements for conceptual progress vary depending on the problem. In the Sun and other stars, for example, the largest scales of motion relevant for momentum transport are likely of order the depth of the convection zone (about 200 million meters); estimates of the smallest relevant length scales vary, but a plausible upper limit is the width of the solar tachocline region, thought to be a few percent of the solar radius (or about 2 Mm). In the MHD context, transport may depend on growing or parasitic modes on much smaller length scales. Together, these imply that the latest generation of stellar convection simulations — with resolutions of roughly $1000^3$ — are only beginning to capture much of the relevant physics. For accretion disks, the range of length scales may be even wider, varying over 10 or more orders of magnitude. In such cases, resolving all scales is not possible; but it also may not be needed. Calculations that have sufficient scale-separation (at least a factor ~ 10-100) between relevant but disparate scales may provide the needed insights into the dynamics. Within the next few years, petascale machines will enable modeling at even higher resolution and correspondingly greater realism. Such machines require extremely high levels of parallelism in the numerical algorithms, and it will be challenging for researchers to exploit these new computer architectures, and the enormous volumes of data that will be produced.

A new generation of astrophysical techniques and telescopes will provide a qualitative advance in our understanding of stellar rotation and accreting systems during the next five to ten years. The Solar Dynamics Observatory will dramatically increase our understanding of solar convection, interior rotation, and magnetism. High-precision photometry on new satellites, such as COROT and Kepler, is providing extremely precise stellar light curves and, in some cases, measurements of many different modes of stellar oscillation. Over the next five years this should provide a breakthrough in our understanding of stellar (differential) rotation as a function of mass and age; in

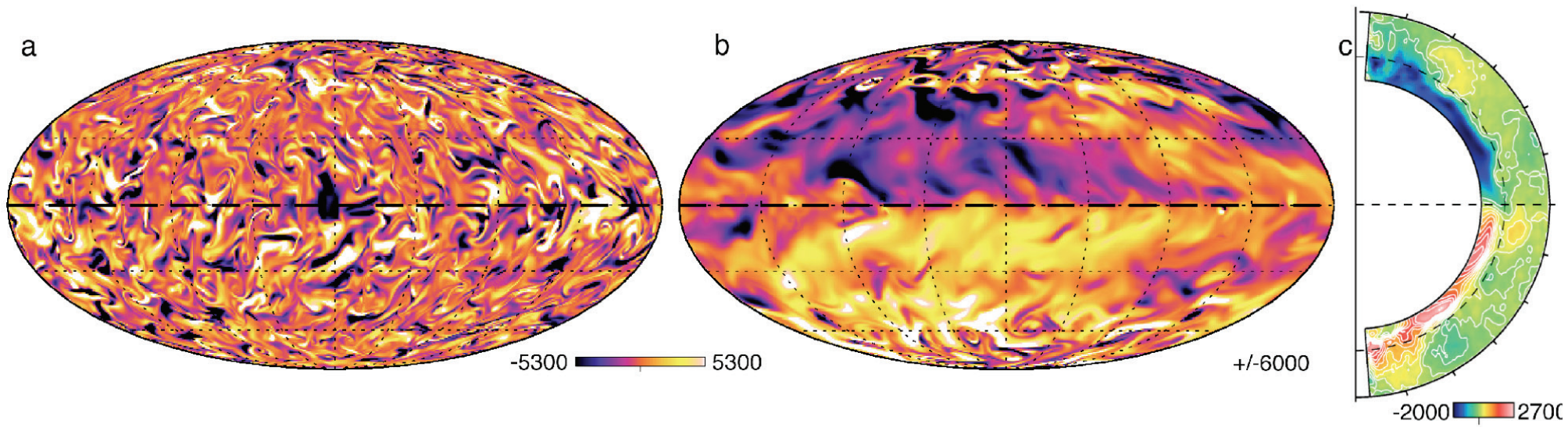

***Magnetic fields realized in a 3-D MHD simulation of turbulent solar convection penetrating into a stably stratified shear layer. Toroidal magnetic fields at one instant time are shown in Mollweide projection on spherical surfaces at (a) mid-depth in the convection zone and (b) in the stable region below. Complex magnetic fields in the convection zone have been stretched and amplified into more orderly structures in the shear layer. (c) Contours in radius and latitude of toroidal fields averaged in time and longitude reveal strong fields of different polarities in the northern and southern hemispheres within the stable region (below curved dashed line). Image courtesy of Matt Browning.***

principle, the rotational diagnostics can also be correlated with surface magnetic activity to better understand the relationship between rotation and magnetic field. These observational results will provide a wealth of data for comparison to models of stellar dynamos, differential rotation, and stellar death. This data will complement the much more detailed information available for the Sun.

Observations of accreting compact objects have traditionally been spatially unresolved. Our inferences are based largely on interpreting the spectrum of emitted radiation and its variation with time. In many contexts (e.g., the observed time variability), theory lags behind observations, and improvements in numerical models likely are required for significant progress. Qualitatively new observational insights will be provided, however, on several fronts. Using advances in interferometry at mm and infrared wavelengths, spatially resolved images of plasma in the vicinity of a black hole will be obtained for a few of the nearest massive black holes in galactic nuclei. Higher spectral resolution observations of X-ray lines from accreting neutron stars and black holes will provide direct constraints on the rotational structure of the underlying disk.

Experiments using rapidly rotating conducting fluids, such as plasmas and liquid metals, can be used to study the instabilities that give rise to angular momentum transport in accretion disks and stars. Instabilities arise from various free energy sources in astrophysical systems, including flow shear, pressure gradients, and current gradients. For each free energy source, there can in principle be hydrodynamic, magnetohydrodynamic, or small-scale plasma instabilities. Although there has been extensive theoretical work assessing which instabilities dominate under different astrophysical conditions (e.g., the MRI), much can be gained from a range of experiments that can produce these instabilities in a laboratory setting. They also may be used as a platform for validating numerical codes used in simulating astrophysical systems. Taken together, experiments and simulations span a broader parameter space than either individually; both will be required for constructing scaling laws that extrapolate to astrophysically significant regimes. Both hydrodynamic and magnetohydrodynamic experiments have been performed in Taylor-Couette flow established between differentially rotating cylinders. This simple device provides a standardized platform for studying shear flow instabilities that give rise to turbulence in rapidly rotating fluids. Spherical Couette devices with differentially rotating spheres also have been employed to study geometries closer to that of planetary and stellar dynamics. Most current experiments use either water (for hydrodynamic experiments) or a liquid metal, in particular, gallium or sodium (for MHD experiments). Proposed experiments in weakly collisional plasmas will significantly increase the range of astrophysically relevant conditions that can be studied experimentally.

A recent breakthrough on the experimental side was in the study of the hydrodynamic stability of differentially rotating Taylor-Couette flow. A purely hydrodynamic transition to turbulence would be a compelling model for cooler collisional accretion disks such as protoplanetary disks, which may be too resistive to sustain transport by magnetic fields. Initial experiments using water seemed to support such a transition to hydrodynamic turbulence, but careful measurements of the momentum transport in follow-up experiments demonstrated that quasi-Keplerian flows up to $Re \sim 10^6$ can be achieved with transport comparable to that due to microscopic viscosity (i.e., little turbulence). This difference in experimental results is attributed to differences in boundary layer control. At such high Reynolds numbers sharp boundary layers form along the container

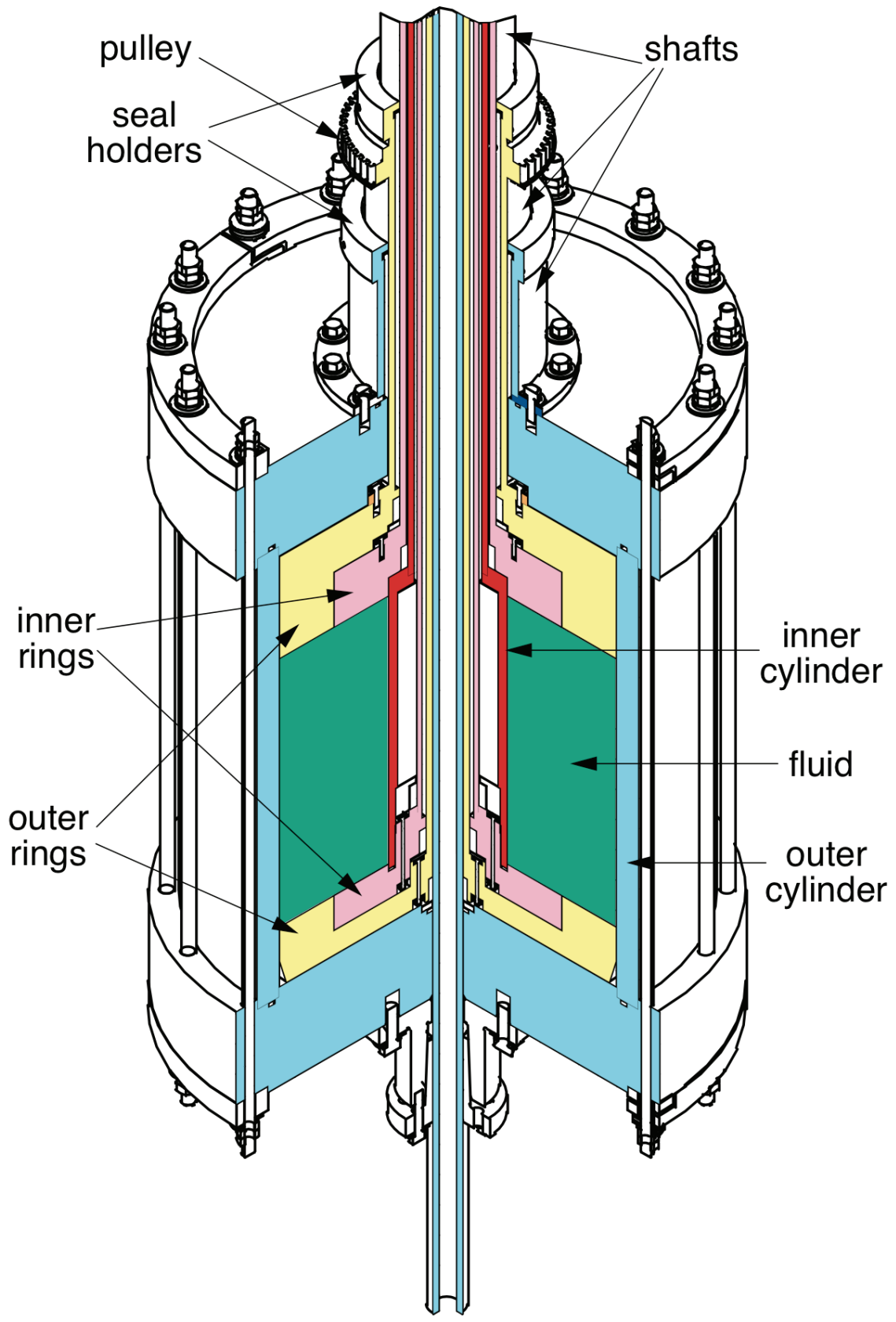

***A rotating fluid laboratory experiment used to study angular momentum transport in hydrodynamics and magnetohydrodynamics. Two novel features distinguish this apparatus from conventional Taylor-Couette experiments. First, secondary circulation is controlled by dividing each endcap into two independently driven rings. A second novel feature is access to rotation profiles similar in Keplerian disks at Reynolds numbers as large as $10^6$ for study of nonlinear hydrodynamic instabilities and at sufficiently large magnetic Reynolds numbers for triggering linear magnetorotational instabilities. Figure taken from H. Ji, M. Burin, E. Schartman, and J. Goodman, "Hydrodynamic turbulence cannot transport angular momentum effectively in astrophysical disks," Nature 444, 343 (2006).***

walls, resulting in a bulk circulation called Ekman flow. The large-scale radial flows are typically unstable and result in turbulence that enhances angular momentum transport. Disrupting or suppressing the boundary layers is required to create conditions in which the effects of bulk flow are not masked by instabilities in this secondary flow. Further use of boundary layer control in experiments may be critical for experimental studies of angular momentum transport.

### Challenges for The Next Five Years

We expect continued progress from the interaction among theory, simulation, and experiments. Focused experiments have the potential to directly study important astrophysical processes in a controlled setting. The plasma physics community has developed advanced numerical algorithms for the study of weakly collisional plasmas; the application of such tools to astrophysics problems could help advance the field substantially. Collaborative use of advanced computing systems would also be of considerable use. In conclusion, we provide a summary of key opportunities over the next five years for the study of angular momentum transport in laboratory experiments, stars, and accretion disks.

### The Study of Angular Momentum Transport in Laboratory Experiments

A challenge for studying angular momentum transport in laboratory experiments or numerical simulations is extrapolating the results to astrophysically relevant parameters. Neither experiments nor simulations can reach the dimensionless plasma parameters appropriate for accretion disks or stars. They do, however, provide valuable insight into the basic processes thought to be responsible for angular momentum transport. Existing liquid metal experiments operate in the regime of dissipative MHD, whereas in real astrophysical plasmas one may need to consider the Hall effect, ambipolar diffusion, anisotropic viscosity, or microinstabilities that limit anisotropy (depending on the problem of interest). Significant new opportunities will be provided by existing or improved liquid metal experiments, and by plasma experiments that span wide magnetic Reynolds and Prandtl numbers, and access low-collisionality weakly magnetized plasmas. These will provide an opportunity to access a wide range of plasma conditions and study transport in astrophysical plasmas in detail.

**The Study of Angular Momentum Transport in Stars**

The differential rotation observed in stars, and probed in detail in the Sun, remains poorly understood. The change in stellar rotation with stellar age is even less well understood, but is crucial for understanding stellar death (via supernovae and gamma-ray bursts), and the properties of neutron stars and black holes. Turbulent convection and large-scale magnetic fields are probably the most important transport processes in stars. Simulations of convection have made some contact with the helioseismic constraints for the Sun, but have not yet been fully tested by observations of other stars or by experiment. The flood of data on stellar variability soon to come from Kepler, the near-term advent of the Solar Dynamics Observatory, and the prospect of a tunable plasma experiment all represent major opportunities for understanding the stellar angular momentum problem.

**The Study of Angular Momentum Transport in Accretion Disks**

In the next five to ten years, observations will — for the first time — directly resolve plasma near the event horizon of a black hole, in our Galactic Center and in the nearby galaxy M87. This is possible through interferometry at radio and (possibly) infrared wavelengths. These observations will constrain the temperature, density, magnetic field, and time-variability of the plasma near the black hole. Such observations can only be understood with the appropriate plasma physics models. Kinetic plasma models in General Relativity, combined with relativistic radiative transfer calculations that include the relevant high-energy radiation processes, can predict the emission from plasma accreting onto a black hole. Although some such tools exist, they do not currently include sufficient plasma physics to make reliable predictions for the observed emission. Doing so is a grand challenge for plasma astrophysics. These systems may provide the best opportunity for astronomical observations to constrain the plasma physics of accretion disks. There also will be enormous public interest in these results.

## IMPACTS

Angular momentum transport plays a critical role in a wide range of astrophysical systems, from stars and planets, to black holes, to the disks of galaxies. In many instances, the problem of angular momentum transport also is fundamentally a problem in plasma astrophysics. It therefore requires advances in basic plasma physics. Progress on understanding the plasma physics of momentum transport would have a major impact on a broad range of astrophysics problems, including how planets form, how the brightest sources of electromagnetic radiation in the universe work (accreting black holes and neutron stars), and how stars form, evolve, and end their lives.

## CONNECTIONS TO OTHER TOPICS

Momentum transport in ionized plasmas is fundamentally a problem in how magnetic fields are created, organized, and destroyed. Understanding momentum transport thus requires understanding other important plasma astrophysical processes, most notably magnetic reconnection, dynamos, and turbulence. For many astrophysical applications, the transport problem also requires moving beyond ideal MHD. In accretion disks around massive stars, and for luminous disks around black holes and neutron stars, the energy in the radiation field can greatly exceed that in the plasma. Understanding transport thus requires understanding the behavior of plasmas under radiation-dominated conditions. By contrast, in disks around stars like the Sun, the fluid is so poorly ionized that non-ideal MHD effects are critical. In many cases, the primary charge carrier in such disks is not even the plasma itself, but rather the dust grains that are the building blocks of rocky planets like the Earth.

The connections between momentum transport and other problems in plasma astrophysics go both directions: theoretical and experimental developments in understanding magnetized rotating plasmas in accretion disks have a significant impact on our understanding of dynamos more broadly. In addition, the large-scale magnetic fields that dominate transport in disks and stars — the origin of which is a central problem in dynamo theory — also create the initial conditions for understanding the collimated magnetized jets these systems can produce. Magnetized outflows from rotating stars and disks may be one of the most important sources of magnetic fields for the universe at large.

# CHAPTER 7:
## DUSTY PLASMAS

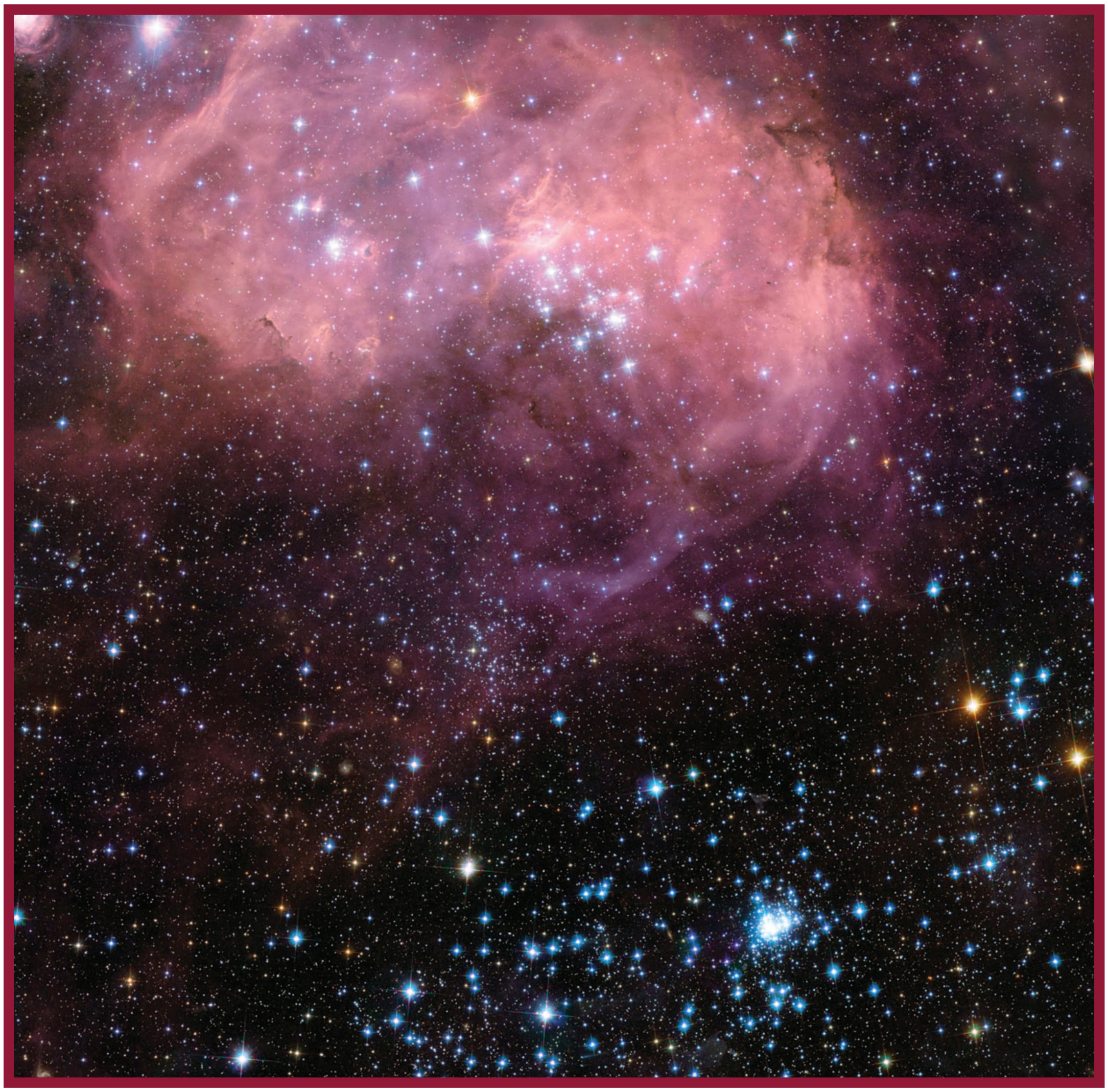

*ON PREVIOUS PAGE*

*NASA/ESA Hubble Space Telescope's Advanced Camera for Surveys (ACS) image of a star-forming region in the Large Magellanic Cloud. (Courtesy of NASA, ESA, and Jesús Maíz Apellániz, Instituto de Astrofísica de Andalucía, Spain.)*

# CHAPTER 7: DUSTY PLASMAS

## INTRODUCTION

In the astrophysical environment, the fraction of ionized particles varies widely from nearly no ionization in cold regions to fully ionized in regions of high temperature. This leads to a wide range of parameters where astrophysical plasmas can exist. While the astrophysical environment is frequently dominated by the presence of the plasma, this plasma is often strongly influenced by and coupled to the presence of embedded particulates (i.e., dust). These dust grains — which range in size from a few nanometers to micron-sized objects — can become either positively or negatively charged due to interactions with the background plasma environment and ionizing radiation sources in the astrophysical environment. Understanding the processes that govern these plasma-particle interactions is critical to the study of astrophysics. The agglomeration and growth of larger particles from single atoms and dust grains leads to the eventual formation of objects so large that gravity becomes the dominant force controlling their subsequent evolution.

In any plasma environment, electromagnetic forces can dominate or at least perturb the charged species. For most systems, these forces dominate the ion and electron dynamics, and perturb charged dust grains. These electromagnetic forces will not directly affect uncharged species (i.e., neutral particles). However, the charged and uncharged particles interact through collisions, which transfer energy and momentum from one species to another.

Furthermore, larger particles (i.e., dust grains) can become both an important source and sink of charged particles. For example, in regions of space with ionizing radiation, photoionization processes at the surface of the grains can lead to the generation of electrons from the surface of the grains as well as the creation of charged dust. Similarly, in regions where the grains have strong collisional interactions with the background plasma particles or neutrals, the grains can become heated to the point of thermionic emission — again leading to the grains becoming a source of electrons for the plasma. At the other extreme, in very cold regions of space, atoms and molecules can become trapped on the surfaces of these dust grains leading to the growth of the particles, as well as becoming a sink for the background gas and plasma environment. The presence of dust can alter the density, energy distribution, and the composition of its plasma environment.

From this wide range of phenomena that can occur in astrophysical plasmas, it is important to consider those processes that can be studied in a laboratory. Such studies will enable new insights and understanding about the astrophysical plasma environment. Three topics have emerged as "grand challenges" for laboratory astrophysics in the area of plasma-particle interactions.

1. How do the dust grains become charged in astrophysical plasmas?

2. How does the plasma influence the growth and breakup of macroscopic particles in astrophysical environments?

3. What is the role of magnetic fields in influencing charged macroscopic particles?

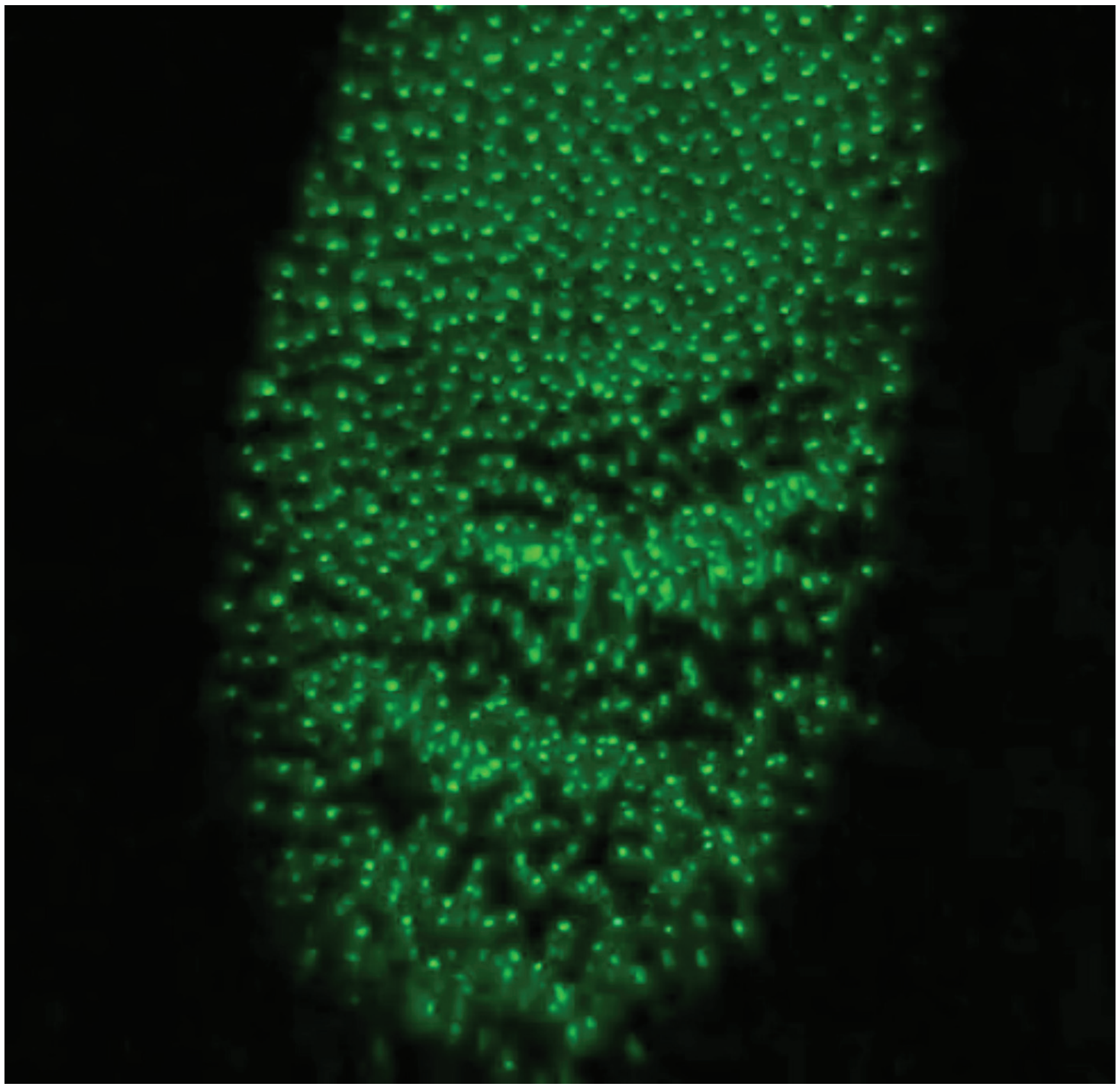

*Image of a laboratory dusty plasma experiment. A dust cloud of 3 micron diameter particles is illuminated using a Nd:YAG laser. Single particles and collective structures can be observed in this image. Courtesy of E. Thomas, Plasma Sciences Laboratory, Auburn University.*

## KEY SCIENTIFIC CHALLENGES

In the context of astrophysical plasmas, three questions are part of a singular, overarching theme about how larger objects — comets, planets, and stars — form from gas and dust clouds. At the same time, these three questions also have a strong connection to tangible issues in modern laboratory plasma science research in areas such as plasma manufacturing and fusion. Each of these challenges is discussed in the following section.

### *Question 1: How do dust grains become charged in astrophysical plasmas?*

Perhaps the dominating question for much dusty plasma research — for laboratory, fusion, and astrophysical plasma environments — is how to define the mechanisms that lead to the charging of the dust grains. In laboratory plasmas, the charging process is primarily due to the collection of ions and electrons from the background plasma. However, in astrophysical plasmas, the charging process will not only be driven by the collection of ions and electrons, but also will be influenced — and possibly dominated by — sources of ionizing radiation, shock-driven heating processes, or interactions with high-energy particles that lead to the production of secondary electrons. Understanding the charging process, particularly in highly collisional environments or regions with large magnetic fields, is an area of research in which much theoretical and experimental work remains to be done.

As a scientific problem, the determination of the charge of dust grains in an astrophysical plasma is quite challenging. First, the material composition of the dust grain is important because it de-termines the work function of the grain. Also, the charge distribution on an individual grain is determined by whether the grain is insulating or conducting. Second, the shape of the grain can determine how much surface area is available for charge to be collected. And third, the size of the grain, with respect to the local Debye length, can have a significant influence on how free charges (ions and electrons) are captured by the grain. These are all issues for which significant theoretical and computational efforts are needed.

In terms of the astrophysical plasma environment, the dust grains can become charged either positively or negatively — depending upon the conditions of the local plasma and radiation environment. Once charged, electromagnetic, as well as gravitational forces, can influence the particles. An often-cited example of this competition between electromagnetic and gravitational forces on the transport of charged grains in the space environment is the formation of the radial spoke structures in Saturn's rings. Ultimately, the dust grains can play an important role in the distribution of charge in astrophysical plasmas because they can transport substantial quantities of charge from one region to another. As such, these grains can be important sources and sinks of plasma.

***Question 2: How does plasma influence the growth and breakup of macroscopic particles in astrophysical environments?***

The growth and agglomeration of particles from clusters to atoms to the size of dust grains or larger is a long-standing problem in astrophysics. The matter that eventually forms the stars, planets, comets, and the other objects in solar systems begins its life as small grains of dust particles. These particles interact with and are influenced by the neutral and plasma particles in the space environment. If these small (nanometer- to micrometer-sized) particles are charged, the strength of the electromagnetic force is much larger than the gravitational force. As a result, precisely how these small particles become charged, the sign of that charge, the number of charges on the particles, and the spatial distribution of charges on these particles all can have a significant influence on the processes that eventually lead to the formation of stars and planets.

Initially, in the gas phase — when atoms first combine to form molecules and then atomic clusters — the growth of particles is believed to be driven by a Brownian motion growth process for particles up to ~100 nm. As particles grow to larger sizes, a number of environmental features begin to have a stronger influence on the growth process. These include the density and temperature of source materials (neutral atoms, plasma ions, and smaller agglomerates) as well as the thermal properties of the grains themselves. If these particles can be become charged, the rate at which further particle growth can occur can be modified.

It is well known that the electrostatic force is many orders of magnitude stronger than the gravitational force. Therefore, in the simplest approximation, two small dust grains that carry even a single, but opposite charge will experience an attractive force that is far greater than the gravitational attraction. By contrast, if those two same particles have the same sign of charge, the electrostatic repulsion will far exceed the gravitational attraction. By simply taking into account the charge state of the grains and how those charged grains are distributed in space, it is immediately obvious that the presence of charged dust can have a significant influence on the evolution of a planetary nebula in an astrophysical environment.

Therefore, current experimental, theoretical, and numerical studies all point to the need to have a better understanding of the initial formation of large-scale dust grains in astrophysical environments. And, because charged grains can have a profound impact on the coagulation of material into these larger grains, having detailed knowledge of the plasma environment and its coupling to the dust is vital.

***Question 3: What is the role of magnetic fields in influencing charged macroscopic particles?***

Like plasmas, magnetic fields are ubiquitous in the universe. For much of the development of astrophysics, magnetic fields have not been considered to play a significant role. However, as the importance of charged particle effects becomes more evident, it has become increasingly necessary to determine if the presence of magnetic fields also influences plasma-particle interactions. In this context, it is vital to determine how the magnetic field may shape the properties of the background plasma, and how the magnetic field may directly influence the charge macroscopic particles themselves.

Although the influence of the magnetic fields on macroscopic particles in astrophysical plasmas has not always been a prominent topic, as early as the 1950s it was recognized that the coupling between the plasma, dust, and magnetic fields is critical to understanding the astrophysical environment. In recent decades, particularly in the context of phenomena within a solar system, such as planetary rings or particles in planetary magnetospheres, it is quite important to include magnetic fields in order to properly reconstruct the dynamical behavior of these systems. Magnetic field effects on the dust particles in astrophysical plasmas can be considered from two aspects. First, how does the presence of a magnetized plasma affect the coupling with the dust? And second, how does the presence of a magnetic field affect the dynamics of the charged grains?

As noted in Questions 1 and 2, the underlying phenomenon that connects the dust grains to the plasma is the charging process. Over the years, there have been many theoretical works that have modeled the charging processes in laboratory and astrophysical plasmas in the presence of magnetic fields. The magnetic field alters the ion and electron fluxes to the grains, and can result in differences in both the final charge of the grains and the distribution of charge on the grains. At the microscopic level, this could alter the agglomeration processes that lead to the growth in particle size.

Presently, it is unclear which astrophysical environment regimes could provide observations of direct magnetic field effects on the charged dust. While there are some theoretical works, there are few direct observations or experimental studies to validate these models. Thus, this is an area of research that is ripe for new scientific discoveries.

## MAJOR OPPORTUNITIES

The three topical areas described in this section represent areas of scientific study that can each stand on their own merit. However, since a major thrust of this work is to make connections between laboratory studies and astrophysical processes, these three areas are particularly well suited to bridge the gap between the lab and space. The underlying issues of dust grain charging, dust grain growth and breakup, and the effect of the magnetic fields are all areas that have strong overlap with current and future research directions from the laboratory plasma community.

***Particle charging:*** Presently, there are a number of dusty plasma laboratory experiments in which the charging of the dust grains is a research program component. However, these studies are primarily focused on ion and electron collection from the background plasma. To extend this work to areas relevant to astrophysical systems, it would be necessary to have dedicated studies that also focus on charging in intense radiation environments.

***Particle growth and breakup:*** In the area of particle growth and breakup, there are important issues relevant to the plasma astrophysics community that have a great deal in common with the industrial plasma processing and fusion research communities. In both of these applied areas, the formation of nanometer and micrometer-sized particles from the gas phase in reactive plasmas is often considered to be a major source of contamination. Nonetheless, the particles formed in these environments share a number of common features with their astrophysical counterparts — namely, the particles were charged while they were in the plasma and large particles form from the coagulation of many smaller particles. To date, there have been few dedicated experiments on the formation of grain aggregation and coagulation processes. To make progress in this area, experimental studies that can simulate specific aspects of the space plasma environment (e.g., choosing the ratio of ion, electron, dust, and neutral gas densities to mirror a particular planetary nebula region) may provide a more complete representation of the processes that occur in nature. Additionally, studies performed in chemically active plasmas that can mimic processes occurring in space environments (e.g., star forming regions) may give insight into the material properties of the dust grains.

***Magnetic field effects:*** Magnetic field effects are an aspect of dust-plasma interaction studies that remains essentially unexplored in the laboratory. Almost all studies to date have been performed without a magnetic field or at magnetic field strengths where only the electrons are magnetized. This is because there are significant technical challenges to building an experiment that can operate in a regime where the electrons, ions, and charged dust can be magnetized (e.g., typically requiring steady-state, multi-Tesla magnetic field strengths and the ability to detect nanometer-sized particles). There is currently one operating, 4-Tesla dusty plasma experiment, and various groups in the community are planning up to three additional experiments to come online within the next five years.

These new experiments offer a unique opportunity to verify and validate which of the various numerical models have properly captured the role of the magnetic field. Moreover, studies with magnetic fields open up entirely new regimes of dust — plasma interactions that have not previously been considered. Experiments on dust transport parallel to the magnetic fields, perpendicular to the magnetic fields (e.g., Hall effects), and parallel to electric fields and perpendicular to magnetic fields (e.g., Pedersen effects) can be investigated. Moreover, the study of fully magnetized plasmas with magnetized dust may allow the study of new wave modes such as dust cyclotron, dust magnetosonic, and dust Alfvén modes.

## IMPACTS AND MAJOR OUTCOMES

Since the laboratory study of dusty plasmas is still in its infancy, there are many possible directions for the field to go and, subsequently, many areas of plasma physics research that could be af-fected. Perhaps the most important impact of this work is gaining an understanding of a plasma in the most "general" state. Because plasmas, dust, and magnetic fields are found throughout the universe — often in combination with each other — the study of astrophysical plasmas should focus on understanding the various complex interactions among these three components. A new generation of dusty plasma laboratory experiments is becoming available that may make it possible to explore aspects of this complex system.

## CONNECTIONS TO OTHER TOPICS

The topic of dusty plasmas in astrophysical plasmas has connections to many other areas of plasma astrophysics research. In particular, studies of radiation from charged dust (Radiative Hydrodynamics group) is particularly important in interpreting data from star forming regions. Additionally, the presence of shocks (Collisionless Shocks and Particle Acceleration group) can "process" dust grains leading to changes in their morphology, material characteristics, and charge state. Finally, there is already experimental evidence that the presence of charged dust grains can modify many existing plasma waves and give rise to new waves and instabilities (Waves and Turbulence group). Mapping these results to the astrophysical plasma environment represents a new area of research for the community.

# CHAPTER 8:
## RADIATIVE HYDRODYNAMICS

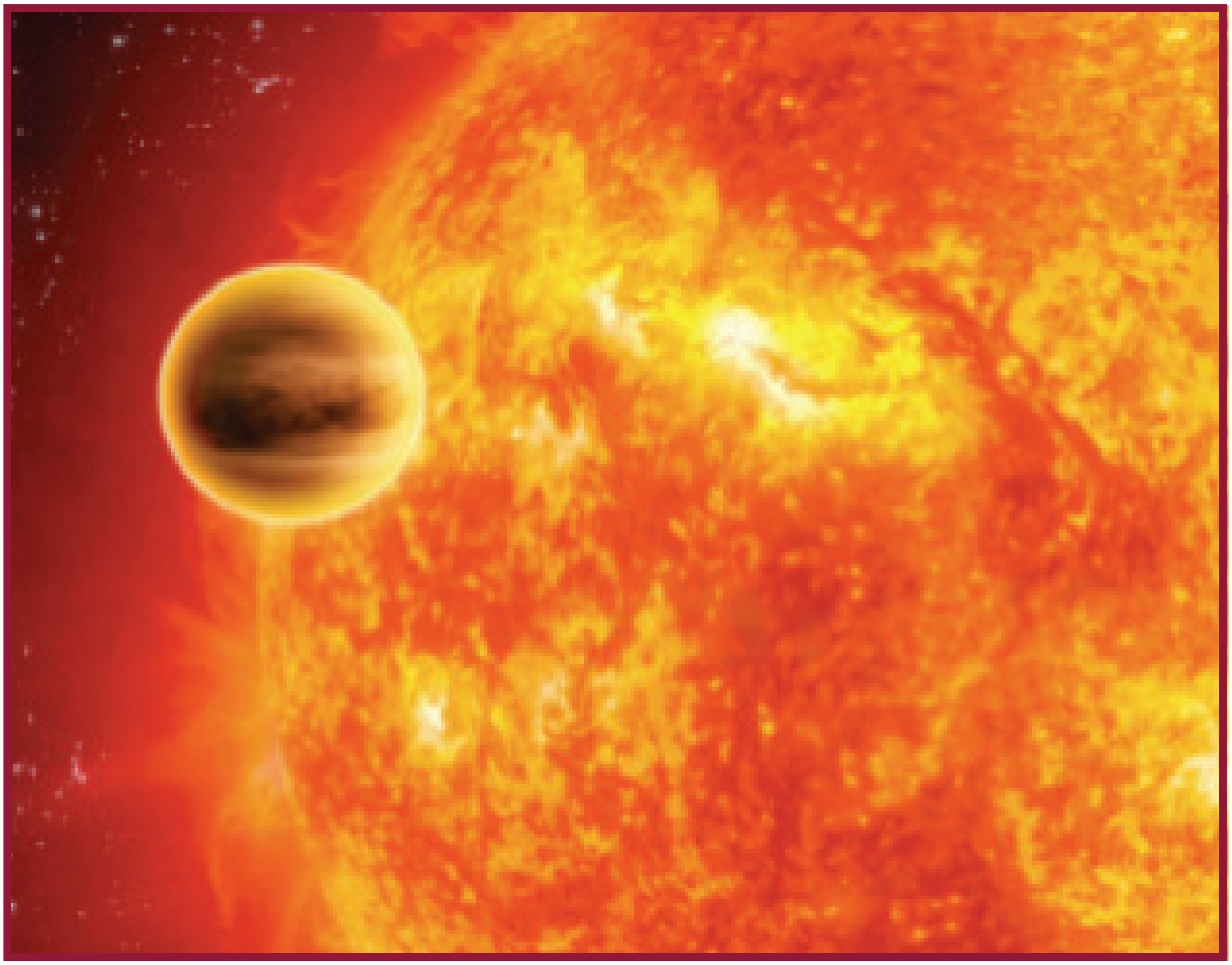

*ON PREVIOUS PAGE*

*Artist's conception of an orbiting exoplanet in close proximity to its parent star.* (*Image courtesy of ESA-C. Carreau.*)

# CHAPTER 8: RADIATIVE HYDRODYNAMICS

## INTRODUCTION

With several high-profile satellite missions, innovative ground-based instrumentation, and new large telescopes, astrophysics stands at the threshold of revolutionary discoveries in the next decade. In most cases, the quality of the scientific output from these efforts inexorably links to a deep understanding of how radiation interacts with matter. The physics of this process is radiation hydrodynamics, an interdisciplinary field of study that includes theoretical, numerical, laboratory, and observational efforts. It spans an enormous range of objects and conditions.

Although astrophysical systems can be quite different from terrestrial ones, there are genuine similarities to laboratory plasmas. For example, in radiation-driven, inertial confinement fusion systems, radiation is the dominant heat carrier. Linked by these qualitative resemblances, the tremendous range of conditions found in astrophysical plasmas can illuminate laboratory plasma physics by exhibiting different physical mechanisms in extreme form. Conversely, for cases in which laboratory plasma conditions can be matched or scaled to those of astrophysical plasmas, we have the rare privilege (in astronomy) of being able to experiment rather than merely observe.

## KEY SCIENTIFIC CHALLENGES

Radiation hydrodynamics plays a dominant role in a wide range of astrophysical phenomena. In this section, we list a number of contemporary astrophysical problems and challenging questions in which radiation hydrodynamics is especially important.

### *How are the largest stars formed?*

Stars are born in dusty molecular clouds. Once the first stars in a region light up, their radiation drives significant forces in the surrounding gas. Young stars can also photoionize and heat the gas around them. The most massive stars have surface temperatures so high that they produce substantial ultraviolet (UV) and X-ray radiation, which in turn can change the size distribution of interstellar dust, changing its total opacity.

There are now stunning examples of accretion disks around forming stars photo-evaporated by nearby young massive stars. Also, there are majestic pillar structures in molecular clouds — structures that are driven by photo-evaporation from young massive stars just turning on in star forming regions, so-called stellar nurseries. Much current research centers on studying how disks of dense gas and dust evolve into planetary systems. Observations have shown that these disks often lack near-infrared emission, a sign that they do not extend all the way to the star but are truncated at some radius — probably at the point where radiation from the star sublimates the dust in the disk.

Radiation also plays a key role in determining how disks transfer angular momentum outward. In areas where radiation penetrates enough to partially ionize the gas, large-scale magnetic fields will couple the gas to the overall motion of the star and disk. This leads to a system where disk material can funnel onto the star along field lines, and be ejected from the system in a collimated, pro-

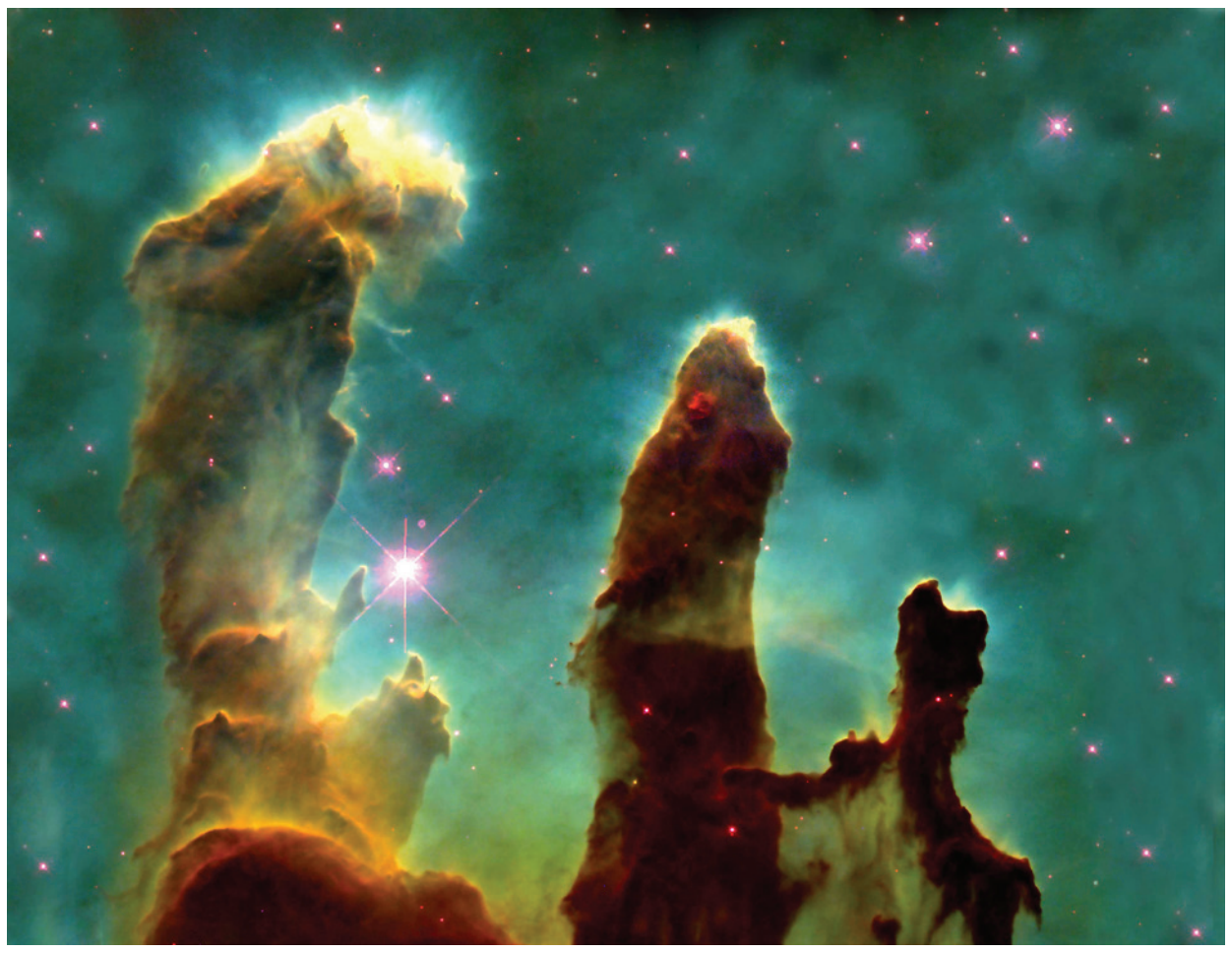

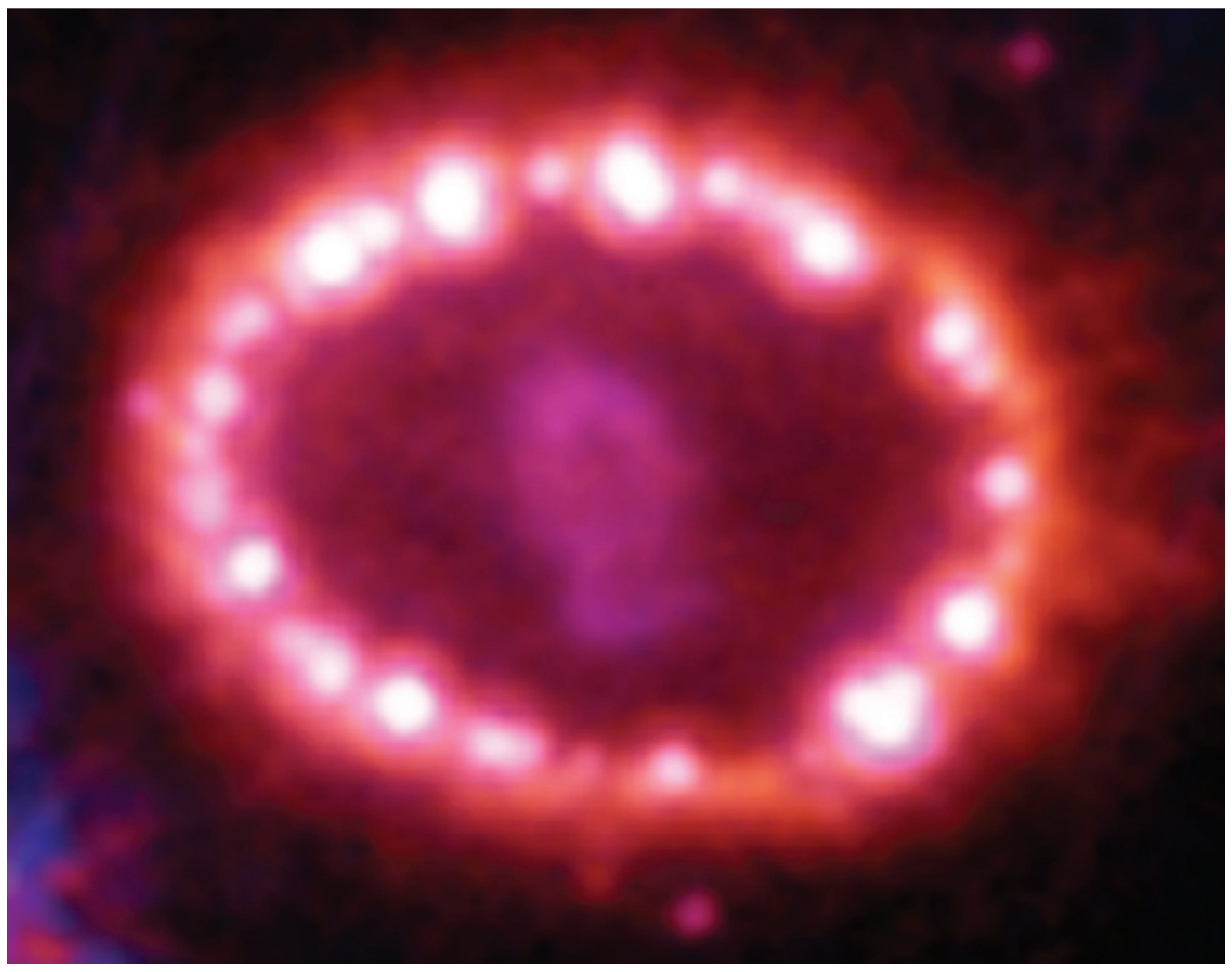

***Images showing areas of plasma astrophysics where radiation hydrodynamics plays a dominant role: (left) the stellar "nursery" known as the Eagle Nebula; (right) remnant from SN1987A. Images courtesy: P. Challis, R. Kirshner (CfA), and B. Sugerman (STScI), NASA; and Hubble Space Telescope, NASA.***

tostellar jet. The physics of the disk-jet system is fundamental, occurring throughout astrophysics in young stars, X-ray binaries, active galactic nuclei, and possibly planetary nebulae. The associated radiation transfer problem is multi-dimensional, frequency-dependent, and depends on the local velocity of the fluid. Moreover, the passage of photons through the moving fluid controls the ionization balance, which determines the line opacity, so both the continuum-transfer and ionization-balance problems must be solved together.

The Sun continues to provide new challenges and opportunities for stellar physics. Helioseismology is the most powerful tool in studies of the solar interior. It enables precise tests of how accurately solar models predict the solar structure and, in particular, how well they predict the interior radiation transport. Increasingly sophisticated analysis of photospheric spectra has led to major revisions in the chemical composition, and solar models now significantly disagree with observations. Is this because even the most refined photosphere models are in error? Is it because the theoretical opacity models lack sufficient accuracy? Or is it because the physical approximations used in the solar models do not capture all the essential science? Stellar interior models require input about properties of stellar matter such as opacities, and those material property models can now, for the first time, be experimentally measured at relevant conditions. The combination of precise observations, detailed models, and new ability to measure the properties of stellar interior matter in laboratories provides an opportunity to answer these questions. The results will refine our picture of the internal structure and chemical composition of the Sun and other stars.

### *How do stars explode as supernovae?*

Massive stars explode by the collapse of their cores, but scientists believe there are several distinct types of core-collapse supernovae. Although the most common are Type II supernovae (collapse of a massive star that has retained its hydrogen-rich envelope), there is recent speculation that still more energetic explosions called "hypernovae" may be the sites of gamma-ray bursts. All supernovae are strong radiation hydrodynamic events, with the radiation pressure exceeding the material pressure for long periods.

The details of explosion mechanisms are far from well understood. It may be that the early neutrino heat transport — which is strongly convective — combined with later dynamics during the explosion are sufficient to explain the observed distribution of ejected material. Alternatively, mechanisms involving rapid rotation and strong magnetic fields in the stellar core may conspire to produce jet-driven explosions. Astronomers are beginning to obtain observations of the emergence of shocks from these stars as they explode. It is also possible that core-collapse explosions in red supergiant stars can create a sufficiently strong shock that the radiation from the shock affects hydrodynamic mixing in the envelope. The models applied to this problem have had limited physical fidelity and no demonstration tests based on experiments. At much later times, when the ejected material has expanded into the interstellar medium, supernova remnants go through a phase in which radiative heat losses from the explosion's strong shock significantly affect the evolving dynamics. When the stellar remnant is a highly magnetized neutron star, such as the Crab Nebula, relativistically strong electromagnetic waves coupled to the plasma can energize the entire remnant in spectacular fashion.

Type Ia Supernovae (SNe-1a) originate from thermonuclear explosions of White Dwarf (WD) stars whose mass has grown to exceed the limiting Chandrasekhar mass. When the WD approaches the critical mass, it triggers a nuclear burning front that then consumes the WD. The maximum luminosity can vary by up to a factor of about 10. There is a universal relation, however, between the absolute brightness and the width of the light curve, which allows SNe-1a to serve as very bright, standardized candles. This property led to the discovery of dark energy. Planned dark energy studies will need calibration of SNe-Ia luminosities at the 2 to 3 percent level. This requires a much improved ability to model SNe-1a light curves. Radiation hydrodynamics plays two important roles. First, deep in the exploding matter, heat is diffused primarily by radiation transport. The magnitude of radiative preheat in advance of the propagating nuclear flame (i.e., radiative preheat) can have a major influence on how the reactions and dynamics proceed. Second, when the shocks move out through the stellar envelope into the circumstellar gas, breakdowns in thermodynamic equilibrium lead to energy transport that is both non-LTE [local thermodynamic equilibrium] and non-local. These effects can significantly alter the details of the light curve. Improved understanding of radiation hydrodynamics will be an important adjunct to future dark energy studies relying on SNe-Ia light curves.

***How do black holes radiate?***

Matter accreting onto black holes made tens of percent of the total light created after the Big Bang. In the accretion flows accounting for the great majority of this output, radiation pressure — rather than gas pressure — is the principal support against the vertical component of gravity where most of the light is made. The fact that the radiation pressure is proportional to the accretion rate when averaged over long timescales relative to the thermal-equilibration timescale implies (in linear theory) an instability in the inflow rate. In other words, our standard picture of accretion disk equilibrium may not describe real disks in nature. Even in regions where gas pressure is greater than radiation pressure, current computational techniques are inadequate for defining the heating-cooling balance that determines the fluid equation of state (EOS).

A still more serious paradox emerges when the accretion rate is high enough that the generated luminosity exceeds the Eddington value. In such a state, radiation forces should disrupt the in-

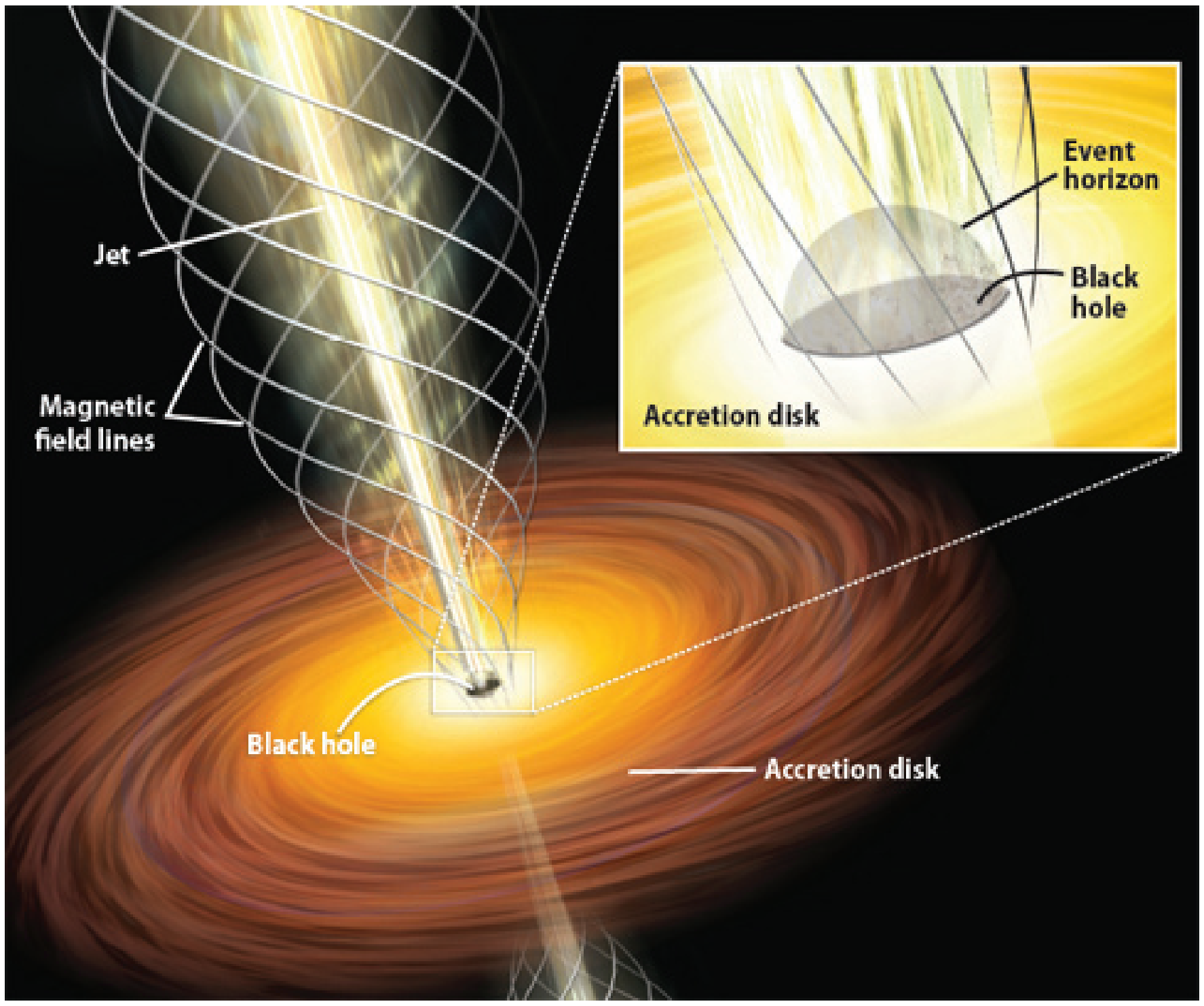

*Artist's concept of an accreting black hole. Image courtesy of Astronomy/Roen Kelly.*

flow. But the accretion rate is controlled by events far from the central gravitating object, so there's nothing to prevent this situation from arising. What happens when it does? Observational evidence suggests that there are black holes in nearby external galaxies accreting at super-Eddington rates, making this problem particularly pointed. Understanding how radiation couples to the disk dynamics is essential to making progress in this area.

Another example of radiation hydrodynamics in accreting black holes is found in the spectra of quasars. These objects are thought to be actively accreting supermassive ($10^8$-$10^9$ solar masses) black holes, each at the center of a galaxy. In about 15 percent of the population, very broad absorption troughs (about 10,000 km/s) are seen to the blue of the systemic red-shift in a number of ionic resonance lines. All quasars likely have these outflows, but they cover only a fraction of solid angle. "Line-locking" matches between features in different line profiles, as well as the approximate match between the photon momentum removed in the absorption trough and the momentum in the outflow, indicates that the driving force is radiation pressure. Precisely because such strong resonance-line absorption is the signature of these outflows, the radiation force is most likely expressed through UV resonance lines in a manner closely analogous to the radiation-driven winds of massive stars. Just as in that case, however, there are formidable technical problems standing in the way of genuine understanding.

### *How does intense stellar radiation affect exoplanet atmospheres?*

In the decade since radial velocity surveys uncovered the first planets around other stars, more than 400 such planets have been discovered, and the numbers keep rising. With such a large sample, astronomers now face the reality that planets populating our galaxy can differ markedly from those in our solar system. The notion that all planetary systems have small rocky planets close to their stars and large, cool, gas giants farther away has yielded to a much richer reality that includes Jupiter-sized planets orbiting closer to their host stars than Mercury does around our Sun. In these cases, the star's intense radiation field will alter the chemistry of these dense planetary atmospheres, determine the cloud structure, and drive large-scale flows in their atmospheres. We have just begun considering how such intense radiation will affect these exoplanet atmospheres.

## MAJOR OPPORTUNITIES

Many of the opportunities in this field arise from improvements in either computational power or experimental facilities. Combining the two — testing simulation codes on idealized problems realizable in the laboratory — promises to be exceptionally fruitful. After code validation and verification, we can be more confident in the results of these simulation codes when they are applied to actual astrophysical situations.

### Opportunities for understanding star formation

Understanding how stars form and the nature of their interior structure has been one of astronomy's most enduring problems. The star formation problem encompasses a vast range of scales and conditions. The collapsing gas sheds its angular momentum, possibly aided by the ejection of protostellar jets, in a complex interaction with magnetic fields. When the new star starts to burn, it radiatively drives the surrounding molecular cloud, possibly triggering additional star birth.

Modeling of these dynamics requires 3-D MHD simulations with radiation, electrons, ions, neutrals, dust, magnetic fields, turbulence, and nuclear burning. The rapid advance of computational power accelerates progress in this area. More powerful algorithms, ultimately checked by comparison with observations and, where possible, laboratory experiments, can shed light in the area. The chief issue in advancing numerical radiation hydrodynamics is creating radiation transfer algorithms both accurate enough to capture the physics and rapid enough to be employed in concert with hydrodynamic or magnetohydrodynamic simulation codes. A variety of improved methods are already foreseeable.

For the first time, we can create experimental conditions to reproduce those in stellar interiors, allowing measurements of complex opacities. These opacities are central to stellar birth, stellar evolution, and the detailed dynamics of stellar interiors. Among the most inspiring and widely recognized examples of radiative hydrodynamics are the majestic columns in the Eagle Nebula. Here, intense UV radiation from a few bright young stars drives deep nonlinear radiative hydrodynamics through the pressure created from photo-evaporation at the molecular cloud surface. Simulations of these nonlinear radiation hydrodynamics are notoriously difficult, yet essential, if we are to understand the dynamics of star-forming regions. Well-scaled experiments would be invaluable to test such simulations. Radiative shocks occur in multiple phases of star formation, such as the ejected protostellar jets and shocks propagating through molecular clouds. Radiative cooling in shocked clumps in molecular clouds may be a trigger for further star formation. Simulations of such radiative effects are difficult and, in most cases, are still untested with experimental data. Facilities now coming on line offer the potential for testing these codes in conditions scaled to the astrophysical context. Dust contributes in a major way to the star formation process. In molecular clouds, it radiates away excess thermal energy, keeping the cloud cold and dense. The interaction of the dust with these strong radiative shocks in the cloud is a major unknown in the dynamics of star formation. Current facilities offer possibilities for conducting experiments to test our understanding of shock and radiation processing of dust.

Recent ground-based observations using adaptive optics have just begun to probe the regions of protostellar jet acceleration. The larger telescopes planned for the next decade, such as the Thirty Meter Telescope (TMT) and Large Magellan Telescope, will further study this phenomenon. Diagnosing conditions within the inner parts of protostellar accretion disks — where radiation plays a key role in the dynamics — is a real possibility. The new Large Millimeter Telescope (LMT) will be able to study molecular emission within disks, another process dominated by radiation. It now is possible to study differential motions within stellar jets using synoptic images from the Hubble Space Telescope (HST). These reveal much about the flow dynamics from these objects. New wide-field infrared arrays, such as the NEWFIRM camera available at the National Observatories at Kitt Peak and Cerro Tololo, make it possible to survey entire regions of massive star formation in a short time. By observing in the light from molecular hydrogen (H2), these images reveal outlines of globules that identify current and future star formation areas. In some cases, the globules show what appear to be spectacular fluid dynamical instabilities caused by intense radiation and winds from nearby massive stars. Additional telescopes that will be highly valuable include the James Webb Space Telescope (JWST), Stratospheric Observatory for Infrared Astronomy (SOFIA), and the Atacama Large Millimeter Array (ALMA).

**Opportunities for understanding supernovae**

Understanding supernovae — both core-collapse and thermonuclear — is a grand challenge. Modeling supernova explosions in their entirety will require 3-D simulations that include gravity, nuclear equation of state, nuclear reaction chains, nuclear burn wave evolution, intense neutrino fluxes, radiation, magnetic fields, turbulence, and (for the light curve) non-LTE, non-local radiation transport. Experiments can contribute in several ways. We can match laboratory conditions to those of stellar interiors to provide test-beds for measuring atomic opacities. We also can measure nuclear reaction rates. Beyond this, we can test 3-D radiation transport codes by matching their boundary conditions and scaling parameters to those achievable in the lab. Strong shock- driven turbulence, with or without embedded magnetic fields, may also be within reach with the latest experimental facilities.

Radiation flow through an expanding atmosphere with a large number of atomic resonance lines is a difficult theoretical problem. This problem is central to understanding stellar winds from massive stars. It is made more challenging by the larger velocity gradients of supernova explosions. Scaled experimental tests of both scenarios may be possible, and would improve our understanding of massive stars and supernovae. Radiative shocks occur in core-collapse supernovae and supernova remnants (SNRs). Radiative shock instabilities are thought to contribute to the crenellated and sinewy structures seen in many SNRs. Yet it is difficult to simulate such radiative effects, which in most cases are untested with experimental data. The destruction of interstellar dust grains by intense fluxes of X-rays and UV radiation from gamma-ray bursts (GRBs) significantly affects our interpretation of GRB light curves and afterglow. There are large theoretical uncertainties in our understanding of this dust destruction. Experiments in this area are likely also possible and would be beneficial in improving our understanding and interpretation of GRB light curves and afterglow.

Observationally, the Swift satellite has been operational for a few years and regularly discovers gamma-ray bursts that are routinely followed up by observatories on the ground and in space. A major effort of the newly launched Fermi Gamma-ray Space Telescope is to measure how the bursts' overall spectral energy distribution evolves with time. Models of this phenomenon rely entirely on how radiation transfers within the burst. As in the above cases, interpreting the results from these missions goes hand in hand with the study of radiation hydrodynamics. The SuperNova Acceleration Probe (SNAP) satellite is a proposed space observatory designed to observe Type-1a supernovae to measure the expansion of the universe. NASA and the Office of High Energy Physics at the U.S. Department of Energy (DOE) will jointly fund and develop the Joint Dark Energy Mission (JDEM). This mission will make precise measurements of the expansion rate of the universe to understand how this rate has changed with time.

**Opportunities for understanding accreting black holes**

Black holes are some of the most fascinating objects in the universe, but modeling their surrounding flows will require 3-D simulations that include intense radiation, strong gravity, magnetic fields, turbulence, and strong radiation forces. The state of the art in global accretion modeling includes 3-D MHD turbulence in full general relativity, but does not incorporate either a realistic equation of state or radiation forces. However, researchers in the field are poised to add these elements as they develop new algorithms and can run their codes on ever-faster computers. Many of the same techniques developed for radiation-MHD simulations in the star-formation context can be readily applied to the black hole accretion problem as well.

Similarly, the same experimental tests will likely benefit codes applicable to the star-formation problem. A key observational diagnostic of accretion flows comes from the profile of the Fe K$\alpha$ emission line, whose emissivity is sensitive to the ionization balance in the accretion disk atmosphere. New laboratory experiments could improve our knowledge of key atomic data (photoionization cross sections and fluorescence yields) necessary for appropriate interpretation of these emission line data.

Observations will play a key role in developing an understanding of the dynamics of accreting black holes. Operating telescopes relevant to this area include the Chandra and XMMNewton X-ray Observatories. Several new telescopes, such as NuStar and GEMS, the very first astronomical X-ray spectropolarimeter, are planned for launch in the next several years. The International X-ray Observatory will likely be launched in the future.

**Opportunities for understanding exoplanet atmospheres**

There are enormous challenges to predicting, measuring, and understanding the nature of exoplanet atmospheres. Making predictions of exoplanet atmospheres requires large-scale computer simulations using 3-D radiation-MHD codes that include electrons, ions, radiation, neutrals, dust, and the couplings between them. These exoplanet atmosphere models must be guided by astronomical observations from next-generation telescopes and satellites that are being planned, under construction and soon to be launched. Such sophisticated code simulations will require extensive verification and validation tests with experimental data. In particular, this will require input data such as UV opacities, radiation-plasma-dust interaction rates, and possibly equation of state. We must develop suitable experimental techniques for this purpose. Dust could play a major role in the astronomical observables from exoplanet atmospheres. In particular, the intense radiation from the parent star will interact dynamically with the dust in the atmospheres to drive the chemistry, climate, and observational signatures. We could begin experiments in this area soon, yet little or no work has been done to date.

Determining the atmospheric properties of exoplanets until recently seemed impossible. Current space missions, however, are now yielding results in this area. The best targets are those where the planetary systems are oriented so that they move in front of the star for part of their orbit, and vanish behind the star on the other side. For these systems, the Hubble Space Telescope has been able to detect absorption lines in the planetary atmospheres as the planet leaves the star. Remarkably, the Spitzer Space Telescope has detected the thermal radiation from the planet by observing in the infrared, where the flux contrast between the star and planet is lower than that in the optical. Many of the main molecular absorptions lie in the infrared, so by performing these measurements as a function of wavelength and orbital phase, one can deduce much about the composition of the planetary atmosphere. When the James Webb Space Telescope launches in 2014, these types of studies will be extended. The JWST will be an orbiting infrared observatory that will complement and extend the discoveries of the Hubble Space Telescope, with longer wavelength coverage and greatly improved sensitivity.

The longer wavelengths will enable the JWST to look much closer at the beginning of time, hunt for the unobserved formation of the first galaxies, and look inside dust clouds where stars and planetary systems are forming today. By then, the exoplanet database will likely include true Earth-like planets, as these are the main focus of the Kepler Mission, which has just begun to produce spectacular and unexpected results. The Kepler mission is specifically designed to survey our region of the Milky Way galaxy to discover hundreds of Earth-size and smaller planets in or near the habitable zone and determine how many of the billions of stars in our galaxy have such planets.

## IMPACT AND MAJOR OUTCOMES

The potential for major impact, both in astrophysics and laboratory physics, comes from the linking of astronomical observations, theory, simulations, and experimental verification and validation. Experiments will lead to improvements in the theories and simulations, which in turn, will motivate new observations, closing this research loop. Astrophysics and astronomy have never had an experimental V&V component strongly coupled to their routine research. Conversely, through a strong coupling to astronomical observations and theoretical interpretation, laboratory experiments will gain important insights into plasma physics at the most extreme conditions.

## CONNECTIONS TO OTHER TOPICS

Radiation hydrodynamics links to many, possibly all, of the other topics: magnetic reconnection, collisionless shocks and particle acceleration, waves and turbulence, magnetic dynamos, interface and shear instabilities, angular momentum transport, dusty plasmas, relativistic plasmas, and jets and outflows.

# CHAPTER 9:

# RELATIVISTIC, PAIR-DOMINATED AND STRONGLY MAGNETIZED PLASMAS

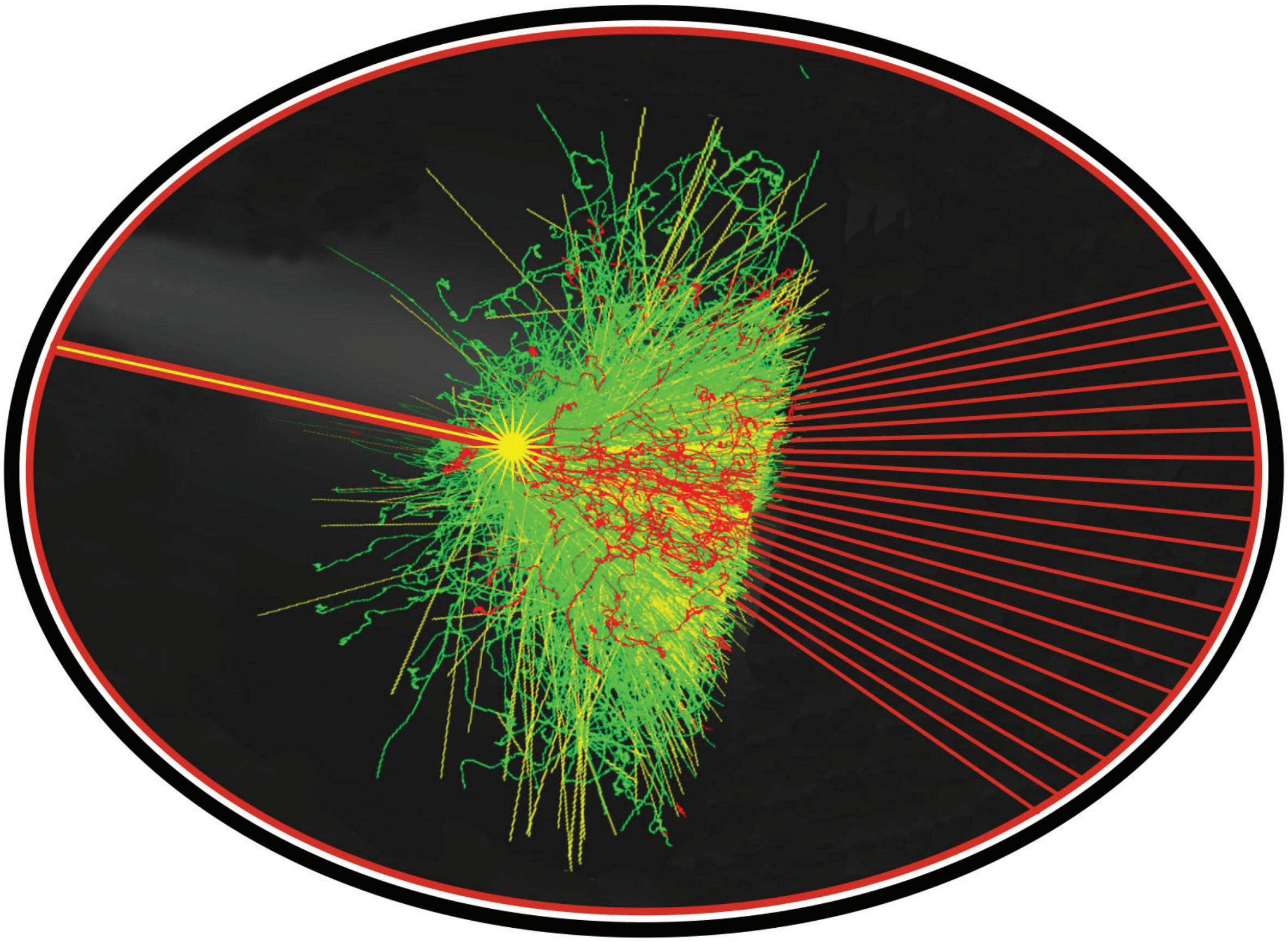

*ON PREVIOUS PAGE*

*Monte-Carlo simulation of electron-positron-gamma-ray shower when 25 MeV electrons accelerated by an ultra-intense laser interact with a gold disk. The laser strikes from the left. Electrons (green) lose energy as they strike the gold nuclei and emit gamma-rays (yellow). Positrons (red) are then produced when the gamma-rays interact with gold nuclei, some of which emerge to the right. (Artwork/simulation by Scott C. Wilks and Newton Elberson, copyright 2010.)*

# CHAPTER 9: RELATIVISTIC, PAIR-DOMINATED AND STRONGLY MAGNETIZED PLASMAS

## INTRODUCTION

Relativistic, pair-dominated, and strongly magnetized plasmas are ubiquitous in the high-energy universe. Their properties, behaviors, and observable manifestations often differ drastically from those of ordinary plasmas. While some theoretical progress has been made in recent decades, such plasmas have rarely been studied experimentally in the laboratory. This landscape is rapidly changing due to the advance in ultra-intense lasers and accelerator beams, which makes the laboratory creation and controlled studies of these plasmas a reality.

## KEY SCIENTIFIC CHALLENGES

Here we discuss the challenges and opportunities in relativistic plasmas, focusing on four main topics:

1. Dissipation of relativistic beams and collisionless shocks.

2. Creation of pair plasmas and ultrastrong magnetic fields.

3. Relativistic jets.

4. Relativistic turbulence and reconnection.

**Dissipation of Relativistic Beams and Collisionless Shocks**

Violent astrophysical phenomena often produce relativistic directed outflows such as pulsar winds (PW), gamma-ray bursts (GRB), jets in active galactic nuclei (AGN), and microquasars. Such outflows are observed because they produce intense radiation throughout the electromagnetic spectrum. For this to happen, bulk flow energy must be efficiently converted into the internal energy of the radiating electrons and magnetic fields. Theory and computer modeling indicate that Weibel-like current instability likely plays an important role when relativistic beams and outflows interact with an unmagnetized plasma. Although Weibel (1959) instability has been well studied theoretically, there have been few dedicated experiments to test it in relativistic beams. Weibel instability also plays a critical role in relativistic collisionless shocks. Shocks are believed to be responsible for the observed radiation of pulsar wind nebula (PWN), GRBs, some AGN jets, and the production of high-energy cosmic rays. Besides dissipating the bulk-flow energy, shocks accelerate non-thermal particles, and generate and amplify magnetic fields. However, we do not fully understand the physics of relativistic collisionless shocks, and the resultant particle acceleration and magnetic field generation.

***Major scientific questions on relativistic beams and shocks include:***

1. How do the growth and saturation of Weibel-like instability depend on the beam size, density, composition, temperature, and Lorentz factor?

2. When is Weibel stabilized or suppressed? When the instability is suppressed (or stabilized) can beams still dissipate efficiently via two-stream electrostatic instability?

3. How does Weibel-generated magnetic turbulence form shocks and accelerate nonthermal particles?

4. What is the radiation output from current-unstable electron beams?

5. How does the shock structure depend on upstream magnetization, composition (pair vs. e-ion), and field geometry?

6. How do relativistic and strongly magnetized shocks differ from nonrelativistic and unmagnetized shocks?

**Creation of Pair Plasmas and Ultrastrong Magnetic Fields**

Relativistic electron-positron (e+e-) pair plasmas are everywhere in the high-energy universe — from the first few seconds of the Big Bang, to pulsar winds, blazars' jets, and GRBs. Transient, thermal pair-equilibrium plasmas also may be present around stellar-mass black holes during gamma-ray flares. Because of their unity mass ratio, pair plasmas behave differently from electron-ion plasmas in many respects. Hence, it is extremely desirable to study pair plasmas in the laboratory, both for the fundamental physics and for astrophysical applications. Neutron star magnetic fields can exceed 100 TG, while white dwarf fields may reach 100 MG. Ultra-intense lasers are now capable of generating transient fields approaching $10^9$G, which overlaps with magnetic white dwarfs and millisecond pulsars. Studying laboratory plasmas with strong fields may demonstrate, for the first time, that we can reproduce the conditions appropriate to the atmospheres of these neutron stars and magnetic white dwarfs in a terrestrial laboratory. Measurements of such plasmas may enable the study of highly dynamical phenomena such as the "photon bubble" instability. They also may permit probes of nonlinear regimes of the Zeeman effect in hydrogenic atoms, as well as "guiding center drift atoms" where the strong field changes electron orbits into $\mathbf{E}\times\mathbf{B}$ drift orbits. Laboratory insights may, in turn, spawn new observational diagnostics of neutron stars and magnetic white dwarfs. Theories of anisotropic radiation and particle transport in strong fields may be meaningfully tested in the laboratory. For example, lasers coupled with gigagauss magnetic fields will allow us to probe the Landau levels using resonant scattering and explore laser cooling of magnetically trapped electrons. Since pairs and strong magnetic fields are both present in neutron star magnetospheres, the simultaneous creation of pairs and strong fields using ultra-intense lasers will provide a unique platform to study neutron star physics.

***Major scientific questions on pairs and strongly magnetized plasmas include:***

1. How do pair plasma kinetics — such as plasma instabilities, wave-particle interactions, particle acceleration mechanisms and radiative processes — differ from e-ion plasmas?

2. Can we create thermal pair-equilibrium plasmas in the laboratory?

3. What are the observable manifestations of pair plasmas besides their annihilation radiation?

4. How do strong magnetic fields alter the atomic structure, ionization, collision, and radiative properties?

5. Can we test novel astrophysical phenomena such as the "photon-bubble" instability in accreting neutron stars in the laboratory?

6. Can we model the magnetospheres of strongly magnetized white dwarfs and neutron stars using laboratory experiments?

### Relativistic Jets

Relativistic jets are long narrow dynamic structures that emanate from compact objects such as stellar mass black holes, neutron stars, and active galactic nuclei. Despite their widespread occurrence, astrophysical jets have many aspects that are not well understood (e.g., how jets are launched and accelerated, why jets are so narrowly collimated, and why jets appear to be extremely stable and straight). Moreover, we do not know the relative abundance of ions, pairs, and Poynting flux; their roles in the jet dynamics, dissipation, and radiation remain to be understood. We can determine only a few jet parameters through observations, so models of jets are still in a primitive stage. However, new observations from Fermi and other observatories will shed important new light on jets.

***Major scientific questions on relativistic jets include:***

1. How are relativistic jets launched and accelerated?

2. Why are relativistic jets so well collimated?

3. Why are they so stable?

4. Do relativistic jets behave differently than non-relativistic jets?

5. What are the differences between Poynting flux and hydrodynamic jets?

6. How do the multi-scale jet regions interact with each other (e.g., via internal shocks, shear layers or reconnection)?

7. How do e+e-pair jets behave differently from e-ion jets?

8. Which kinetic dissipation and radiation mechanisms are important in jets?

### Turbulence and Reconnection in Relativistic Plasmas

Many astrophysical plasmas exist in a strongly turbulent state, where the local properties such as density and electromagnetic fields experience quasi-random fluctuations driven by large-scale forces, due to the motion of macroscopic bodies or plasma elements. Turbulent motion dissipation leads to plasma heating, the generation of magnetic fields, and the acceleration of particles to super-thermal energies. Thus, we must understand turbulent processes to interpret many astrophysical observations, from solar flares to black hole accretion disks. Though these environments have very different plasma parameters, the turbulent processes can be generally grouped into several categories: magnetohydrodynamics (MHD) (collisional) turbulence; Whistler/Hall turbulence; shock-and-reconnection-generated turbulence; turbulence in collisionless plasmas; and turbulence in strongly magnetized plasmas (sigma greater than 1). The new features added to this list are relativistic plasma temperatures, and flow speeds and pairs. As turbulence cascades down to the kinet-ic level, eventually it forms many thin current sheets. Hence, the dissipation of turbulence at the kinetic level is inseparable from that of reconnection and current sheet instabilities.

***Sample scientific questions on relativistic turbulence and reconnection include:***

1. Alfvén/Whistler/Hall turbulence. As the turbulent cascade propagates to smaller scales, the typical frequencies of fluctuating electromagnetic fields may become high enough so that ions stop responding to them. What are the properties of the turbulence in this case?

2. Turbulence in collisionless, high-beta plasmas. In most relativistic regimes, Coulomb collision times are much longer than the cyclotron and plasma oscillation periods. How does dissipation proceed on kinetic scales?

3. Turbulence in strongly magnetized plasmas (e.g., corona of magnetars, AGN and GRB jets), where the energy density of the magnetic field exceeds plasma energy density, including rest mass. In all of these turbulence types the key questions are: what are the spectra and anisotropic properties of the fluctuating quantities on different scales, and what are the spectra of particles accelerated by a Fermi-type diffusion mechanism and DC-electric field generated in current sheets and reconnection sites? This is an especially promising route in a magnetic-dominated plasma, where most of the energy is stored in the magnetic field.

4. How do current sheet dissipation and reconnection proceed in relativistic, pair, and strongly magnetized plasmas?

## MAJOR OPPORTUNITIES

The study of relativistic plasmas will benefit from advances in theory and computer simulation, as well as from new experiments using lasers and particle beams. Below we highlight some emerging experimental opportunities.

The ability to perform relativistic beam dissipation and shock experiments in the laboratory will provide critical new information on complex shock physics and cosmic particle acceleration, and allow the calibration of computer codes. Intense lasers and accelerators are ideal tools for generating relativistic beams and collisionless shocks in the laboratory. A collisionless ambient plasma may be first created using multi-MJ-class pulse-power machines, or MJ-class long pulse lasers to heat a sufficiently large volume of hydrogen plasma to multi-keV temperatures so that it becomes collisionless. Then a short-pulse laser of intensity greater than $10^{20}$Wcm$^{-2}$ can be used to deliver a strong shock in this overdense plasma. While the shock generated by such collisions will be mildly relativistic (≲0.1c), it can propagate through a large volume of plasma to be studied in detail. Such mildly relativistic shocks are relevant to the afterglows of GRBs and blazar jets and microquasars. Alternatively, multiple short-pulse lasers can create colliding multi-MeV electron-positron jets. If these jets can be generated with sufficient density and column density — so that the interaction region is larger than the plasma skin depth — they may be able to form ultra-relativistic shocks with Lorentz factors greater than 50, which are relevant to the emissions of AGN, GRB, and PWN. Particle beams produced by conventional accelerators also may be able to drive relativistic shocks after undergoing current filamentation instability. Current filamentation experiments are underway and being planned.

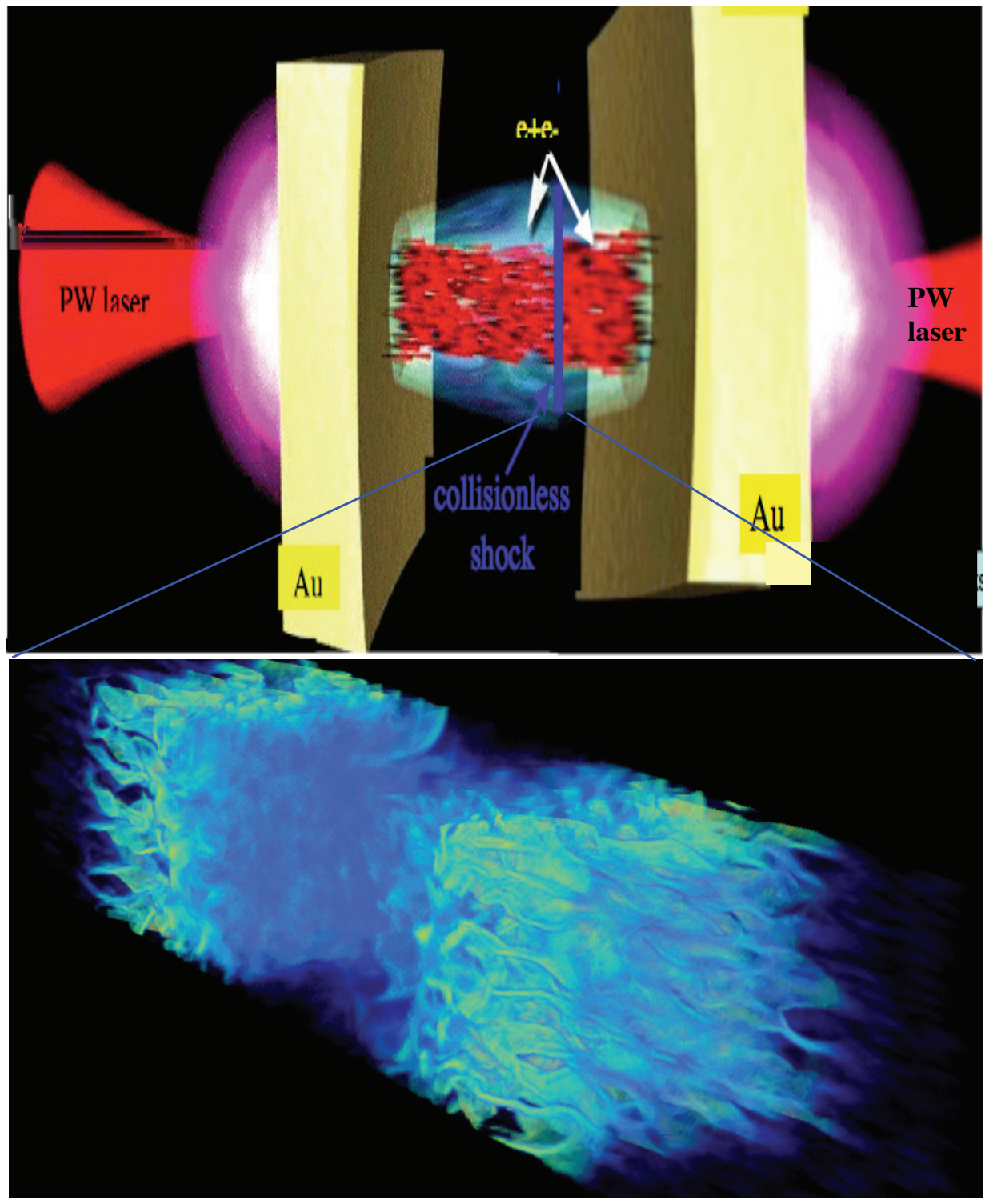

***(Top) Artist conception of a relativistic shock launched by the head-on collision of two multi-MeV e+e- pair jets created by two high-energy PW lasers irradiating mm-thick gold targets similar to the previous figure (artwork adapted by Edison Liang 2010). (Bottom) Visualization of magnetic energy density in a PIC simulation of the relativistic collisionless shocks created by the colliding pair jets. Magnetic energy inside each of the shocks is generated by Weibel instability and displays characteristic filamentation. Image courtesy of Anatoly Spitkovski 2010.***

Diagnostic development is a critical need for studying relativistic Weibel instability. The Weibel instability may be probed indirectly using the jitter radiation emitted by hot electrons passing through the self-generated turbulent magnetic field. A more direct approach is to image the transition radiation emitted when the electron and positron filaments emerge from a target. The magnetic fields created by Weibel may be probed by deflecting proton beams created by another intense laser. Computationally, 3-D Particle-in-cell (PIC) simulations of Weibel must be extended to much larger space-and-time domains to replicate realistic laboratory conditions, not to mention astrophysical settings. This will require larger supercomputers (100,000 CPUs, terascale to exascale memory system), faster, memory-efficient algorithms, as well as smarter graphics and visualization software to handle the correspondingly large amount of data generated in the simulations. In addition to lasers with intensities greater than $10^{20}$W.cm$^{-2}$, magnetized shock experiments will require greater than 10 MG pulse magnets. Relativistic shocks also must be diagnosed on picosecond time scales to measure in situ temperature, density, magnetic field, pair fraction, particle spectra, and radiation output with high-spatial resolution. Computationally, we must link and merge various multi-physics codes to perform end-to-end simulations of realistic shock experiments, including 3-D MHD, PIC, and hybrid codes, plus post-processing codes to model the radiation. DOE supercomputers should be made available to academic groups working on Weibel instability and shocks. We should form an international team of leading academic groups and experimentalists at the laser facilities to integrate astrophysical observation, theory, simulation, and laboratory experiments.

Lasers with intensity greater than 1.4x$10^{18}$Wcm$^{-2}$ irradiating solid, high-Z (e.g., Au) targets can be used to create e+e-pairs in a high-Z target, via the Trident and Bethe-Heitler processes. Recent experiments using a laser irradiating millimeter-thick gold targets created an estimated $10^{11}$ pairs in a picosecond, with an estimated in situ pair density greater than $10^{16}$cm$^{-3}$. Later experiments produced even higher pair yields, and showed that the pairs have a quasi-thermal energy distribution.

Future experiments should produce a much higher pair yield and density. Conventional accelerators can also produce GeV $e^+/e^-$ bunches with densities in the $10^{17}cm^{-3}$ range. To replicate radiation-dominated neutron star accretion columns in the laboratory, we require radiation temperatures of the order of 1 keV at densities of order $10^{-3}$g/cc and magnetic fields ~0.1-1 gigagauss, which would be required to prevent transverse expansion of photon bubbles and confine the plasma to flow in one direction. Ultra-strong fields can be generated by relativistic laser interactions (greater than $10^{18}$ W/cm$^2$) due to currents produced by supra-thermal electrons accelerated in the evanescent region of the laser plasma. This magnetic field is in the azimuthal direction about the laser axis, and the peak field extends for about an anomalous skin depth into the plasma, near the critical surface during the actual time of the picosecond laser pulse and high-density plasma interaction. Such fields cannot be measured with conventional techniques such as the Faraday rotation.

The needs of future pair plasma experiments include:

1. Dedicated facilities with multiple kJ-class PW lasers.

2. Diagnostic development for measuring in situ pair densities and temperature, and positron to ion ratio.

3. Techniques for trapping and cooling dense pair plasmas.

4. Techniques to accelerate and collimate pairs to form pair jets.

Computationally, we need to link 3-D plasma codes with particle physics codes to perform self-consistent, end-to-end simulations of pair-creation experiments. We should formulate a coordinated program to create relativistic pair plasmas, and study their astrophysical and technological applications. Such a program should involve astrophysicists, positron physicists, accelerator physicists, and laser experimental teams at DOE facilities.

Experimental work is presently behind theory in the creation of superstrong magnetic fields. Precise magnetic field measurements are most critical to identify the important mechanisms and to verify predictions. To date, the highest measured fields (up to 0.7 gigagauss) have been inferred by laser plasma interactions at $10^{20}$W/cm$^2$ using polarization measurements of scattered radiation. Pending lasers with intensities up to $10^{23}$W/cm$^2$ should be able to generate magnetic fields greater than several gigagauss, which should allow a more systematic study of high-field physics. The use of charged particle probe beams (e.g., protons) may allow high-resolution measurement of the strength and dynamics of the magnetic fields. To examine the "photon bubble" instability requires the co-location of a nanosecond high-energy laser system with a PW-level short-pulse laser to produce the large B-fields, plus another high-power, short-pulse laser to generate a particle probe beam. Several facilities capable of such experiments are presently in operation and others are under construction. To model pulsar magnetospheres, we need the co-location of pair plasmas and superstrong fields. Such experiments may be pursued with multiple lasers, using some of the laser beams to create the pairs while using other beams to create the superstrong magnetic fields.

Already underway are several laboratory experiments on magnetized jets using advanced pulsed power magnetic technology, as well as laboratory experiments on unmagnetized hydrodynamic jets using high-energy density laser technology. However, there is no detail characterization of most jet parameters. Relativistic jets of electrons and hybrid electron-pair plasmas have been generated using ultra-intense short-pulse lasers and conventional accelerators. Recent experiments demonstrated that laser-created positrons can be efficiently accelerated to high Lorentz factors by sheath electric fields to form a relativistic jet.

The next decade offers a remarkable opportunity for developing an understanding of relativistic jets, which have been an enigma for many decades. This is because new experimental facilities can replicate essential features of astrophysical jets, new computer codes can solve the complex systems of equations characterizing jets, and new telescopes can observe jets with higher energy and resolution than ever before. To take advantage of this opportunity, we propose establishing a National Center for Astrophysical Jet Studies. This center would support the experimental, numerical, and observational studies now underway at a number of institutions, place these efforts under one roof, and coordinate them by holding regular workshops. By promoting a synergism of the institutions now working on jets, the center would greatly accelerate the rate of answering and addressing the questions above.

If we inject relativistic electron-positron or MeV proton jets generated by intense lasers into a plasma with a steep density gradient, or if we can create multiple relativistic colliding shocks using multiple laser beams, they may be able to generate sufficient plasma turbulence to address some of these questions. Furthermore, if we can create magnetic fields of opposite polarities in the ambient plasma prior to the jet-shock interactions, the MHD turbulence created by the jet-jet and jet-plasma interactions may induce or enhance current sheet dissipation and magnetic reconnection. This will help address the outstanding question about whether turbulence can dissipate and nonthermally heat electrons, positrons, and ions by enhancing the current sheet dissipation and magnetic reconnection. The physics of current sheet dissipation and reconnection in relativistic and pair plasmas is important to black hole accretion, the pulsar wind "sigma problem," gamma-ray bursts, and many other high-energy astrophysics phenomena.

We need multiple intense lasers to create the colliding shocks or particle beams capable of generating relativistic plasma turbulence. At present only one laser facility may possess such capability, when proposed short-pulse kJ beams are completed. In addition to the lasers, pulsed magnets of greater than 10 MG fields with reversing polarities will be needed to create intense current sheets. Diagnosis of the magnetic field, particles, and waves generated by the turbulence will be major challenges. New diagnostic techniques will be needed before meaningful measurements can be made. Computationally, we must bridge the huge gap between MHD and PIC simulations before we can confidently explore the cascade from the MHD to the kinetic scale. Three-dimensional reconnection simulations have recently reached a major threshold where electron-positron plasmas can be meaningfully studied with sufficient mode numbers for both the kink and tearing instabilities. However, realistic electron-ion simulations in 3-D reconnection are still in their infancy and must await even faster and larger supercomputers. To study particle acceleration in turbulence and reconnection, we need to keep track of only the most energetic particles. Finally, since radiative cooling may be energetically important in strongly magnetized relativistic plasmas, we need to include radiative damping terms in PIC codes for some turbulence and reconnection applications.

## IMPACTS AND CONNECTIONS TO OTHER TOPICS

The physics of beam dissipation and collisionless shocks provides the foundation for understanding the most energetic phenomena of the universe, from gamma-ray bursts to high-energy cosmic rays. This subject will benefit from close interactions with the NASA community. Results of laboratory experiments should be rapidly communicated to astrophysicists analyzing Fermi data. These plasma phenomena are also critical to future technologies such as the Fast Ignition approach to inertial fusion.

Both pair plasmas and gigagauss magnetic fields are important frontiers of high-energy astrophysics. The creation of such plasmas in the laboratory will allow us to explore these most exotic regimes in astrophysics. Potential technological applications of such plasmas have not been thoroughly explored, but clearly would be transformative. Funding agencies should strongly support and encourage the study of this new regime of plasma physics.

Relativistic jets are only the extreme version of jets, from young stellar objects to white dwarfs and neutron stars. The study of relativistic jets is therefore intimately connected to that of non-relativistic jets. There should be considerable synergism between both kinds of jets.

Relativistic turbulence and reconnection are extensions of conventional turbulence and reconnection. Advances in their understanding will undoubtedly benefit the study of this topic.

# CHAPTER 10:
## JETS AND OUTFLOWS

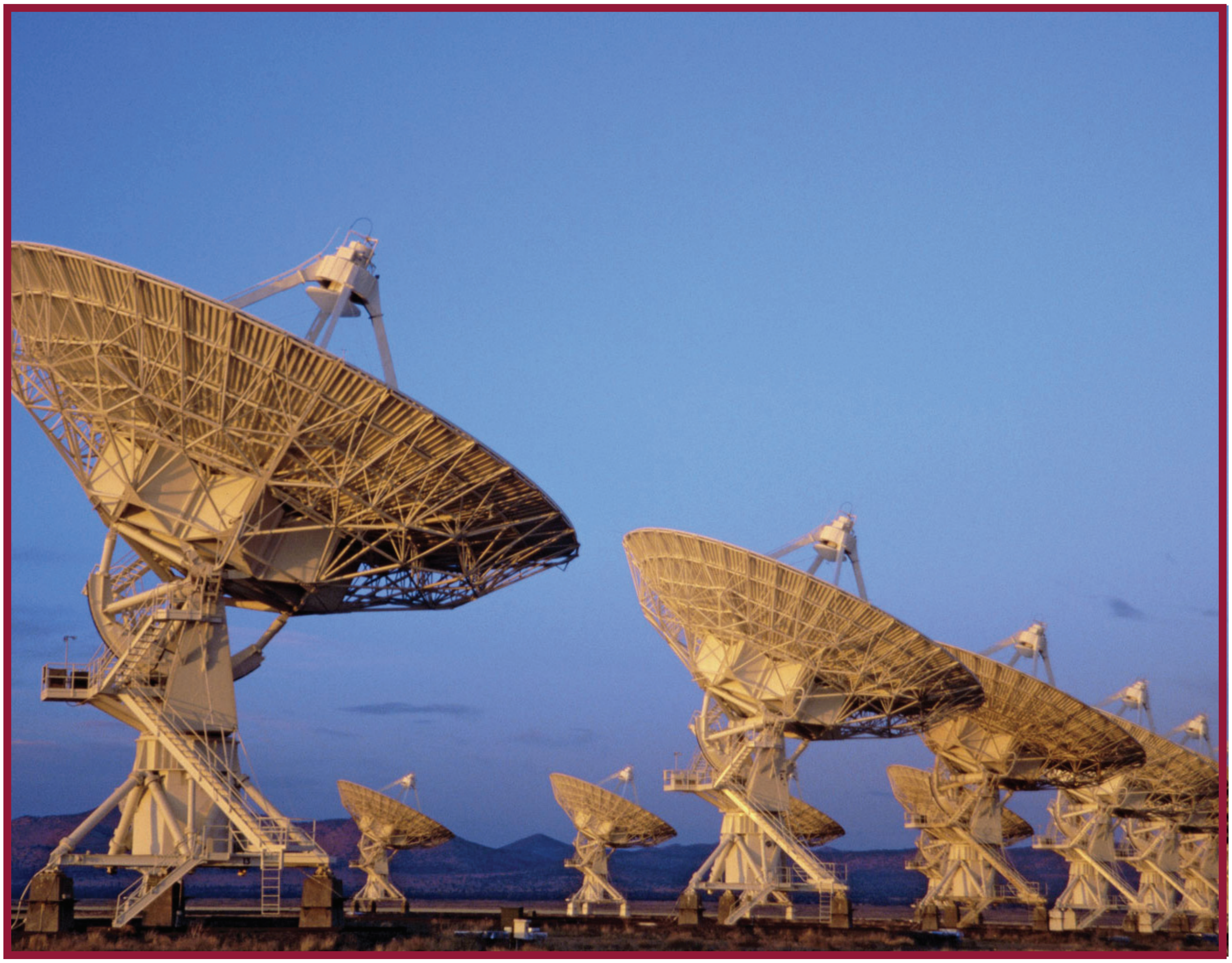

ON PREVIOUS PAGE

*The Very Large Array, a radio telescope system located on the plains of San Agustin, near Socorro, N.M.* (Image courtesy of the National Radio Astronomy Observatory, a facility of the National Science Foundation.)

# CHAPTER 10: JETS AND OUTFLOWS

## INTRODUCTION

Fast, axially collimated outflows, or "jets," are ubiquitous in astrophysics. They flow from most classes of compact objects that spin and/or accrete matter from their surroundings, which range from young stellar objects (YSOs), neutron stars, and stellar mass black holes (BH) to super massive black holes (SMBH) in the centers of galaxies. Compactness leads to high gravitational potential. Rapid spin provides a second free energy source. The axial symmetry of spin allows energy, momentum, and angular momentum to concentrate and be transported away along the symmetry axis, typically in oppositely directed jet pairs. Some jets are probably matter dominated (highly ionized plasmas), while others may consist primarily of electromagnetic energy, and some a blend (electron-positron pairs are involved in some high-energy settings). The jet speed can range from a few tens of kilometers per second to close to the speed of light. The total jet energy can be a large fraction of the gravitational energy released during the host objects' formation (up to a fraction of their rest mass energy), and can be sufficient to "de-spin" the host compact object.

We also now realize that jets play an important feedback role in the evolution of their host systems. Jets from protostars energize the parent, star-forming molecular cloud, possibly regulating the rate and efficiency of star formation. There is ample evidence that jets from SMBHs in galactic nuclei both energize the nearby interstellar plasma and, for those in galaxy clusters, have an impact on the intracluster medium (ICM) and the inter-galactic medium (IGM). This contributes to the extra-galactic magnetic fields and cosmic rays, including ultra-high-energy cosmic rays (UHECRs), neutrinos, and gamma rays. Jets and lobes serve as a useful calorimeter of the non-thermal energy component in the overall cosmic energy flow.

Many fundamental challenges remain for the understanding of jets and outflows, including:

1. Plasma conditions of jets are not well known. What are jets made of? More and better measurements are needed around the jet "engine," in the jets, and at the jet termination.

2. The range of scales is vast. As an example, the jet originates from a SMBH at approximately 1 astronomical unit ($10^{13}$ centimeters) scale but extends to megaparsecs ($10^{24}$ centimeters). There are even smaller scales associated with the collisionless nature of jet plasmas. This vast scale separation makes it extremely challenging to build a coherent theory from engine to termination.

3. There is a lack of plasma physics understanding. Plasma physics processes govern the energy transfer among gravitational, kinetic, thermal, and magnetic-electric components, and particles. How are jets accelerated and collimated? How do magnetic fields behave? Why are jets stable? How does large-scale fluid motion "communicate" with small-scale kinetic processes? Will relativistic effects change the physics drastically? What are the dissipation processes to produce 10 TeV electrons and $10^{20}$ eV cosmic rays?

Jets and outflows therefore present a suite of important laboratories to test our understanding of plasma physics. Such knowledge conversely provides part of the necessary physics underpinning for understanding the dynamics and evolution of the universe. Thus, astrophysical jets and outflows offer a natural laboratory for a concerted study by the astrophysics and plasma communities. Substantial progress can only be made when these two communities work together; when observation, laboratory experiments, theory, and simulation communities have built strong ties; and when the infrastructure (for research collaboration and funding) is conducive for such efforts.

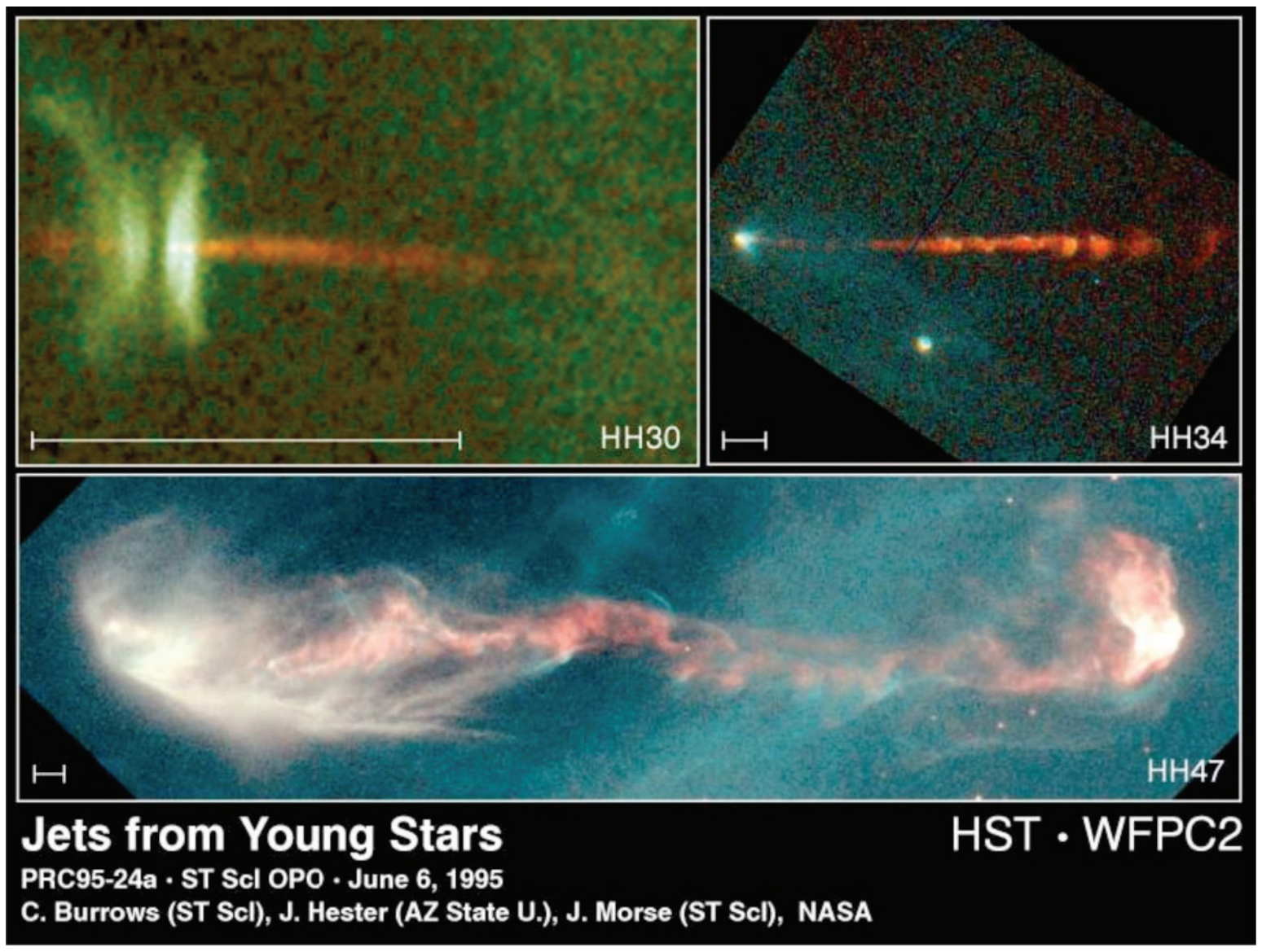

***Three examples of YSO jets on different scales. In each case, the scale bar corresponds to a length scale of 1000 AU, and the image is oriented so that the jet source is on the left. The compact source is obscured from direct view in each case, but its light is scattered from the walls of the outflow cavities (white light). For HH30, there is an obvious disk surrounding the central object, seen edge on.***

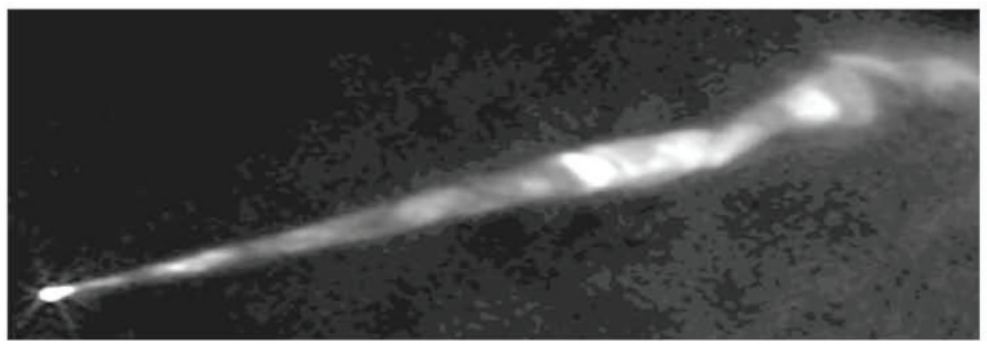

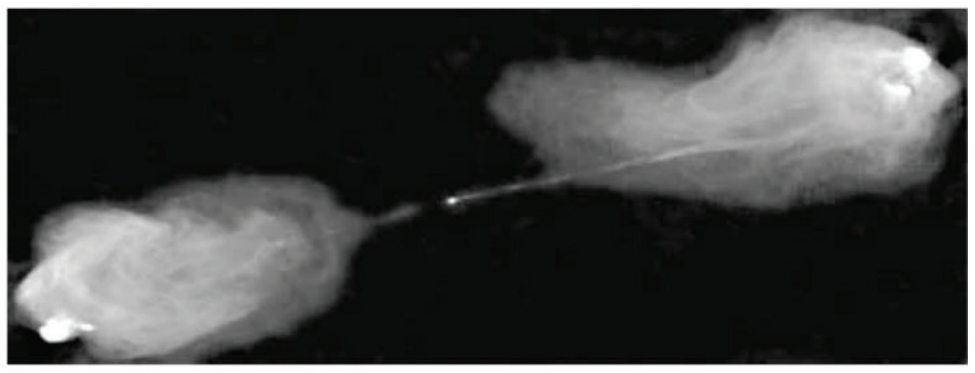

***The left image (M87), 2-kiloparsec scale, shows a jet which is beginning to be seriously disrupted by a growing helical instability. (Courtesy of Owen, F.N., NRAO.) The right image (Cyg A), 40-kiloparsec scale, shows two jets that have NOT gone unstable or ever seriously disrupted. (Courtesy of Carilli, C.L., NRAO.)***

## KEY SCIENTIFIC CHALLENGES

### *Challenge I: Physics of Jet Launch and Acceleration*

Due to the difficulty of direct observations of the disks and central objects (young stellar objects, neutron stars, and black holes), it is challenging to constrain the initial condition for jet production. For YSOs, we need improved measurements of magnetic field strengths and orientations on all scales, especially the launching regions. The Atacama Large Millimeter Array (ALMA) may be able to measure B fields on ~1AU scales in the nearest sources. High spectral resolution spectroscopy, combined with high spatial resolution, allows us to probe the physical properties of outflowing material (e.g., bulk velocity, density, temperature, shock velocity, and ionization state). Measurements of jet rotation are also important for determining the launching region and mecha-nism. For extragalactic sources, the VLBI arrays are beginning to probe close to the central BH accretion disk "engine" where the jets are launched. Recent VLBI results reveal time-varying helical magnetic structures that characterize the jet-launching region on scales as small as one light year. Magnetic field strengths and plasma densities are such that Faraday rotation effects are readily seen at these compact dimensions, so that quasi-3-D imaging of the jet launching regime can be undertaken. This will be important in addressing questions such as whether jets carry significant currents, as well as the jet power in different components (kinetic vs. Poynting).

It has become possible during the past decade to perform General Relativistic magnetohydrodynamics (MHD) simulations of disks around a BH while investigating the launching of relativistic jets. Many questions, however, remain: If the jets are magnetically mediated, where do the large-scale fields come from? What controls the advection, diffusion, and dynamo of B field in magnetorotational instability (MRI) disks ($\beta$>1) and the non-MRI disk corona ($\beta$ <1, where $\beta$ is thermal pressure over magnetic pressure)? During the jet launching, is the acceleration primarily by magneto-centrifugal effects, "magnetic pressure" effects, or non-magnetic forces? What determines the mass loading and thermal energy input at the outflow base? Under what conditions do magnetically dominated jets occur rather than hydromagnetic jets? What do jets consist of, electrons and ions, or $e^{\pm}$ pairs? What determines the bulk Lorentz factors and energy fluxes of jets? And how do these quantities depend on the BH spin, the disk magnetic field, and the accretion rate? What is the role of the angular momentum outflow of jets and winds? The plasma physics understanding (including relativistic MHD) in these questions is clearly needed to make progress.

During the past decade, substantial progress also has been made using laboratory plasma experiments to model some aspects of the physics of astrophysical jets. If the dynamical scaling constraints are properly applied, one can reproduce astrophysical processes on some segment of their time-evolution. Many of the astrophysical systems are well described by ideal magnetohydrodynamics, due to very large values of Reynolds ($LV/\nu$), magnetic Reynolds ($LV/\eta$), and Peclet numbers ($LV/\alpha$, where $\nu$, $\eta$, $\alpha$ are viscosity, resistivity, and thermal conductivity, respectively) — primarily because of the very large spatial scales involved. Recently, it has become possible to include dynamically significant magnetic fields and to achieve sufficiently high magnetic Reynolds numbers and high Alfvénic Mach numbers in some laboratory jet experiments. Some questions under investigation include: How does the Lorentz force convert electrical power into directed flows? Where and how does the jet acceleration take place? What is the quantitative scaling showing how different combinations of electrical and material inputs (voltage, current, field topology, and mass source morphology) produce different types of flows (e.g., low-mass density with high velocity, high-mass density with low velocity, and degree of collimation)?

### *Challenge II: Physics of Jet Collimation, Propagation and Termination*

For YSOs, launching happens on scales from ~1 stellar radius to ~1AU, while collimation appears to occur on size scales of ~100AU. It is unknown whether collimation is primarily due to the internal outflow properties or to an interaction with the environment. Time series of images and spectra, spaced a few to several years apart, allow us to probe the interaction of outflows with their environments, as well as internal outflow-outflow interactions. These studies are useful for determining whether the clumpiness and variability observed in jets is primarily due to variations in jet launching source, instabilities in the flow, or interactions with environment. This will also put constraints on the amount of feedback from forming stars on the parent molecular cloud. For extragalactic sources, imaging polarimetry with VLA already shows evidence for helical magnetic field structures on kiloparsec scales. The tentative measurement of a jet current is ~$10^{18}$ amperes, and the spatially averaged plasma $\beta$ could be as low as $10^{-5}$ at currently available resolutions. In the intracluster medium, X-ray observations of jet-blown cavities have allowed the first direct measures of the energy deposition by BH-driven jets. Radio observations (e.g., VLA and Expanded VLA) of the morphology of jets and lobes provide valuable constraints on the composition of jets and the physics of jet-ICM interaction. They, in turn, provide unique probes into the physical and dynamical properties of the ICM.

Some fundamental questions regarding jet collimation, propagation, and termination still remain. Are jets collimated by the magnetic hoop force, pressure of external medium, or by uncollimated outer disk winds? What are the key parameters of the jets that control the global stability of the jet to kink and other instabilities? How will the current-driven kink instability change in relativistic MHD? What processes will govern the energy conversion from B-field dominated limit to kinetic energy? What is (are) the mechanism (s) for efficiently accelerating leptons in situ to energies sufficient to give TeV radiation (shock acceleration, reconnection, KH instability, etc.)? Do UHECRs come from BH, disk, jet, and lobe systems? How important are the jets for energy and momentum feedback into the interstellar medium (ISM) and for driving turbulence in molecular clouds? How will active galactic nucleus (AGN) jets excite turbulence in the ICM and IGM? Will AGN provide significant magnetic fields and cosmic rays to the ICM and IGM? What are the quantitative feedback effects of SMBH jets on the cosmic structure formation?

Existing laboratory experiments can produce highly collimated jets that become kink unstable at a critical length and, in certain cases, manifest internal shocks, knots, and even disconnection of the jet from the source. Existing experiments can thus be used in the near term to address a set of questions discussed above, including: What collimates the jets? What is the jet stability with sufficiently strong axial currents (magnetic fields)? What mechanisms can suppress the kink instability in jets? How does the jet interact with the ambient medium? Is there intrinsic connection between the physics of radio jets and lobes, and laboratory spheromak and reversed-field pinch experiments?

## MAJOR OPPORTUNITIES

### Establishing New Infrastructure for Jet Research

The study of astrophysical jets is an active research area in astronomy and astrophysics, and it is ripe for progress with significant inputs from plasma physics. Nearly all current major observatories (e.g., Fermi, Chandra, Spitzer, Pierre Auger, HST, VLT, EVLA, VLBA/VLBI) contribute strongly to understanding the nature of jets. It has been well established that jets play an important role in determining the physical condition of the interstellar medium, the intra-cluster medium, and the inter-galactic medium. Furthermore, jets could be closely related to the origin of ultra-high energy cosmic rays. The last decade also saw major advances on two fronts: the use of laboratory experiments to study jets, and sophisticated multi-dimensional general relativistic MHD simulations of jets. Yet we do not yet have an intellectual grasp on these systems. We urgently need an understanding from the plasma physics point of view of jets. Presently, jet research is fragmented (relatively little collaboration among observers, experimentalists, theoreticians, and people doing simulations). Different funding agencies, while supporting jet research in certain ways, have different priorities and constraints.

We propose the formation of a consortium on astrophysical jets, with funding for research and regular workshops. This consortium would bring together interested astronomers and plasma scientists, and take advantage of recent breakthroughs in parallel numerical simulations and laboratory experiments. Such a consortium must be supported by multiple agencies (e.g., DOE, NSF, and NASA). We especially encourage direct contact between members of the different communities, via in-person visits or attendance at small meetings. This encourages the growth of new ideas and intuition much more efficiently than the traditional transfer of information via journals and large conferences. Specific suggestions include: 1) seeking funding for the development of research networks, such as the very successful JetSet network created in the EU; 2) having members of the computational or observational communities make extended visits to experimental facilities and collaborate in the operation and interpretation of experiments; 3) encouraging collaboration and cooperation between the plasma community and the radio astronomy community (to fully utilize available resources on observations of magnetic fields and jets); and 4) holding focused workshops on specific topics (such as why jets are stable), summer and winter schools, and joint seminars, etc.

**Opportunities from Radio Astronomy**

New capabilities of radio astronomical instruments could play a pivotal role in revealing the plasma processes in extragalactic AGN jets. The jet-launching region near a galaxy's SMBH is accessible to VLBI and on the kiloparsec-to-megaparsec (kpc-Mpc) scale phase of SMBH-powered jets. Cross-jet angular resolution is critical. This requires imaging interferometers spanning a few hundred kilometers, with full polarization capability at wavelengths up to about 30 centimeters. For probing the launching region, we need logistics and additional equipment — particularly short wavelength receivers — to reach as close as possible to the SMBH accretion disk zone. For the required sensitivity in future VLBI imaging, the exceptional size of the Arecibo telescope and the Green Bank Telescope are key existing U.S. resources. For kpc-Mpc scales, the imaging capability of the National Radio Astronomy Observatory's (NRAO) new Expanded Very Large Array (EVLA) is well developed in the needed direction, except for its limited maximum baseline of 35 kilometers. To resolve the jets transversely at centimeter wavelengths requires a 10 to 30 times larger maximum baseline. A design and proposal have been produced for such an instrument, the "EVLA2," but this appears to be inactive. In the absence of a future U.S.-based EVLA2, a European array of similar dimensions might conceivably be combined with the 35-kilometer EVLA to achieve similar capabilities.

**Understanding Why Jets Are Stable Over Long Distances**

One of the most challenging aspects of understanding jets is why they are stable over long distances (the extent of the jets can be greater than $10^{10}$ times the size of the engine). Since stability is closely tied with global jet dynamics, this is one parameter space that observations (having enough resolution), experiments, and theory and simulations can jointly address. The challenges include better constraints on the jet composition on large scales and well-developed theory incorporating more detailed measurements and observations. We believe that through a close collaboration among observation, experiment, and theory and simulation, significant progress can be made in this area, including modeling the jet energetics, stability, and morphologies in well-characterized background environments (e.g., radio jets and lobes in galaxy clusters).

## IMPACTS AND MAJOR OUTCOMES

Jets and outflows can be very high $\beta$ (e.g., the ICM or IGM, and possibly BH-driven jets far from their origin). Alternatively, they can be very low $\beta$ (e.g., BH-driven jets near their origin). Jet and outflow systems often involve high-speed flows. Examples are relativistic flows in jets and pulsar winds; supersonic or super-Alfvénic flows in jets; and supersonic or super-Alfvénic turbulence in the ICM and IGM. The astrophysical and laboratory plasma communities must work together more closely for mutual benefit, to understand which plasma processes are important, and how they operate in different parameter spaces. We know that jets from YSOs and SMBHs play significant roles in regulating their surroundings. Consequently, jets affect our understanding of processes such as star formation, galaxy formation, and the physics of the ICM and IGM — all issues of paramount importance in astrophysics.

## CONNECTIONS TO OTHER TOPICS

The study of jets naturally brings together a number of topics discussed in this Report. For example, the existence and critical role of magnetic fields in disks around stars and BHs — and in facilitating jet launching — are closely related to the dynamo and angular momentum transport processes. Particle acceleration processes such as collisionless shocks and reconnection determine the production of high-energy particles (perhaps UHECRs) and photons (observed up to 10 TeV). The relativistic speeds observed in jets, and the inferred high magnetization, call for studies of relativistic plasmas in extreme parameters. The accretion into SMBHs that sometimes leads to jet formation can also produce some of the most prodigiously luminous radiators in the universe where radiation hydrodynamics play an essential role. Accretion disks, and jets and lobes, are believed to be inherently turbulent. Turbulence could strongly influence the accretion process and energy conversion from magnetic fields to particles. On large scales, jet propagation and stability begs for the understanding of shear instability.

# APPENDIX A: ACRONYMS AND ABBREVIATIONS

| | |
|---|---|
| **2-D** | Two-dimensional |
| **3-D** | Three-dimensional |
| **ATST** | Advanced Technology Solar Telescope |
| **AGN** | Active Galactic Nuclei |
| **ALMA** | Atacama Large Millimeter Array |
| **ARC** | Astrophysics, Relativity and Cosmology research group, University of Montana |
| **AU** | Astronomical unit |
| **AUGER** | The Pierre Auger Observatory in Argentina |
| **B** | Magnetic field |
| **BH** | Black hole |
| **CARMA** | Combined Array for Research in Millimeter-wave Astronomy in California |
| **CERN-GEANT4** | A particle physics simulation code |
| **Cerro Tololo** | An inter-American observatory |
| **CHANDRA** | An X-ray observatory by NASA |
| **CLUSTER** | A space mission by the European Space Agency |
| **CFHT** | The Canada-France-Hawaii Telescope in Hawaii |
| **Cm** | Centimeter |
| **CMBPol** | The Cosmic Microwave Background Polypropylene Lens |
| **CMEs** | Coronal mass ejections |
| **COROT** | A space telescope by France and international partners |
| **DC** | Direct current |
| **DNS** | Direct numerical simulations |
| **DOE** | Department of Energy |
| **DSA** | Diffusive shock acceleration |
| **EOS** | Equations of state |
| **ESA** | European Space Agency |
| **EU** | European Union |
| **EVLA** | Expanded Very Large Array in New Mexico |
| **e-ion** | Electron-ion |
| **e+e-** | Positron-electron |
| **eV** | Electron volt |
| **ExB drift** | E cross B drift of charged particles across magnetic field by a perpendicular electric field |
| **FACET** | Facility for Advanced aCcelerator Experimental Tests at SLAC |
| **Fe** | Iron |
| **G** | Gauss |
| **g/cc** | Gram per cubic centimeter |
| **Geotail** | A space mission by Japan in collaboration with NASA |
| **GBT** | Radio telescope in West Virginia |

## ACRONYMS AND ABBREVIATIONS (continued)

| | |
|---|---|
| **GEMS** | Gravity and Extreme Magnetism Small Explorer |
| **GeV** | Giga-electron volts |
| **GRB** | Gamma-ray burst |
| **HED** | High Energy Density |
| **Hercules** | A galaxy cluster |
| **HESS** | High Energy Stereoscopic System telescope array, South Africa |
| **High-Z** | Elements with a large number of protons in the nucleus |
| **HII** | A region in space where recent star formation has recently occurred |
| **Hinode** | A solar observatory satellite by Japan in collaboration with the U.S. and U.K. |
| **HST** | Hubble Space Telescope, NASA |
| **IBEX** | Interstellar Boundary Explorer, NASA |
| **ICM** | Intracluster medium |
| **IGM** | Intergalactic medium |
| **II** | Interchange instability |
| **ISEE** | International Sun-Earth Explorer spacecraft, Smithsonian Astrophysical Observatory/NASA |
| **ISM** | Interstellar medium |
| **JAXA-CSA SCOPE** | Japan Aerospace Exploration Agency-Canadian Space Agency joint space mission |
| **JDEM** | Joint Dark Energy Mission, NASA and DOE |
| **JetSet** | A research network in the EU to study astrophysical jets |
| **JWST** | James Webb Space Telescope, NASA |
| **KH** | Kelvin-Helmholtz instability |
| **Kepler** | A space mission, NASA/Ames Research Center |
| **keV** | Kilo-electron volts |
| **Kitt Peak** | A national observatory, NOAO |
| **kJ** | kiloJoule |
| **km(/s)** | Kilometer per second |
| **Kpc** | Kiloparsec |
| **LANL** | Los Alamos National Laboratory |
| **LLNL** | Lawrence Livermore National Laboratory |
| **LMT** | Large Millimeter Telescope, Mexico |
| **LSST** | Large Synoptic Survey Telescope (planned), Chile |
| **LTE** | Local thermodynamic equilibrium |
| **M87** | A giant, elliptical galaxy |
| **Mach number** | Ratio of the speed of an object to the speed of sound in the surrounding medium |
| **MeV** | Mega-electron volts |
| **MFE** | Mean field electrodynamics |
| **MFE** | Magnetic fusion energy |
| **Mg** | Magnesium |
| **MG** | MegaGauss |

## ACRONYMS AND ABBREVIATIONS (continued)

| | |
|---|---|
| **MHD** | Magnetohydrodynamics |
| **MIT** | Massachusetts Institute of Technology |
| **Mm** | Megameter |
| **MMS** | Magnetospheric and MultiScale mission, NASA |
| **Mpc** | Megaparsec |
| **MRI** | Magnetorotational instability |
| **NASA** | National Aeronautics and Space Administration |
| **NTF** | Nevada Terawatt Facility (laser) |
| **NEWFIRM** | NOAO Extremely Wide Field Infrared Imager |
| **NIF** | National Ignition Facility, Lawrence Livermore National Laboratory |
| **NIF-ARC** | NIF's Advanced Radiographic Capability |
| **Nm** | Nanometer |
| **NOAO** | National Optical Astronomy Observatory |
| **NRAO** | National Radio Astronomy Observatory |
| **NSF** | National Science Foundation |
| **NuStar** | Nuclear Spectroscopic Telescope Array, NASA |
| **OFES** | Office of Fusion Energy Sciences, DOE |
| **Omega** | A laser facility for high energy density physics, University of Rochester |
| **PAMELA** | Payload for Antimatter Matter Exploration and Light-Nuclei Astrophysics, Italy |
| **PI** | Principal investigator |
| **PIC** | Particle in cell |
| **PPPL** | Princeton Plasma Physics Laboratory |
| **PW** | Pulsar wind |
| **PW lasers** | Petawatt lasers |
| **PWN** | Pulsar wind nebula |
| **RT** | Rayleigh-Taylor instability |
| **RFP** | Reversed Field Pinch |
| **RHESSI** | Reuven Ramaty High Energy Solar Spectroscopic Imager, a satellite explorer, NASA |
| **RMI** | Richtmyer-Meshkov Instability |
| **SDO** | Solar Dynamics Observatory, NASA |
| **SGRs** | Soft Gamma Repeaters |
| **SLAC** | Stanford Linear Accelerator |
| **SLAMS** | Short Large Amplitude Magnetic Structure |
| **SMA** | Submillimeter Array, Smithsonian Astrophysical Observatory |
| **SMBH** | Super Massive Black Holes |
| **SN (SNe)** | Supernova (Supernovae) |
| **SNAP** | SuperNova Acceleration Probe |
| **SNL** | Sandia National Laboratories |
| **SNR** | Supernova Remnants |
| **SOFIA** | Stratospheric Observatory for Infrared Astronomy, U.S. and Germany |

## ACRONYMS AND ABBREVIATIONS (continued)

| | |
|---|---|
| **SOHO** | Solar and Heliospheric Observatory, NASA |
| **Solar Orbiter** | ESA spacecraft |
| **Solar Probe Plus** | A NASA mission |
| **Spitzer Space Telescope** | A NASA Jet Propulsion Laboratory telescope |
| **Square Kilometer Array** | An international radio telescope |
| **STEREO** | A NASA spacecraft for studying the Sun in 3-D |
| **Sweet-Parker Model** | An early theory for magnetic reconnection |
| **Swift Satellite** | A NASA gamma-ray burst mission |
| **Taylor-Couette Flow** | A fluid flow driven between concentric cylinders |
| **TeV** | Tera-electron Volts |
| **TG** | TeraGauss |
| **THEMIS** | Time History of Events and Macroscale Interactions during Substorms, a NASA mission |
| **Texas Petawatt** | A laser experiment at the University of Texas-Austin |
| **Titan** | A laser facility at the Lawrence Livermore National Laboratory |
| **TMT** | Thirty Meter Telescope to be built in 2018 |
| **Type-II Supernova** | Collapse of a massive star that has retained its hydrogen-rich envelope |
| **UHECRs** | Ultra-high Energy Cosmic Rays |
| **UV** | Ultraviolet |
| **UVCS** | UltraViolet Coronagraph Spectrometer, SOHO |
| **V&V** | Validation and Verification |
| **VLA** | Very Large Array in New Mexico, NRAO |
| **VLBI** | Very-Long-Baseline Interferometry |
| **VLT** | Very Large Telescope, European Southern Observatory |
| **Voyagers 1 and 2** | Twin spacecraft to study Jupiter and Saturn, NASA |
| **W/cm** | Watts per centimeter |
| **WD** | White Dwarf star |
| **Weibel Instability** | A plasma instability |
| **Whistler/Hall** | Whistler waves and Hall effects |
| **X-class flares** | Large solar flares |
| **X-line** | A line through which magnetic reconnection occurs in 2-D |
| **XMM-Newton** | X-ray Multi-Mirror Mission-Newton, an ESA space mission |
| **YSO** | Young Stellar Object |
| **Z Machine** | A pulsed power facility at Sandia National Laboratories |

# APPENDIX B: WORKSHOP PARTICIPANTS AND WORKING GROUP MEMBERS

SPIRO ANTIOCHOS, NASA Goddard Space Flight Center
JONATHAN ARONS, University of California
JAMES BAILEY, Sandia National Laboratories
STUART BALE, University of California, Berkeley
MATTHEW BARING, Rice University
TONY BELL, University of Oxford
PAUL BELLAN, Caltech
AMITAVA BHATTACHARJEE, University of New Hampshire
ERIC BLACKMAN, University of Rochester
STANSLAV BOLDYREV, University of Wisconsin-Madison
ALLEN BOOZER, Columbia University
JOSH BRESLAU, Princeton Plasma Physics Laboratory
MATTHEW BROWNING, Canadian Institute for Theoretical Astrophysics
DAVID BURGESS, Queen Mary University of London
TROY CARTER, University of California, Los Angeles
FAUSTO CATTANEO, University of Chicago
CHI-KWAN CHAN, Harvard University
MORRELL CHANCE, Princeton Plasma Physics Laboratory
T.K. CHU, Princeton Plasma Physics Laboratory
STIRLING COLGATE, Los Alamos National Laboratory
RAMANATH COWSIK, Washington University
STEVEN CRANMER, Harvard-Smithsonian Center for Astrophysics
WILLIAM DAUGHTON, Los Alamos National Laboratory
CHARLES DERMER, Naval Research Laboratory
PATRICK DIAMOND, University of California, San Diego
SETH DORFMAN, Princeton Plasma Physics Laboratory
WILLIAM DORLAND, University of Maryland
GEORGE DOSCHEK, Naval Research Laboratory
JAMES DRAKE, University of Maryland
ERIC EDLUND, Princeton Plasma Physics Laboratory
PHIL EFTHIMION, Princeton Plasma Physics Laboratory
JAN EGEDAL, Massachusetts Institute of Technology
JEAN EILEK, New Mexico Tech

## WORKSHOP PARTICIPANTS AND WORKING GROUP MEMBERS (continued)

CARY FOREST, University of Wisconsin-Madison
GUOYONG FU, Princeton Plasma Physics Laboratory
THOMAS GARDINER, Sandia National Laboratories
CHRISTOPHE GISSINGER, Princeton University
MELVYN GOLDSTEIN, NASA Goddard Space Flight Center
PETER GOLDREICH, Institute for Advanced Study
JEREMY GOODMAN, Princeton University
NIKOLAI GORELENKOV, Princeton Plasma Physics Laboratory
GREG HAMMETT, Princeton Plasma Physics Laboratory
PATRICK HARTIGAN, Rice University
ROBERT HEETER, Lawrence Livermore National Laboratory
PETER HOEFLICH, Florida State University
MASAHIRO HOSHINO, University of Tokyo
JOHN P. HUGHES, Rutgers University
THOMAS INTRATOR, Los Alamos National Laboratory
HANTAO JI, Princeton Plasma Physics Laboratory
JAY JOHNSON, Princeton Plasma Physics Laboratory
J. R. JOKIPII, University of Arizona
THOMAS JONES, University of Minnesota
IGOR KAGANOVICH, Princeton Plasma Physics Laboratory
JUSTIN KASPER, Harvard-Smithsonian Center for Astrophysics
EUN-HWA KIM, Princeton Plasma Physics Laboratory
MARK KOEPKE, U.S. Department of Energy
JULIAN KROLIK, Johns Hopkins University
PHILIPP KRONBERG, Los Alamos National Laboratory
KARL KRUSHELNICK, University of Michigan
RUSSELL KULSRUD, Princeton Plasma Physics Laboratory
CAROLYN KURANZ, University of Michigan
ALEX LAZARIAN, University of Wisconsin-Madison
SERGEY LEBEDEV, Imperial College
JEONGWOO LEE, New Jersey Institute of Technology
MARTIN LEE, University of New Hampshire
HUI LI, Los Alamos National Laboratory
EDISON LIANG, Rice University
ROBERT LIN, University of California, Berkeley

## WORKSHOP PARTICIPANTS AND WORKING GROUP MEMBERS (continued)

RICHARD LOVELACE, Cornell University
VYACHESLAV LUKIN, Naval Research Laboratory
IAN MANN, University of Alberta, Canada
SEAN MATT, NASA Goddard Space Flight Center
WILLIAM MATTHAEUS, University of Delaware
LORIN MATTHEWS, Baylor University
JONATHAN MCKINNEY, Stanford University
ROBERT MERLINO, University of Iowa
AARON MILES, Lawrence Livermore National Laboratory
PATRIC MUGGLI, University of Southern California
CHRISTOPH NIEMANN, UCLA
MARK NORNBERG, University of Wisconsin-Madison
GILES NOVAK, University of Chicago
ERDEM OZ, Princeton Plasma Physics Laboratory
MARTIN PESSAH, Institute for Advanced Study
TAI PHAN, University of California, Berkeley
CYNTHIA PHILLIPS, Princeton Plasma Physics Laboratory
MARC POUND, University of Maryland
ANNICK POUQUET, The National Center for Atmospheric Research
STEWART PRAGER, Princeton Plasma Physics Laboratory
ELIOT QUATAERT, University of California, Berkeley
YEVGENY RAITSES, Princeton Plasma Physics Laboratory
ALESSANDRA RAVASIO, CNRS Luli
MARTHA REDI, Princeton University
BRUCE REMINGTON, Lawrence Livermore National Laboratory
YANG REN, Princeton Plasma Physics Laboratory
JOHN RHOADS, Princeton Plasma Physics Laboratory
AUSTIN ROACH, Princeton Plasma Physics Laboratory
MARIENE ROSENBERG, University of California, San Diego
DMITRI RYUTOV, Lawrence Livermore National Laboratory
STEVE SABBAGH, Columbia University
RAVI SAMTANEY, Princeton Plasma Physics Laboratory
WILTON SANDERS, NASA
JOHN SARFF, University of Wisconsin-Madison
YASUHIKO SENTOKU, University of Nevada

## WORKSHOP PARTICIPANTS AND WORKING GROUP MEMBERS (continued)

URI SHUMLAK, University of Washington
LUIS SILVA, Instituto Superior Tecnico
DAVE SMITH, University of Wisconsin-Madison
PAUL SONG, University of Massachusetts Lowell
ERIK SPENCE, Princeton Plasma Physics Laboratory
ANATOLY SPITKOVSKY, Princeton University
JAMES STONE, Princeton University
EDWARD THOMAS, Auburn University
ANDREY TIMOKHIN, University of California, Berkeley
DMITRI UZDENSKY, University of Colorado at Boulder
MARCO VELLI, Jet Propulsion Laboratory
MASAAKI YAMADA, Princeton Plasma Physics Laboratory
JONGSOO YOO, Princeton Plasma Physics Laboratory
MICHAEL ZARNSTORFF, Princeton Plasma Physics Laboratory
JIAN ZHOU ZHU, University of Maryland
ELLEN ZWEIBEL, University of Wisconsin-Madison
ANDREW ZWICKER, Princeton Plasma Physics Laboratory

# APPENDIX C: WORKSHOP AGENDA

## MONDAY, JANUARY 18

8:00 am – Registration

8:30 am – Welcome, Purpose, Format, and Expected Outcome (S. Prager and H. Ji, PPPL)

9:00 am – Magnetic Reconnection (Chair: James Drake, U. Maryland)

James Drake, U. Maryland: Introduction – 15 min.

Masaaki Yamada, PPPL: The Rate of Reconnection – 15 min.

Jan Egedal, MIT: The Onset Problem – 15 min.

William Daughton, LANL: Cross Scale Coupling in Large Systems – 15 min.

Dmitri Uzdensky, U. Colorado: Reconnection in Extreme Environments – 15 min.

James Drake, U. Maryland: Heating and Particle Acceleration – 15 min.

James Drake and Masaaki Yamada: Moderated Discussions Including A Possible National Reconnection Experiment – 30 min.

11:00 am – Break and Registration

11:30 pm – Waves and Turbulence (Chair: Amitava Bhattacharjee, U. New Hampshire)

Amitava Bhattacharjee, U. New Hampshire: Introduction and Overview – 10 min.

Stanislav Boldyrev, U. Wisconsin – Madison: Nature and Properties of Turbulence Cascades – 20 min.

William Matthaeus, U. Delaware: Dissipation and Particle Acceleration and Heating – 20 min.

12:20 pm Lunch – PPPL Cafeteria

1:20 pm – Waves and Turbulence (Continued)

Troy Carter, UCLA: Roles of Laboratory Experiments and Observations – 15 min.

Amitava Bhattacharjee, U. New Hampshire: Turbulence in Inhomogeneous Plasmas – 15 min.

Amitava Bhattacharjee, U. New Hampshire: Connections to Other Topics – 10 min.

Amitava Bhattacharjee, U. New Hampshire: Moderated Discussions – 30 min.

2:30 am – Collisionless Shock and Particle Acceleration (Chair: Martin Lee, U. New Hampshire)

Martin Lee, U. New Hampshire: Introduction – 12 min.

Randy Jokipii, U. Arizona: Heliospheric Shocks – 12 min.

Robert Lin, UC Berkeley: Shocks in Corona and Solar Energetic Particles – 12 min.

Anatoly Spitkovsky, Princeton U.: Supernova Remnant and Other Shocks – 12 min.

Ramanath Cowsik, Washington U. St. Louis: Acceleration and Propagation of Galactic Cosmic Rays – 12 min.

3:30 pm Break

## WORKSHOP AGENDA (continued)

4:00 pm – Collisionless Shock and Particle Acceleration (continued)

David Burgess, Queen Mary U. London: Shock Simulations – 15 min.

Christoph Niemann, UCLA: Laboratory Shock Experiments – 15 min.

Martin Lee, U. New Hampshire: Moderated Discussions – 30 min.

5:00 pm – Public Comments (Chair: Masaaki Yamada, PPPL)

5:30 pm – Breakout Discussions

Magnetic Reconnection – Room B318

Collisionless Shock – Director's Conference Room (3rd floor)

Waves and Turbulence – Theory Conference Room

Magnetic Dynamos – Room A118

Interface and Shear Instabilities – Auditorium

Angular Momentum Transport – Display Wall Room

Dusty Plasmas – Room S213

Radiative Hydrodynamics – Procurement Conference Room

Relativistic Plasmas – Room B233

Jets – Room B252

7:00 pm – Reception and Dinner at Salt Creek Grille – Princeton

## TUESDAY, JANUARY 19

8:00 am – Magnetic Dynamos (Chair: Ellen Zweibel, U. Wisconsin – Madison)

Ellen Zweibel, U. Wisconsin – Madison: Introduction and Observations – 30 min.

Cary Forest, U. Wisconsin – Madison: Laboratory Experiments – 20 min.

Fausto Cattaneo, U. Chicago: Theory and Numerical Simulations – 20 min.

Eric Blackman, U. Rochester: Broader Impacts – 20 min.

Ellen Zweibel, U. Wisconsin – Madison: Moderated Discussions – 30 min.

10:00 am Break

10:30 am – Interface and Shear Instabilities (Dmitri Ryutov, LLNL)

Ian Mann, U. Alberta: Space Plasma Prospective – 20 min.

Marc Pound, U. Maryland: Astrophysical Prospective – 20 min.

Aaron Miles, LLNL: Supernovae and Supernova Remnants – 10 min.

Carolyn Kuraz, U. Michigan: High-Energy-Density Laboratory Experiments: – 20 min.

Uri Shumlak, U. Washington: Astrophysical Connections from Magnetic Confinement Research – 20 min.

Dmitri Ryutov, LLNL: Moderated Discussions – 30 min.

## WORKSHOP AGENDA (continued)

12:30 pm Lunch – PPPL Cafeteria

1:30 pm – Angular Momentum Transport (Chair: Eliot Quataert, UC Berkeley)

Eliot Quataert, UC Berkeley: Introduction and Overview – 25 min.

James Stone, Princeton U.: Scientific Opportunities on Accretion Disks – 20 min.

Matthew Browning, Canadian Inst. Theoretical Astrophysics: Scientific Opportunities on Stars – 20 min.

Mark Nornberg, U. Wisconsin – Madison: Laboratory Opportunities – 15 min.

Eliot Quataert, UC Berkeley: Relations to Other Topics and Broader Impacts – 10 min.

3:30 pm Break

4:00 pm – Dusty Plasmas (Chair: Edward Thomas, Auburn U.)

Edward Thomas, Auburn U.: Introduction – 15 min.

Astrophysical Issues – 20 min.

Opportunities and Summary – 15 min.

Moderate discussions – 30 min.

5:30 pm – Public Comments (Chair: Greg Hammett, PPPL)

7:00 pm – Banquet at Prospect House – Princeton University

## WEDNESDAY, JANUARY 20

8:00 am – Radiative Hydrodynamics (Chair: Bruce Remington, LLNL)

Julian Krolik, John Hopkins U.: Theory and Astrophysical Relevance – 30 min.

Patrick Hartigan, Rice U.: Observational Relevance – 30 min.

Bruce Remington, LLNL: Experimental Relevance – 30 min.

Bruce Remington, LLNL: Moderated Discussions – 30 min.

10:00 pm Break

10:30 am – Relativistic, Ultra-strongly Magnetized, and Pair Plasmas (Chair: Edison Liang, Rice U.)

Edison Liang, Rice U.: Introuduction and Overview – 12 min.

Luis Silva, Lisbon U.: Relativistic Beams: Generation, Dissipation and Connection to Shock Physics – 15 min.

Masahiro Hoshino, U. Tokyo: Reconnection in Relativistic and Strongly Magnetized

Plasmas and Their Radiation – 15 min.

Anatoly Spitkovsky, Princeton U.: Numerical Simulations of Collisionless Shocks – 12 min.

Igor Kaganovich, PPPL: Weibel Instability and Collisionless Shocks – 12 min.

Andrey Timokhin, UC Berkeley: Numerical Models of Pulsar Magnetosphere – 12 min.

Patric Muggli, USC: High Brightness Particle Beams – 12 min.

Edison Liang, Rice U.: Moderated Discussions – 30 min.

## WORKSHOP AGENDA (continued)

12:30 pm Lunch – PPPL Cafeteria

1:30 pm – Jets and Outflows Including Structure Formation (Chair: Hui Li, LANL)

Phil Kronberg, LANL: Observations of AGN Jets – 15 min.

Richard Lovelace, Connell U.: Theory and Simulations Black Hole Jets – 15 min.

Sean Matt, U. Virginia: Jets from Young Stellar Objects – 20 min.

Paul Bellan, Caltech: Jet Experiments – 20 min.

Hui Li, LANL: Jet Impact on Structure Formation, Key Issues and Connections to Other Topics – 20 min.

Hui Li, LANL: Moderated Discussions – 30 min.

3:30 pm Break

4:00 pm – Public Comments (Chair: Jay Johnson, PPPL)

4:30 pm – Breakout Discussions

## THURSDAY, JANUARY 21

8:00 am – Updates from Each Topic (Chair: Jeremy Goodman, Princeton U.)

James Drake, U. Maryland: Magnetic Reconnection – 15 min.

Martin Lee, U. New Hampshire: Collisionless Shock and Particle Acceleration – 15 min.

Amitava Bhattacharjee, U. New Hampshire: Waves and Turbulence – 15 min.

Ellen Zweibel, U. Wisconsin – Madison: Magnetic Dynamos – 15 min.

Dmitri Ryutov, LLNL: Interface and Shear Instability – 15 min.

Eliot Quataert, UC Berkeley: Angular Momentum Transport – 15 min.

Edward Thomas, Auburn U.: Dusty Plasmas – 15 min.

Bruce Remington, LLNL: Radiative Hydrodynamics – 15 min.

Edison Liang, Rice U.: Relativistic, Ultra-strongly Magnetized, and Pair Plasmas – 15 min.

Hui Li, LANL: Jets and Outflows Including Structure Formation – 15 min.

10:30 am Break

11:00 am – Cross-Cutting Physics Themes and Research Tools (Chair: Hantao Ji, PPPL)

12:00 pm – Discussions on Next Steps (Chair: Stewart Prager, PPPL)

1:00 pm Lunch – PPPL Cafeteria

# APPENDIX D: WORKING GROUP MEMBER TABLE

| CHAPTER 1 Magnetic Reconnection | CHAPTER 2 Collisionless Shocks and Particle Acceleration | CHAPTER 3 Waves & Turbulence | CHAPTER 4 Magnetic Dynamos | CHAPTER 5 Interface & Shear Instabilities | CHAPTER 6 Angular Momentum Transport | CHAPTER 7 Dusty Plasmas | CHAPTER 8 Radiative Hydro-dynamics | CHAPTER 9 Relativistic, Pair-dominated and Strongly Magnetized Plasmas | CHAPTER 10 Jets and Outflows |
|---|---|---|---|---|---|---|---|---|---|
| J. Drake Maryland (lead) | M. Lee New Hampshire (lead) | A. Bhattacharjee New Hampshire (lead) | E. Zweibel Wisconsin (lead) | D. Ryutov LLNL (lead) | E. Quataert Berkeley (lead) | E. Thomas Auburn (lead) | B. Remington LLNL (lead) | E. Liang Rice (lead) | H. Li LANL (lead) |
| S. Antiochos GSFC | R. Jokipii Arizona (co-lead) | S. Bale Berkeley (co-lead) | F. Cattaneo Chicago (co-lead) | co-lead M. Pound Maryland | M. Browning CITA (Toronto) | L. Matthews Baylor | J. Bailey SNLA | J. Arons Berkeley | P. Bellan Caltech |
| W. Daughton LANL | T. Bell Oxford, UK | S. Boldyrev Wisconsin | E. Blackman Rochester | C. Kuranz Michigan | G. Hammett PPPL | R. Merlino Iowa | P. Hartigan Rice | M. Baring Rice | J. Eilek NM Tech |
| J. Egedal MIT | D. Burgess Queen Mary, UK | T. Carter UCLA | C. Forest Wisconsin | I. Mann Alberta, Canada | M. Nornberg Wisconsin | M. Rosenberg UCSD | R. Heeter LLNL | C. Dermer NRL | T. Jones Minnesota |
| A. Lazarian Wisconsin | R. Cowsik Washington, St. Louis | S. Cranmer CfA | G. Novak Chicago | A. Miles LLNL | J. Stone Princeton | P. Song UML | P. Hoeflich Florida State | M. Hoshino Tokyo | J. Kasper CfA |
| R. Lin Berkeley | T. Intrator LANL | P. Diamond UCSD | A. Pouquet NCAR | U. Shumlak U. of Washington | | | J. Hughes Rutgers | K. Krushelnick Michigan | P. Kronberg LANL |
| T. Phan Berkeley | R. Lin Berkeley | B. Dorland Maryland | J. Sarff Wisconsin | | | | J. Krolik JHU | Y. Sentoku U Nevada | S. Lebedev Imperial College |
| D. Uzdensky Colorado | C. Niemann UCLA | P. Goldreich IAS | | | | | | L. Silva Lisbon | R. Lovelace Connell |
| M. Yamada PPPL | A. Spitkovsky Princeton | W. Matthaeus Delaware | | | | | | M. Velli JPL | S. Matt Virginia |